\documentclass[twoside,12pt]{article}
\usepackage{epsfig}
\usepackage{indentfirst}
\usepackage{multirow}

\def\Journal#1#2#3#4{{#1} {#2} (#4) #3 }

\def\NPA{{\em Nucl. Phys.} A}

\def\PLB{{\em Phys. Lett.} B}

\def\PRL{\em Phys. Rev. Lett.}

\def\PRA{{\em Phys. Rev.} A}

\def\PRC{{\em Phys. Rev.} C}
\def\PRB{{\em Phys. Rev.} B}

\newcommand{\be}{\begin{equation}}
\newcommand{\ee}{\end{equation}}
\newcommand{\bea}{\begin{eqnarray}}
\newcommand{\eea}{\end{eqnarray}}
\newcommand{\nn}{\nonumber}
\begin{document}

\title{ \vspace{1cm} Effective density functionals beyond mean field}
\author{Marcella Grasso,$^{1}$ \\
\\
$^1$Institut de Physique Nucl\'eaire, CNRS-IN2P3, \\ 
Universit\'e Paris-Sud,
Universit\'e Paris-Saclay, 91406 Orsay, France}
\maketitle
\begin{abstract} 
I present a review on non relativistic effective energy--density functionals (EDFs). An introductory part is dedicated to traditional phenomenological functionals employed for mean--field--type applications and to several extensions and implementations that have been suggested over the years to generalize such functionals, up to the most recent ideas. 
The heart of this review is then focused on density functionals designed for beyond--mean--field models. Examples of these studies are discussed. Starting from these investigations, some illustrations of {\it{ab--initio}}--based or  
{\it{ab--initio}}--inspired functionals are provided. Constructing functionals by  building bridges with 
{\it{ab--initio}} models 
represents an extremely challenging and timely objective. This 
will eventually reduce/eliminate the empirical character of EDFs and link them with the underlying theory of QCD. 
Conclusions are presented in the last part of the review.  
\end{abstract}
\tableofcontents

\section{Introduction}
The nuclear energy--density--functional (EDF) theory globally provides a  satisfactory frame for the description of several properties of finite nuclei (covering the whole nuclear chart), of nuclear matter (from symmetric to pure neutron matter at densities close to the saturation point), as well as of nuclear systems located in the crust of neutron stars. Structure and reaction analyses are successfully performed since decades with traditional effective density functionals. 

Starting from their birth in the 70s, with the first mean--field (MF) calculations carried out for the ground state of spherical nuclei, nuclear EDF theories have evolved over time following two main directions: on the one hand, a continuous and intense effort was devoted to generalize and implement the employed effective density functionals. Several directions explored along this direction will be illustrated in the following sections. On the other hand, the necessity to formulate and develop more sophisticated beyond--mean--field (BMF) models has become over the years a clear evidence. Nowadays, several EDF BMF models are available for structure and reaction applications. Mutual exchanges between practitioners working on each of these two directions (functionals and models) have been extremely beneficial and have led to significant progress and achievements in both sectors. 
These mutual exchanges have also opened the doors to a new challenging direction which represents the main topic of the present review, that is the generalization of EDF functionals and the introduction of a new generation of functionals designed for BMF applications. Related to this, the construction of {\it {ab--initio}}--based or {\it{ab--initio}}--inspired  EDF functionals became an important goal for rendering EDF functionals less empirical, linking them with the underlying theory of QCD, and, correspondingly, increasing their reliability for predicting properties of nuclei far from observation. 

In the past years, special 
attention was also devoted to clarify and rigorously identify the existing links between nuclear EDF theories and the density functional theory (DFT) developed for electronic many--body systems and employed since decades in quantum chemistry and solid--state physics.  

The present review focuses in particular on non--relativistic EDF developments and does not cover the achievements and progress obtained within covariant EDF theories. 

The review is organized as follows. Section \ref{edfmf} provides a general overview of EDF theories based on the MF approximation and presents the most currently used traditional phenomenological functionals as well as some illustrations of new ideas which were recently proposed for generalizing the functionals in the MF context. Section \ref{edfbmf} deals with BMF EDF. After a schematic overview on several available BMF models, the importance of going beyond traditional effective functionals is underlined. Some examples of new--generation effective functionals tailored for BMF models are given in Sec. \ref{newgene} for nuclear matter and finite nuclei. The last topic addressed in this review is the link between EDF and {\it {ab--initio}} models. New functionals inspired from effective field theories (EFTs) are presented in Sec. \ref{abinitio}. Applications to nuclear matter and neutron drops are illustrated. 
Other examples of {\it{ab--initio}}--inspired functionals are also discussed. 
Finally, a summary is presented in Sec. \ref{conclu}.  

\section{The nuclear energy--density--functional (EDF) theory based on the mean--field (MF) approximation \label{edfmf}}
\subsection{\it General overview and links with the density functional theory (DFT) \label{over}}
\begin{figure}[tb]
\begin{center}
\begin{minipage}[t]{8 cm}
\epsfig{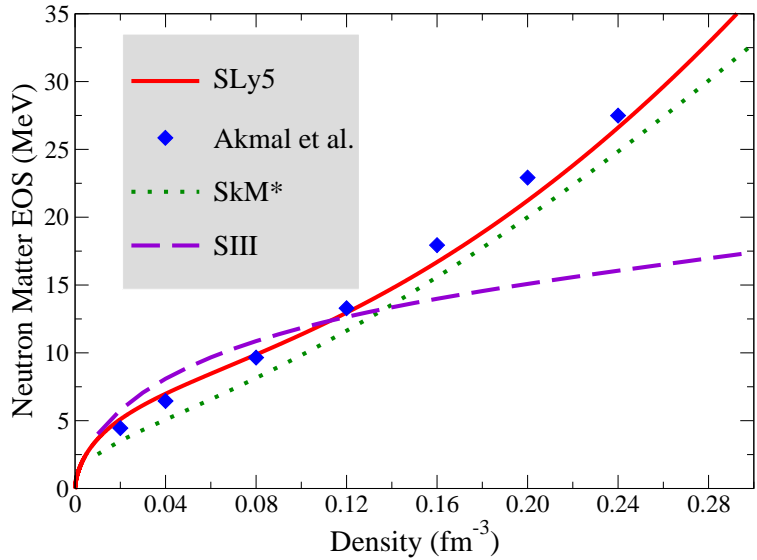}
\end{minipage}
\begin{minipage}[t]{16.5 cm}
\caption{Skyrme MF EOSs for neutron matter obtained with the parametrizations SLy5 (solid red line), SkM* (dotted green line), and SIII (dashed violet line) compared to the EOS by Akmal et al (blue diamonds) \cite{akmal}. \label{akmaleos}}
\end{minipage}
\end{center}
\end{figure}

The most currently employed effective interactions in EDF theories are the 
phenomenological Skyrme \cite{sk1,sk2,vau} and Gogny \cite{gogny1,gogny2} forces. 
The first nuclear MF calculations employing such  effective interactions were
 carried out in the 70s for the description of the ground state of even--even closed--shell nuclei in spherical symmetry. From these effective interactions, density functionals were derived in the MF approximation. For each of the two effective forces, around ten parameters were adjusted with MF calculations to reproduce chosen observables of selected nuclei. This resulted in the birth of the traditional EDF theory where the parameters of the used functionals are adjusted empirically, for example on binding energies and radii of chosen nuclei, on selected single--particle splittings, as well as on the equation of state (EOS) of nuclear matter and on some of its properties, such as the saturation point of symmetric matter. As a first step, only properties of symmetric nuclear matter were used in the fitting protocols. Later, also the EOS of neutron matter was included in the adjusting procedures. 
In the 90s, with the advent of new projects for the construction of next--generation facilities (allowing for the 
production of more exotic isotopes), 
 this was done with the clear objective of improving the quality of the functionals in the treatment of neutron--rich nuclei \cite{lyon1,lyon2,lyon3,d1n}. 
As an illustration of this, Fig. \ref{akmaleos} shows three MF Skyrme EOSs computed for neutron matter and compared to the microscopic Akmal et al. EOS of Ref. \cite{akmal}. One observes that the most recent SLy5 
parametrization \cite{lyon1,lyon2,lyon3} is the one leading to the most satisfactory EOS for neutron matter whereas the SkM* \cite{skm} curve systematically sligthly overbinds neutron matter and the SIII \cite{siii} EOS strongly deviates from the Akmal results at densities larger than 0.12 fm$^{-3}$. 

In the 70s, a new era started in nuclear physics where EDF theories were extensively used and were able to globally provide a very satisfactory agreement in reproducing numerous experimental data. Reference \cite{negele82} is a review containing the achievements of the first decade of EDF applications in nuclear structure and dynamics. It should be mentioned that the first time--dependent EDF models were applied in Ref. \cite{bonche} and the first EDF applications in astrophysics to treat the nuclear systems located in the crust of neutron stars were done in the pioneering work of Negele and Vautherin \cite{negelevautherin}.  
In the successive decades, more sophisticated BMF models, being able to explicitly account for correlations between nucleons (compared to the simpler MF approaches) were also developed in the EDF framework, employing in general the same density functionals used for MF calculations. Reference \cite{bender} is a very exhaustive review of the EDF theory and its applications to ground and excited states. 

A peculiarity of both Skyrme and Gogny interactions is the presence in their expressions of a zero--range two--body density--dependent term which is repulsive and which is important for correctly describing binding energies and radii of finite nuclei, the effective mass of matter, as well as the equilibrium point of symmetric matter. This term was for the first time included by Vautherin and Brink in Ref. \cite{vau} to replace the zero--range three--body term present in the original version of the Skyrme interaction \cite{sk1,sk2}. 
It was first included with a power of the density equal to 1. 
The interaction was in this way strongly simplified and the use of a two--body force instead of the original three-body term made feasible  in practice MF calculations for several spherical nuclei. This density dependence is a crucial point in EDF theories. The presence of a density dependence implies that this type of effective interactions cannot rigorously represent the interaction term of a Hamiltonian. This also means that we cannot 
rigorously speak about Hartree-Fock (HF) or Hartree-Fock-Bogoliubov (HFB) models in the MF description of nuclear ground states in those cases where such functionals are used. 
A related point is also the (possible) fractional power in the density--dependent term. Most of the existing parametrizations have indeed a fractional--power density dependence which leads to a better description of nuclear properties and of the EOS of nuclear matter, for instance of the nuclear incompressibility modulus in nuclear matter and of the breathing compression excited modes in finite nuclei \cite{blaizot}. The density dependence and its possible fractional power can however generate several instabilities and pathologies in BMF calculations \cite{be,du,la,doba} as will be discussed later.  

Another type of effective interaction also containing a zero--range density--dependent term (and which is less currently employed) was obtained by Nakada from the effective M3Y-Paris interaction \cite{bertsch} including a density--dependent term and modifying some parameters \cite{nakada1,nakada2}.  

The MF equations which are solved for nuclear systems in EDF theories show very strong similarities with the Kohn-Sham equations \cite{ks} used in condensed matter physics and in chemistry in the DFT theoretical scheme \cite{dreizler,parr,tsuneda}. The DFT was formulated for many--electron systems and is based on the Hohenberg-Kohn theorem \cite{hok1,hok2}. Despite such analogies between these two areas of the many--body physics it is important however to stress that there are fundamental differences. 
First, in the EDF framework most of the density functionals are derived from a Hamiltonian (we still call it Hamiltonian despite the ambiguity generated by the density--dependent term). Some exceptions exist and work in the spirit of a DFT--like theory was done by the community as will be discussed in the following in this review. 

Another important difference consists in the fact that the Hohenberg-Kohn theorem was demonstrated for a system localized by an external potential whereas atomic nuclei are self--bound systems. The DFT theory cannot thus be directly applied to finite nuclei and extensions are required. For this reason, ten years ago there has been an intense effort in the community with the objective of extending the Hohenberg-Kohn theorem to the nuclear case \cite{engel,giraud1,giraud2,giraud3,messud1,messud2} and it was shown that the theorem can be demonstrated for the so--called internal or intrinsic densities.  

Finally, symmetry breaking (generating correlations at the MF level) and the restoration of broken symmetries are extensively used in the EDF framework whereas these concepts are much less employed in the DFT theory. 
There are however studies along this direction which are carried out for example in quantum chemistry. Symmetries were analyzed for instance in the framework of  the so--called constrained--path Monte Carlo approach \cite{shi2013}. Symmetry restoration through projection techniques was discussed in 1D \cite{rodri2013} and 2D \cite{rodri2012,rodri2014} Hubbard models  and for molecular systems \cite{scuse2011,jimenez2012}.  

 A symmetry is broken when the functional does not have all the symmetries of the Hamiltonian,  for example in the description of superfluidity in open--shell nuclei. This is achieved by including pairing correlations in the MF scheme, for instance with HFB or HF + BCS models. This is an example of spontaneous symmetry breaking (U(1) symmetry in gauge space) which corresponds to a ground state having a not well defined particle number (the number of particles is conserved only on average).  
Restoration of broken symmetries is carried out in most cases in the framework of BMF models where additional correlations with respect to the leading order of the Dyson expansion (MF--type truncation of the many--body perturbation theory) \cite{fetter} are also included.   

To conclude this subsection, it is worth mentioning Ref. \cite{drut} where the first steps and the open problems in the attempts of constructing an {\it{ab--initio}} nuclear DFT were reported. In particular, the relevance of this ambitious work for improving existing empirical functionals was undelined. This will be discussed in Sec. \ref{abinitio}.

\subsection{\it Traditional phenomenological functionals (and beyond) for MF applications \label{mffunc}}

\subsubsection{\it Traditional functionals}

In its currently used form, the zero--range Skyrme interaction reads
\bea
\nn
V(\bf{r_1},\bf{r_2})&=& t_0(1+x_0 P_{\sigma}) \delta({\bf{r}}) + \frac{1}{6}t_3
(1+x_3P_{\sigma}) \rho(\bf{R})^{\alpha}  \delta(\bf{r}) \\
&+& \frac{1}{2}  t_1(1+x_1 P_{\sigma}) [{\bf{P}}^{'2} \delta({\bf{r}}) +  \delta({\bf{r}}) {\bf{P}}^{2} ] +   t_2 (1+x_2 P_{\sigma}) \bf{P}^{'} \cdot \delta(\bf{r}) \bf{P} \\
\nn
&+& iW_0 {\bf{\sigma}} \cdot [ \bf{P}^{'} \times \delta(\bf{r}) \bf{P} ],
\label{skyrmeint}
\eea 
where the second term of the first line describes the density--dependent contribution, the second line contains the $s$--wave ($t_1$) and $p$--wave ($t_2$) velocity--dependent or gradient terms, and the third line is the spin--orbit contribution. The used notation is: 
\bea
\nn
{\bf{r}}&=&{\bf{r_1}}-{\bf{r_2}}, \; {\bf{R}}=\frac{1}{2} ({\bf{r_1}}+{\bf{r_2}}), \\
{\bf{P}}&=&\frac{1}{2i}({\bf{\nabla_1}}-{\bf{\nabla_2}}), \\
\nn
\sigma&=&\sigma_1+\sigma_2, \; P_{\sigma}=\frac{1}{2}(1+\sigma_1 \cdot \sigma_2).
\label{conven}
\eea
In the adopted notation, $\bf{P}^{'}$ is the complex conjugate (acting on the left) of ${\bf{P}}$, $\sigma$ are the spin matrices, and $P_{\sigma}$ is the spin--exchange operator; $t_i$, $x_i$ ($i=0,1,2,3$), the power of the density dependence $\alpha$, and the spin--orbit strength $W_0$ are parameters to adjust (10 parameters in total). 

The Gogny interaction in its currently used form reads
\bea
\nn
V(\bf{r_1},\bf{r_2})&=& \sum_{i=1,2} [W_i+B_i P_{\sigma}-H_i P_{\tau} - M_i P_{\sigma} 
P_{\tau}] e^{-r^2/\mu_i^2} \\
&+&t_3
(1+x_3P_{\sigma}) \rho(\bf{R})^{\alpha}  \delta(\bf{r}) \\
\nn
&+& iW_0 {\bf{\sigma}} \cdot [ \bf{P}^{'} \times \delta(\bf{r}) \bf{P} ], 
\label{gognyint}
\eea 
where, compared to the Skyrme force, the $t_0$ and the velocity--dependent ($t_1$ and $t_2)$ parts are replaced by a central finite--range contribution described by two gaussians with different ranges. $P_{\tau}$ is the isospin--exchange operator and $W_i$, $B_i$, $H_i$, $M_i$, $\mu_i$ ($i=1,2$), $t_3$, $x_3$, $\alpha$, and $W_0$ are adjustable parameters (14 parameters in total).
When superfluid nuclei are treated, the same interaction is used in the Gogny case for both the MF and the pairing channels of the HFB equations whereas, for the Skyrme case, the Skyrme parametrization is used only in the MF channel and a different interaction is employed in the pairing channel, in most cases a density--dependent zero--range interaction. Some exceptions exist for the Skyrme case  \cite{skp}. 

From these interactions, functionals may be derived. As an illustration, in the MF approximation, where the ground state is a Slater determinant and only the first order of the Dyson equation is taken into account, the density functional (also called Hamiltonian density) corresponding to the first line of the Skyrme force illustrated in Eq. (1) reads 
\bea
\mathcal{H}_{t_0} + \mathcal{H}_{t_3} = \frac{1}{4}t_0[(2+x_0)\rho^2-(2x_0+1)
(\rho_p^2+\rho_n^2)] 
+  \frac{1}{24}t_3 \rho^{\alpha} [(2+x_3)\rho^2-(2x_3+1)
(\rho_p^2+\rho_n^2)], 
\eea
where $p$ and $n$ stand for proton and neutron, respectively, and $\rho$ is the total nucleon density, sum of the proton $\rho_p$ and neutron $\rho_n$ densities. The total energy of the system associated with this part of the functional is calculated as 
\be
\nn
E_{(t_0,t_3)}= \int (\mathcal{H}_{t_0} + \mathcal{H}_{t_3} ) d^3{\bf{r}}.
\ee
The Skyrme functional is in general a functional of the nucleon density $\rho({\bf{r}})$, the 
kinetic energy density $\tau({\bf{r}})$, and the spin--orbit density ${\bf{J}}({\bf{r}})$ \cite{rs,meyer}. For most of the existing Skyrme parametrizations,  the terms of the functional depending on the square of the spin--orbit density (the so--called ${\bf{J}}^2$ terms) \cite{meyer} were neglected when the parameters were adjusted and have thus to be excluded when MF calculations are carried out. In some cases, for instance for the SLy5 parametrization, these terms were included when the parameters were fitted and have thus to be taken into account in practical calculations. 

In the MF approximation, the Skyrme functional leads to the following EOS for symmetric matter, 
\begin{equation}
\frac{E_{sym}}{A}=\frac{3}{10}\frac{\hbar ^{2}}{m}\left( \frac{3\pi
^{2}}{2}\right) ^{\frac{2}{3}}\rho ^{\frac{2}{3}}+\frac{3}{8}t_{0}\rho +%
\frac{3}{80}\left( \frac{3\pi ^{2}}{2}\right) ^{\frac{2}{3}}\Theta _{S}\rho
^{\frac{5}{3}}+\frac{1}{16}t_{3}\rho ^{\alpha +1},
\label{eos1sm}
\end{equation}%
where $\Theta _{S}=3t_{1}+t_{2}(5+4x_{2})$ and $A$ is the number of nucleons. For neutron matter, the
MF EOS has the form
\begin{equation}
\frac{E_{neutr}}{N}=\frac{3}{10}\frac{\hbar ^{2}}{m}\left( 3\pi
^{2}\right) ^{\frac{2}{3}}\rho ^{\frac{2}{3}}+\frac{1}{4}t_{0}\rho (1-x_{0})+%
\frac{3}{40}\left( 3\pi ^{2}\right) ^{\frac{2}{3}}(\Theta _{S}-\Theta
_{V})\rho ^{\frac{5}{3}}+\frac{1}{24}t_{3}\rho ^{\alpha +1}(1-x_{3}),
\label{eos1nm}
\end{equation}%
where $\Theta _{V}=t_{1}(2+x_{1})+t_{2}(2+x_{2})$ and $N$ is the number of neutrons.  

A tensor non--central zero--range interaction appears in the original version of the Skyrme force \cite{sk1,sk2} but was in practice neglected in almost all the MF calculations carried out up to ten/fifteen years ago, with some exceptions \cite{stancu77}. 
In more recent applications, the Skyrme interaction was enriched by this contribution (that modifies the functional by new terms depending on the spin--orbit density), either by adjusting the tensor parameters on top of existing Skyrme parametrizations \cite{colo, brink,zou,brown,moreno,grasso2013,grasso2015} or by globally fitting new parametrizations \cite{lesi1,lesi2}. In Ref. \cite{grasso2013} the spin--orbit parameter was also readjusted and, in addition,  
Gogny--type interactions were considered. This fitting procedure was inspired by Refs. \cite{zale1,zale2}. 

For finite--range interactions, a Yukawa tensor term is included in the forces proposed by Nakada \cite{nakada1,nakada2}. A first attempt to include a gaussian tensor--isospin term in the Gogny interaction was done by Otsuka and collaborators \cite{otsu} (all the parameters of the force were refitted). A finite--range tensor--isospin term was also inluded on top of existing Gogny parametrizations by Co' and collaborators \cite{co}. In a more recent work, inspired by the study of Onishi and Negele \cite{oni}, a pure tensor term of gaussian form was added to the Gogny force together with a tensor--isospin term \cite{angu2012}. The neutron--proton and like--particle tensor effects could be in this way separately identified. 
The reader may refer to Ref. \cite{sagawa2014} for a review on the tensor force in MF--based calculations. 

Other modifications of traditional functionals have been proposed over time. We mention for instance the extension allowing for two parameters instead of one in the spin--orbit part of the Skyrme interaction for the correct description of the so--called isotope shifts in the Pb region. A different dependence on the neutron and proton densities was introduced in the spin--orbit potential compared to the standard Skyrme case \cite{rei}.
 
The most recent ideas for generalizations of functionals in the MF context are illustrated in the following subsections. 

\begin{figure}[tb]
\begin{center}
\begin{minipage}[t]{8 cm}
\epsfig{file=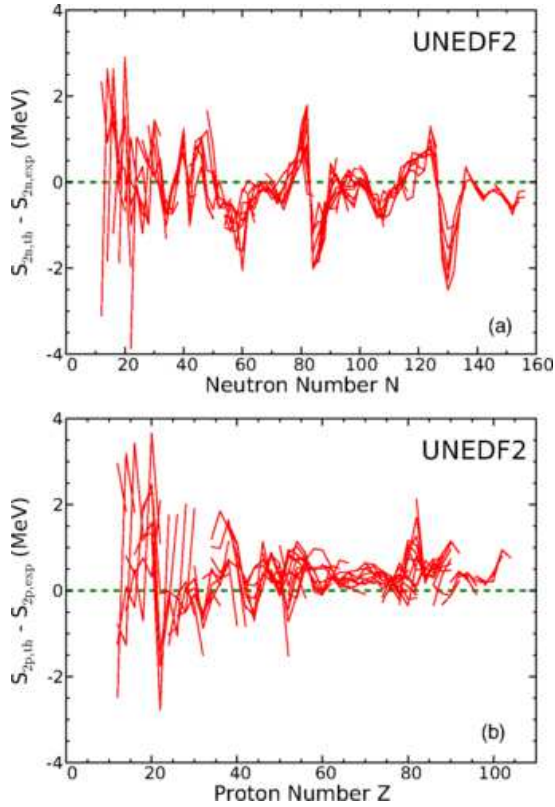,scale=0.6}
\end{minipage}
\begin{minipage}[t]{16.5 cm}
\caption{Difference between the theoretical and the experimental two--neutron $S_{2n}$ (a) and two--proton $S_{2p}$ (b) separation energies. Taken from Ref. \cite{korteleinen2014}. \label{S2}}
\end{minipage}
\end{center}
\end{figure}

Recently, an intense work was dedicated to the optimization procedures of Skyrme--type interactions and their impact on the predictive power, 
 in the quest of universal functionals ideally suited for structure and dynamics studies \cite{bertsch2007,furn2011,nam2012,bogner2013,computing,fidipro,korteleinen2012,korteleinen2014}. 
Figure \ref{S2} displays for example one of the results of such optimization schemes that generated the UNEDF2 functional \cite{korteleinen2014}. The figure shows 
the difference between the theoretical predictions provided with the functional UNEDF2 and the experimental data for the two--neutron $S_{2n}$ and the two--proton $S_{2p}$ separation energies. The authors of Ref. \cite{korteleinen2014} observed that, in spite of the fact that new experimental constraints had been included in the optimization procedure, the quality of the UNEDF2 functional was only slightly enhanced compared to previous Skyrme--type parametrizations. Consequently, their main conclusion was that the standard Skyrme functional 
could not be further ameliorated and that 
new forms of functionals had to be  formulated and explored. 

\subsubsection{\it Recent ideas for generalizing functionals. Dealing with density--dependent terms  \label{3body}}

Let us first concentrate on the zero--range density--dependent term, present in both Skyrme and Gogny interactions. From a practical point of view, this term is definitely necessary in these functionals to correctly describe for instance the saturation point of symmetric matter. On the other side, it is known that, owing to this term, the Hamiltonian cannot be rigorously defined. Pathologies were analyzed in the restoration of broken symmetries (projection methods) within the EDF theory (see, for instance, Refs. \cite{be,du,la,doba}). It was shown that some of these pathologies are related to violations of the Pauli principle due to the spurious self--interaction produced by the density--dependent pseudo--Hamiltonian. It is interesting to note that  spuriosities related to the self--interaction are well known in the DFT framework \cite{pauli1,pauli2,pauli3} and had already been discussed long time ago in nuclear physics \cite{stringari}. Another source of problems is related to possible non--integer powers in the density dependencies \cite{du,doba}. 

\begin{figure}[tb]
\begin{center}
\begin{minipage}[t]{9 cm}
\epsfig{file=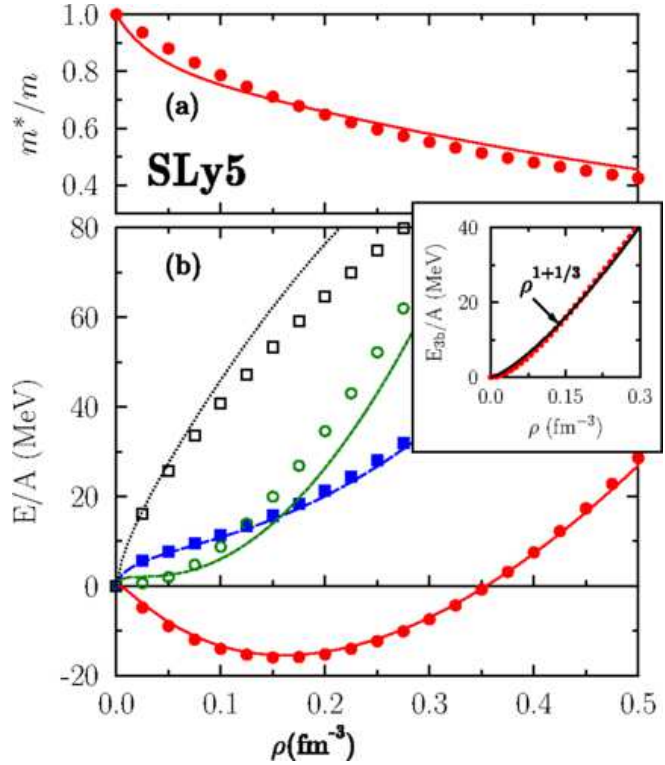,scale=0.57}
\end{minipage}
\begin{minipage}[t]{16.5 cm}
\caption{(a) Isoscalar effective mass as a function of the density. (b) EOSs for symmetric (red), neutron (blue), spin--polarized (green), and spin--polarized neutron (black) matter. The symbols describe the SLy5 benchmark curves. The full lines are the curves generated by the new functional introduced in Ref. \cite{labe}. The inset represents the contribution to the EOS of symmetric matter coming from the new three--body term. The dotted line is a curve of the type $\rho^{4/3}$ adjusted on it (see text). 
Taken from Ref. \cite{labe}.  \label{semicontact}}
\end{minipage}
\end{center}
\end{figure}

All this will be discussed in Subsec. \ref{patho2} where concrete examples of such pathologies (finite steps and divergences) will be illustrated and specific solutions for regularizing them will be presented (for instance, the work of Refs. \cite{la,hupin}). 
Apart from such {\it{ad--hoc}} regularization procedures tailored for the BMF context, 
other more general directions to overcome the problems of dealing with density--dependent terms were proposed in the literature, not specifically designed for MF or BMF frameworks. In Ref. \cite{sadou}, for example,  
the zero--range density--dependent term is replaced by a much more complicated three--body contribution with up to two derivatives. In Ref. \cite{labe}, the zero--range density--dependent term is replaced by a semicontact three--body interaction and applications to nuclear matter are discussed.  
Figure \ref{semicontact} shows for example the results obtained using this semicontact three--body interaction for the isoscalar effective mass (a) and the EOSs (b) of symmetric, neutron, spin--polarized symmetric, and spin--polarized neutron matter. The symbols refer to the curves produced by the SLy5 parametrization, which are used as a benchmark for the adjustment of the parameters. The lines describe the curves generated by the new (refitted) interaction. One may notice in particular the good quality of the EOS of symmetric matter where the equilibrium point is correctly described despite the absence of a density--dependent term in the interaction. The inset of the figure shows the three--body contribution to the EOS of symmetric matter. A function of the form $\rho^{4/3}$ was adjusted on this curve (symbols) which mimics a density--dependent contribution with $\alpha=1/3$. 

\begin{figure}[tb]
\begin{center}
\begin{minipage}[t]{8 cm}
\epsfig{file=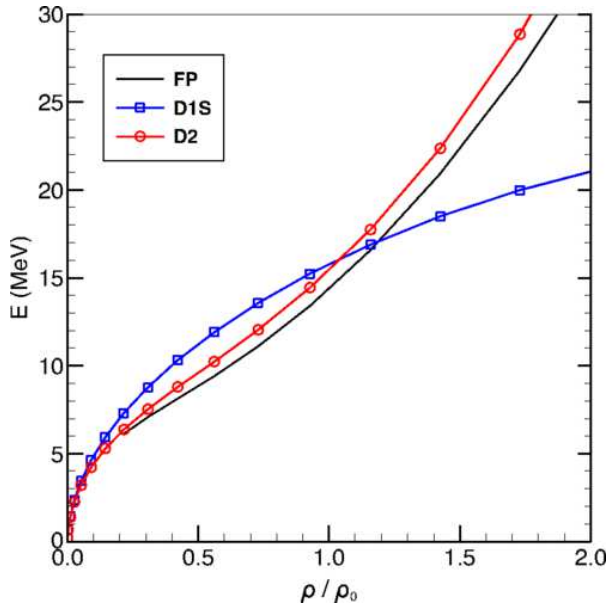,scale=0.45}
\end{minipage}
\begin{minipage}[t]{16.5 cm}
\caption{MF EOSs for neutron matter calculated with the parametrizations D2 and D1S. 
These curves are compared to the FP EOS of Ref. \cite{fp}.
Taken from Ref. \cite{chappert2015}. \label{cha}}
\end{minipage}
\end{center}
\end{figure}

In other studies, the accent was put on the zero range of the density--dependent term (instead of on its density dependence), which is responsible for ultraviolet divergences in many BMF models (Subsec. \ref{patho1}). For the Gogny case, this term and the spin--orbit contribution are the only two terms responsible for divergences in BMF applications due to their zero range (because all the other terms of the interaction have a finite range). Chappert and collaborators  \cite{chappert2015,pillet2017} proposed a new Gogny interaction, D2, where the density--dependent term has a finite range, 
\bea
V^{D2}_{dens}=(W_3 + B_3 P_{\sigma} - H_3 P_{\tau} -M_3 P_{\sigma} P_{\tau} )
\frac{e^{-\frac{({\bf{r_1}}-{\bf{r_2}})^2}{\mu_3^2}}}{(\mu_3 \sqrt{\pi})^3} \frac{\rho^{\alpha}(\bf{r_1})+ \rho^{\alpha}(\bf{r_2})}{2}.
\label{chapper}
\eea
The number of parameters of the density--dependent term becomes 6 (instead of 3 for the traditional zero--range form). In the fitting protocol, as for the previous parametrization D1N \cite{d1n}, also the EOS of neutron matter was included and the Friedman-Pandharipande (FP) EOS of Ref. \cite{fp} was used as a guide for the fit. Figure \ref{cha} shows the corresponding EOS for neutron matter compared to the older D1S curve \cite{berger} and to the microscopic results by Friedman and Pandharipande. One may observe that the quality of the neutron matter EOS is 
appreciably improved by the new interaction D2. 

\subsubsection{\it Recent ideas for generalizing functionals. Introducing higher--order derivatives of the density}
Several years ago, suggestions for generalizing functionals were proposed based on the density--matrix expansion (DME) \cite{car2008,car2010,car2010-2,doba2010,gebre2010,gebre2011,stoi2010,erler2010,rai2011}. 
The DME was originally introduced in the 70s by Negele and Vautherin \cite{negele72,negele75}. 
In Ref. \cite{car2008} a generalized functional was introduced constrained only by symmetry properties where the DME was used as a guide for formulating an expansion in orders of derivatives of the density: a functional containing derivatives of the density up to sixth order was constructed and called N$^3$LO EDF. 
This may be regarded as a work developed in the spirit of the DFT because such a functional was directly introduced and not deduced from an effective interaction. 
Later, the authors of Ref. \cite{rai2011} derived the corresponding expansion of the effective interaction (called pseudopotential) in relative momenta up to N$^3$LO. In this scheme where the order is defined in terms of the derivative terms present in the functional, a simple contact force corresponds to the leading order (LO) and the standard Skyrme force corresponds to NLO (up to second order in the derivatives of the density). As can be easily imagined, a huge number of parameters is present in the  N$^3$LO functional and the adjustment procedure of such coupling constants is extremely complicated.  Regularized pseudopotentials based on delta interactions were introduced in Ref. \cite{dob2012} (gaussian form for the regularization) and   finite--range N$^3$LO pseudopotentials were discussed in Ref. \cite{rai2014}. Due to their finite range, they have a natural cutoff at high momenta and are thus suited for possible future BMF calculations (see Subsec. \ref{patho1} where ultraviolet divergences in BMF applications owing to zero--range forces are discussed). 
Several numerical instabilities were however found in these first practical applications of DME--generated functionals \cite{stoi2010}. We will describe in the last part of this review a recent implementation were these limitations could be overcome and cured and a DME--generated functional constrained by local chiral interactions was presented \cite{navarro}.  

\begin{figure}[tb]
\begin{center}
\begin{minipage}[t]{11 cm}
\epsfig{file=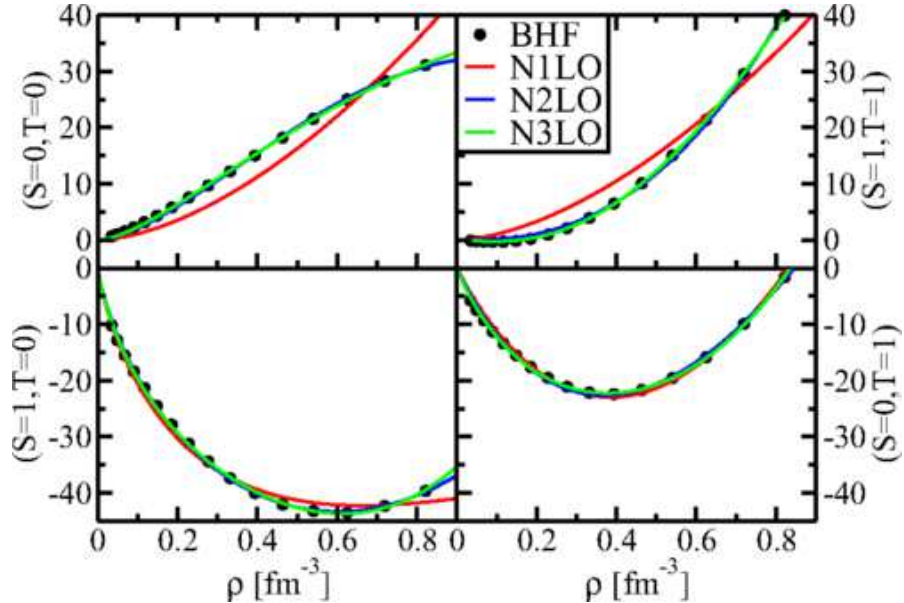,scale=0.67}
\end{minipage}
\begin{minipage}[t]{16.5 cm}
\caption{ Potential energies per particle $\nu^{(S,T)}$ (in MeV) for each $(S,T)$ channel adjusted order by order on BHF results of Ref. \cite{baldo}. 
Taken from Ref. \cite{davesne2015}. \label{bhf}}
\end{minipage}
\end{center}
\end{figure}

Inspired by these studies, 
new gradient terms in N$^l$LO pseudopotentials were also investigated in Ref. \cite{davesne2013}. 
In Ref. \cite{davesne2016} an alternative guide to construct such expansions (with respect to the guide provided by the DME) was discussed: it was shown that a momentum expansion of a finite--range interaction of any form leads to the N$^l$LO pseudopotentials. 
In a first exploratory study, the properties of nuclear matter were analyzed with N$^2$LO and N$^3$LO pseudopotentials and first sets of parameters were adjusted order by order to reproduce available {\it{ab--initio}} pseudodata \cite{davesne2015}. 
The N$^l$LO pseudopotential, denoted by $v_{Sk}$, was written as the sum of a central $v_C$, a tensor $v_T$, and a spin--orbit $v_{LS}$ term \cite{becker}, 
\be
v_{Sk}=v_C+v_T+v_{LS},
\label{carte}
\ee
where the central term $v_C$ contains the $t_0$, $x_0$, $t_1$, $x_1$, $t_2$, and $x_2$ standard Skyrme parameters. They are denoted as $t_0^{(0)}$, $x_0^{(0)}$ (LO), $t_1^{(2)}$, $x_1^{(2)}$, $t_2^{(2)}$, and $x_2^{(2)}$ (NLO). 
This term contains also four additional parameters for the  N$^2$LO pseudopotential, 
 $t_1^{(4)}$, $x_1^{(4)}$, $t_2^{(4)}$, $x_2^{(4)}$, and four additional parameters for the 
N$^3$LO pseudopotential, 
 $t_1^{(6)}$, $x_1^{(6)}$, $t_2^{(6)}$, $x_2^{(6)}$. 
The EOS of symmetric matter can be written as 
\be
\frac{E}{A}=\frac{3 \hbar^2}{10 m} k_F^2 + \sum_{S,T} \nu^{(S,T)}, 
\label{eosST}
\ee
where $k_F$ is the Fermi momentum and the functions $\nu^{(S,T)}$ are the potential energies per particle projected onto spin/isospin $(S,T)$ channels. 
As a first step, the authors of Ref. \cite{davesne2015} did not include any density--dependent term and could adjust the coupling constants of the N$^2$LO and N$^3$LO pseudopotentials to reproduce in a satisfactory way the four $\nu^{(S,T)}$ functions (using {\it{ab--initio}} curves as a benchmark \cite{baldo,hebeler,bogner,rotival}), but obtained a very low effective mass, $m^*/m=$ 0.4. 
By adding a density--dependent term of a standard Skyrme form and by readjusting all the parameters a more acceptable effective mass of 0.7 could be obtained. 
Figure \ref{bhf} shows the fits of the NLO (standard Skyrme),  N$^2$LO, and N$^3$LO pseudopotentials performed to reproduce the Br{\"u}ckner-HF (BHF) curves of Ref. \cite{baldo} (represented by the dots in the figure). The quantities which are displayed are the potential energies per particle $\nu^{(S,T)}$ of Eq. (\ref{eosST}) as a function of the density in the four $(S,T)$ channels.
It can be observed that these NLO adjustments are of much better quality compared to those presented in Ref. \cite{lesi1}. The reason is that additional constraints on properties of finite nuclei were also included in Ref. \cite{lesi1} whereas here only nuclear matter properties are taken into account in the fitting procedure. 

More recently, a first parametrization of the N$^2$LO pseudopotential, called SN2LO1, was introduced, where also properties of finite nuclei were included in the adjustment of the parameters \cite{becker2017}. One has to say that, despite the larger number of parameters, such a pseudopotential has a predictive quality which is comparable to that of the standard SLy5 functional. However, one important difference is that, whereas SLy5 presents an instability in the spin channel, this new parametrizations does not present any unphysical instabilities since it was derived by employing the so--called linear response formalism \cite{pastore}. 

\subsubsection{\it Ideas for generalizing functionals. Introducing DFT--like functionals}

Several ideas for generalizing nuclear functionals in the genuine spirit of DFT have been proposed and investigated over the last decades. Different forms of functionals were introduced in this way, strarting from the work of  
 Fayans \cite{fayans1,fayans2} who proposed a local EDF whose main contributions are a volume part for the description of homogeneous matter and a surface part for the treatment of finite nuclei. More recently, along a similar direction,  
Baldo et al. proposed the so--called BCPM (Barcelona-Catania-Paris-Madrid) functional \cite{baldo1,baldo2}. The starting point of this functional are BHF calculations for nuclear matter (both pure neutron and symmetric) \cite{baldo3}. Isovector pairing correlations are also included \cite{garrido} together with a spin--orbit term, a surface term, rotational energy corrections \cite{egido} and center--of--mass energy corrections \cite{butler}.  

\begin{figure}[tb]
\begin{center}
\begin{minipage}[t]{11 cm}
\epsfig{file=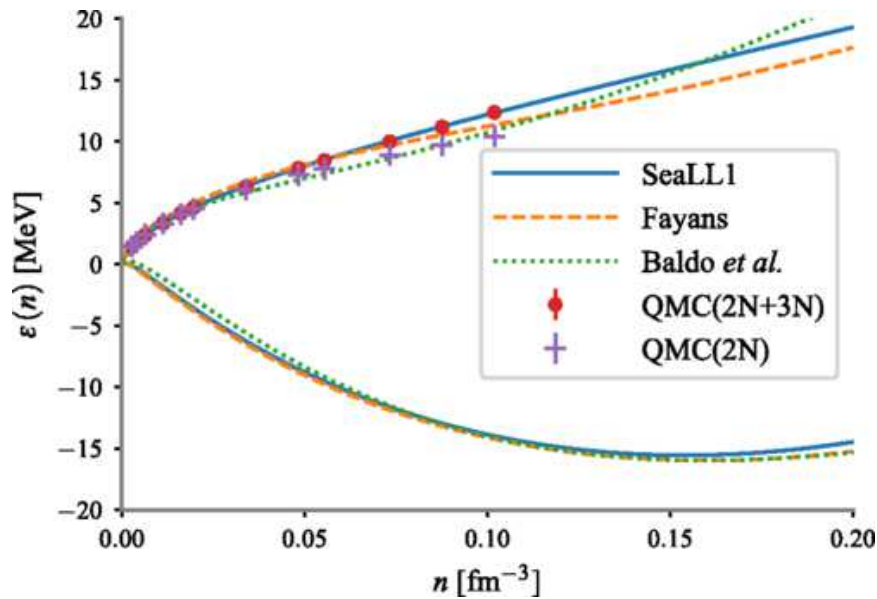,scale=0.65}
\end{minipage}
\begin{minipage}[t]{16.5 cm}
\caption{Energy per particle for neutron and symmetric matter as a function of the density $n$ produced by the SeaLL1, the Fayans \cite{fayans1,fayans2}, and the Baldo et al. \cite{baldo1,baldo2} functionals. For comparison, also the QMC results of Ref. \cite{wla} are shown.
Taken from Ref. \cite{forbes}.  \label{compabaldo}}
\end{minipage}
\end{center}
\end{figure}

In a more recent work, Bulgac and collaborators investigated along the direction of constructing a minimal phenomenological functional, containing seven adjustable parameters, called SeaLL1 \cite{forbes}. The starting point of this minimal functional is based on a generalization of the liquid drop model \cite{weiz}. The functional is also tailored for describing shell effects, pairing correlations, and the slope of the symmetry energy (thus, its density dependence). 
As an example, Fig. \ref{compabaldo} shows the EOSs of neutron and symmetric matter (the density is denoted by $n$) produced by the SeaLL1, the Fayans \cite{fayans1,fayans2}, and the Baldo et al. \cite{baldo1,baldo2} functionals. For comparison, also the QMC results of Ref. \cite{wla} are shown. One may notice the good quality of the SeaLL1 results in  reproducing the EOS of neutron matter compared to the QMC points obtained with 2$N$ plus 3$N$ interactions.  

We mention also the recent work by Dobaczewski \cite{dobaletter}, where a DFT formulation is strictly followed to define a strategy for costructing {\it{ab--initio}}--derived functionals. This development will be illustrated in Sec. \ref{dme}.

\section{Beyond the MF in the EDF \label{edfbmf}}
\subsection{\it Schematic overview on a selection of BMF models}
Overcoming the MF approximation can be done in several ways and, in each case, selected correlations are included in the theoretical scheme. Roughly speaking, one may consider four types of methods based on: (i) the introduction of a state which is the overlap of several trial states (configuration mixing for the restoration of broken symmetries); (ii) the explicit coupling of single--particle degrees of fredom with collective degrees of freedom; (iii) the explicit coupling of single--particle degrees of freedom with multiparticle--multihole configurations; (iv) the explicit introduction of a correlated ground state (instead of the HF state) in RPA--based extensions. Some examples are schematically mentioned below. 

(i) The Generator Coordinate Method (GCM) \cite{rs,hill,bonche1990} is based on the use of a state which is a superposition of reference generating functions $|\Phi (q) \rangle$, where $\{q\}$ represent the generator coordinates. This method is the BMF approach which is usually adopted for 
performing the restoration of broken symmetries and, in addition, it also includes explicit correlations in the theoretical framework through the use of such a superposition of generating functions. A special case of the GCM model is provided by projection techniques. Symmetry--restoration configuration--mixing calculations are discussed for instance in Refs. \cite{egido,bendergcm}. Numerous examples of applications dedicated to the restoration of the number of particles may be found in the literature (see for example Refs. \cite{hupin,heneen,angu2001,samyn,robledo}). 

(ii) Particle--vibration-- or particle--phonon--coupling models are extensively used in nuclear physics and are based on the coupling of single--particle degrees of freedom with collective motions (phonons). This produces a rescaling of the single--particle spectrum as well as a modification of the collective spectrum which allows for an explicit description of widths and fragmentation. Several methods are based on this approach, such as the particle--(or quasiparticle--)phonon--coupling models (nuclear field theory) of Refs. \cite{bertsch1,borti77,colo94,colo98,colo2001,niu2016}, the quasiparticle--phonon model of Refs. \cite{soloviev,giai98,seve2017,loiudice2012}, or the models based on the time--blocking approximation \cite{kame,tse2007,tse2016}. 

In the spirit of describing in a unified way single--particle and collective motions, it is also important to mention the BMF models which are based on the Bohr-Mottelson collective Hamiltonian \cite{bohr52,bohr69,bohr75,bohr76}. In particular, the five--dimensional quadrupole collective Hamiltonian describes both rotations and quadrupole vibrations at the same time. Detailed list of references and the most recent applications to quadrupole shape dynamics may be found in Ref. \cite{matsu}. 

(iii) Single--particle degrees of freedom may also be explicitly coupled to multiparticle--multihole configurations and this represents an alternative way of introducing the fragmentation of collective excitations. Multiparticle--multihole configuration--interaction calculations have been developed based on the Skyrme \cite{bordeaux,bordeaux2} and the Gogny \cite{mpmh1,mpmh2} interactions. A recent self--consistent version developed for the Gogny case is also available \cite{mpmhself1,mpmhself2}. 

In the second RPA (SRPA) model \cite{pro,ya,taka}
 the excited modes are defined as superpositions of standard RPA 1 particle-1 hole (1p1h) configurations plus higher--order 2 particle-2 hole (2p2h) configurations. 
The presence of the 2p2h configurations in the construction of the excited states  
and their coupling with 1p1h configurations produce BMF effects that allows for the description of the damping properties of excitation modes related to their spreading widths. 
Several applications have been done in the past decades \cite{schwe1984,dro1986,nishi1995,lacroix2004,tohyama2012}, up to the most recent calculations with large cutoff values in the 2p2h configurations and without truncations in the matrix to
 diagolalize \cite{papa2009,papa2010,gamba2010,gamba2012,gajpg,gamba2015,epja}.  

(iv) A well known violation of the Pauli principle is generated in RPA--based models by the use of the so--called quasiboson approximation. This approximation amounts in practice to evaluating the various matrix elements in the HF ground state instead of the correlated ground state (on which the model is in principle constructed). This drawback is corrected in several existing extensions of RPA--based models where a correlated ground state is explicitly taken into account in the formalism in several ways. Renormalizing factors depending on differences of occupation numbers between hole and particle states are introduced. Some examples of extended RPA--based models along this direction are illustrated in Refs. \cite{catara1989,volpe1999,samba1999,grasso2000,grasso2002,dukelsky1990,dukelsky1996,raduta1998,
simkovic2000,delion2005,jemai2013,ga2010,tohy2016}. 

\subsection{\it Necessity to construct BMF--tailored functionals}

Two reasons indicating the necessity of overcoming traditional effective functionals when BMF models are used are discussed in the next three subsections: first, a risk of double counting of correlations exists when traditional effective functionals are used and, on the other side, instabilities, pathologies, and ultraviolet divergences may occur in some BMF applications. 

\subsubsection{\it Overcounting of correlations}
It was already underlined in this review that most of the phenomenological effective density functionals which are currently used in EDF theories are adjusted on properties of finite nuclei and nuclear matter with MF calculations. Such a truncation in the Dyson expansion corresponds to the simplest many--body scheme based on the independent--particle approximation and on the use of a ground state which is a Slater determinant. This means that, as far as the many--body approximation is concerned, correlations beyond the MF approximation are completely excluded. Nevertheless, it turns out that correlations which are missing in the model are implicitly contained in the functionals because the values of their coupling constants are adjusted to reproduce obsevables measured for nuclear systems.   

A first consequence of this is a risk of overcounting of correlations in all cases where the MF approximation is overcome and the functional is not modified. It is impossible to know which correlations  (and to what extent)  are included in an implicit way when the parameters are adjusted performing MF calculations.  
The use of the same functionals in BMF theoretical schemes, where selected correlations are explicitly included in the theoretical scheme, becomes thus questionable. The possible associated double counting of correlations cannot be quantified and the predictive power of the model may be strongly impacted. 

This aspect was extensively discussed in the literature. One natural way to handle the problem would be to adjust the coupling constants of the functional by performing calculations at the same level of approximation as the one employed in the model for which the functional is intended. Illustrations of this procedure will be discussed in Sec. \ref{newgene}. Other procedures have been designed {\it{ad hoc}} for specific BMF models. One example is the subtraction method that was introduced by Tselyaev in Ref. \cite{tse2007} for the quasiparticle--time--blocking  approximation (QTBA). This procedure was initially designed  to handle the existence of spurious states in the excitation spectra computed with the QTBA model. The subtraction indeed ensures that the response function has zero--energy poles corresponding to the spurious states. In Ref. \cite{tse2007} Tselyaev noticed also that, by the same procedure, the static correlations (which are already included implicitly in the interaction) are subtracted. The double counting of such correlations is in this way avoided leaving visible only the genuine effect of the BMF correlations induced by the employed model.
The subtraction procedure corresponds to a kind of renormalization of the functional which is intended to remove all the correlations introduced with the adjustment procedure of its coupling constants.

\begin{figure}[tb]
\begin{center}
\begin{minipage}[t]{8 cm}
\epsfig{file=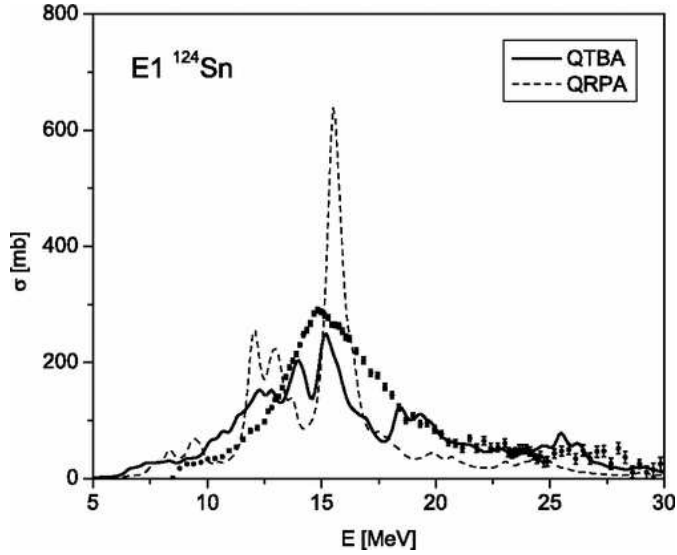,scale=0.5}
\end{minipage}
\begin{minipage}[t]{16.5 cm}
\caption{E1 cross sections calculated for $^{124}$Sn with the subtracted QTBA (full line) and the QRPA (dashed line). The experimental points (black circles) are extracted from Ref. \cite{fultz}. 
Taken from Ref. \cite{lit2007}.) 
\label{litvi124sn}}
\end{minipage}
\end{center}
\end{figure}

Examples of applications of the subtraction procedure within the QTBA model in the non relativistic framework may be found for instance in Refs. \cite{lit2007,tse2009,avdeenkov}. 
As an illustration of these applications, Fig. \ref{litvi124sn} shows the E1 photoabsorption cross section calculated for the nucleus 
$^{124}$Sn with the subtracted QTBA, compared to the cross section obtained with the quasiparticle RPA (QRPA). Also the experimental values taken from Ref. \cite{fultz} are shown (black circles). These calculations are not self--consistent and are based on a phenomenological Woods-Saxon potential. One may observe the genuine BMF effect in the spreading of the excitation mode which has a width in much better agreement with the experimental distribution, compared to the RPA case. 

 In Ref. \cite{tse2013}, the subtraction procedure was discussed for extended RPA models, such as the SRPA scheme. 
The subtraction procedure is formulated by explicitly requiring that, in extended RPA models, the static polarizability is equal to the RPA static polarizability. 
It is known that the static polarizability in a given model (RPA or extended RPA) is equal to $-2m_{-1}$, where $m_{-1}$ is the inverse energy--weighted momentum of the strength distribution calculated in the same model (dielectric theorem \cite{bohigas}). In the spirit of a DFT theory, one may say that correlations are fully included in the DFT--like RPA calculation (and correlations are indeed implicitly included in the numerical values of the coupling constants of the used functionals). From this, one may infer that the RPA  inverse energy--weighted momentum corresponds to the exact static polarizability. 
Eliminating the double counting thus amounts to requiring that the static polarizability in any extended RPA model must be equal to the RPA static polarizability. 

Let us consider for example the case of the SRPA model. It is possible to write the SRPA equations as RPA equations with energy--dependent matrix elements. Such matrix elements are the sum of RPA matrix elements plus an energy--dependent contribution (energy--dependent self--energy \cite{papa2010}). The condition that the static polarizability must be the same as in RPA is realized in practice by imposing that these energy--dependent matrix elements are equal to the corresponding RPA ones in the static limit (energy--dependent self--energy calculated at zero energy). This is required to cancel the double counting of the static correlations. 

The SRPA equations written in a compact form read  
\begin{equation}\label{eq_srpa}
\left(\begin{array}{cc}
  \mathcal{A} & \mathcal{B} \\
  -\mathcal{B}^{*} & -\mathcal{A}^{*} \\
\end{array}\right)
\left(%
\begin{array}{c}
  \mathcal{X}^{\nu} \\
  \mathcal{Y}^{\nu} \\
\end{array}%
\right)=\omega_{\nu}
\left(%
\begin{array}{c}
  \mathcal{X}^{\nu} \\
  \mathcal{Y}^{\nu} \\
\end{array}%
\right),
\end{equation}
where
\begin{equation}
\mathcal{A}=\left(\begin{array}{cc}
  A_{11'} & A_{12} \\
  A_{21} & A_{22'} \\
\end{array}\right), \quad
\mathcal{B}=\left(\begin{array}{cc}
  B_{11'} & B_{12} \\
  B_{21} & 0\\ 
\end{array}\right),
\end{equation}
\begin{equation}
\label{xandy}
\mathcal{X}^{\nu}=\left(\begin{array}{cc}
  X_{1}^{\nu} \\
   X_{2}^{\nu} \\
\end{array}\right),
~~~~\mathcal{Y}^{\nu}=\left(\begin{array}{cc}
  Y_{1}^{\nu} \\
   Y_{2}^{\nu} \\
\end{array}\right);
\end{equation}
1p1h and 2p2h configurations are denoted by 1 and 2, respectively.
The $11^{\prime}$ block represents
the standard RPA $A$ and $B$ matrices.  The 12  and 21 blocks
describe the 1p1h-2p2h and 2p2h-1p1h couplings, respectively, and the $22^{\prime}$ block represents the 2p2h sector of the matrix. 

By writing the SRPA equations as
RPA--type equations with energy--dependent matrices 
$A_{11^{\prime}}(\omega)$ and $B_{11^{\prime}}(\omega)$ one has,  
\begin{eqnarray}
\label{arpa}
A_{11^{\prime}} (\omega) = A_{11^{\prime}}+\sum_{2,2^{\prime}} A_{12} (\omega + i \eta - A_{22^{\prime}})^{-1} 
A_{2^{\prime}1^{\prime}} - 
\sum_{2,2^{\prime}} B_{12} (\omega + i \eta + A_{22^{\prime}})^{-1} 
B_{2^{\prime}1^{\prime}} \,,\\
B_{11^{\prime}} (\omega) = B_{11^{\prime}} + \sum_{2,2^{\prime}} A_{12} (\omega + i \eta - A_{22^{\prime}})^{-1} 
B_{2^{\prime}1^{\prime}} -  
\sum_{2,2^{\prime}} B_{12} (\omega + i \eta + A_{22^{\prime}})^{-1} 
A_{2^{\prime}1^{\prime}}\,. \nonumber
\label{brpa} 
\end{eqnarray}
$B_{12}$ and $B_{21}$ vanish in cases where a genuine Hamiltonian is used (no density dependence in the interaction) and 
appear in the equations only for density--dependent forces owing to the rearrangement terms \cite{gajpg}. 
Let us introduce the energy--dependent quantities $E_{11^{\prime}}(\omega)$ and $F_{11^{\prime}}(\omega)$ (energy--dependent self--energy corrections),
\begin{eqnarray}
E_{11^{\prime}} (\omega) =  \sum_{2,2^{\prime}} A_{12} (\omega + i \eta - A_{22^{\prime}})^{-1} 
A_{2^{\prime}1^{\prime}}  - 
\sum_{2,2^{\prime}} B_{12} (\omega + i \eta + A_{22^{\prime}})^{-1} 
B_{2^{\prime}1^{\prime}}\,, \\
F_{11^{\prime}} (\omega) = \sum_{2,2^{\prime}} A_{12} (\omega + i \eta - A_{22^{\prime}})^{-1} 
B_{2^{\prime}1^{\prime}} - 
\sum_{2,2^{\prime}} B_{12} (\omega + i \eta + A_{22^{\prime}})^{-1} 
A_{2^{\prime}1^{\prime}}\,. \nonumber
\end{eqnarray}
The subtraction procedure consists in subtracting $E_{11^{\prime}}(0)$ and
$F_{11^{\prime}}(0)$ (which represent the static parts) from $A_{11^{\prime}}(\omega)$ and
$B_{11^{\prime}}(\omega)$, respectively,
\begin{equation}
A^S_{11^{\prime}} (\omega)= A_{11^{\prime}} (\omega) - E_{11^{\prime}}(0),
\label{sub1} 
\end{equation}
\begin{equation}
B^S_{11^{\prime}} (\omega)= B_{11^{\prime}} (\omega) - F_{11^{\prime}}(0), 
\label{sub2} 
\end{equation}
where the superscript $S$ stands for ``subtracted". 
Coming back to the usual 
energy--independent SRPA equations, the subtracted matrices are written as 
\begin{eqnarray}
\label{eq:absf}
\mathcal{A}^S=\left(\begin{array}{cc}
  A_{11'}+ \sum_{2,2'} A_{12} (A_{22'})^{-1}A_{2'1'} + \sum_{2,2'} B_{12} (A_{22'})^{-1} B_{2'1'} & A_{12} \\
  & \\
  A_{21} & A_{22'} \\
\end{array}\right)\,, \\[.3cm]
\mathcal{B}^S=\left(\begin{array}{cc}
  B_{11'} + \sum_{2,2'} A_{12} (A_{22'})^{-1} B_{2'1'} + \sum_{2,2'} B_{12} (A_{22'})^{-1} A_{2'1'}  & B_{12} \\
  &  \\
  B_{21} & 0\\ 
\end{array}\right)\,. \nonumber
\end{eqnarray}
We denote the subtracted SRPA model with the acronym SSRPA. 
Several applications have been recently done with the SSRPA model to investigate the dipole spectrum and the dipole polarizability in $^{48}$Ca \cite{gamba2018}, giant quadrupole centroids and widths in a selection of nuclei going from $^{30}$Si to $^{208}$Pb \cite{vasseur}, as well as a beyond--mean--field effect on the enhancement of the effective mass induced in the SSRPA model \cite{grasso2018}. 

It is interesting to note that the subtraction term produces a rescaling of the RPA--type matrix elements of the type $A_{11'}$ and $B_{11'}$ (for the latter case, this occurs only if there are rearrangement terms). In particular, the diagonal matrix elements $A$ are modified. Since such matrix elements contain the single--particle excitation energies (individual excitation energies defined as differences between the energies of particle and hole states), this may be regarded as a rescaling of the single--particle excitation energies. This mimics a modification of the single--particle spectrum and may be put in relation with  BMF effects on the effective mass \cite{grasso2018}. 

\subsubsection{\it Instabilities (imaginary solutions) and ultraviolet divergences in BMF applications \label{patho1}}

In several BMF models instabilities, pathologies, and cutoff dependencies (ultraviolet divergences) occur and, in many cases, these problems are related to the employed effective density functional. 

Let us first deal with the instabilities that may be found in beyond--RPA models, such as the SRPA model, and which corrrespond to the presence of imaginary solutions in the excitation spectrum. 
  Tselyaev demonstrated in Ref. \cite{tse2013} that the subtraction procedure guarantees not only the cancellation of the double counting of correlations, but also the stability of the excitation spectra (real and positive excitation energies). 
In the case of the RPA model, the stability condition is related to the Thouless theorem \cite{thou1,thou2} and it was shown in Ref. \cite{tse2013} that the Thouless theorem may be generalized to extended RPA models by applying the same subtraction procedure previously introduced for the QTBA approximation. 

Reference \cite{papa2014} deeply investigated the stability condition in the SRPA model in terms of the Thouless theorem. The Thouless theorem states that, if the HF state represents the minimum of the Hamiltonian, then the RPA stability matrix is positive semidefinite \cite{thou1,thou2}, where the RPA stability matrix $\mathcal{S}^{RPA}$ is defined as 
\begin{displaymath}
\mathcal{S}^{RPA}=\left(\begin{array}{cc}
  A & B \\
  B^* & A^* \\
\end{array}\right),
\end{displaymath}
and $A$ and $B$ are the usual RPA matrices \cite{rs}. The fact that the stability matrix is positive semidefinite guarantees real solutions from the RPA diagonalization. Now, the Thouless theorem cannot be extended to SRPA. If the HF is the minimum of the Hamiltonian, it is not possible to prove that the SRPA stability matrix is positive semidefinite. With several formal developments and numerical tests, Papakonstantinou  \cite{papa2014} showed that the violation of the stability condition may indeed generate in SRPA imaginary solutions, positive excitation energies with negative norms and negative excitation energies with positive norms, as well as spurious states at finite energy. In addition, the violation of the stability condition is responsible for the anomalous (and unphysical) huge downwards shift of the excitation energies generated in the standard SRPA model with respect to the RPA spectrum. This unphysical shift is independent of the used interaction \cite{papa2009,papa2010,gamba2010,gamba2012}
and of the system under study. For example, it was also found in Ref. \cite{ga2009} where the authors analyzed dipole excitations (dipole plasmons) in metallic clusters. 

\begin{figure}[tb]
\begin{center}
\begin{minipage}[t]{8 cm}
\epsfig{file=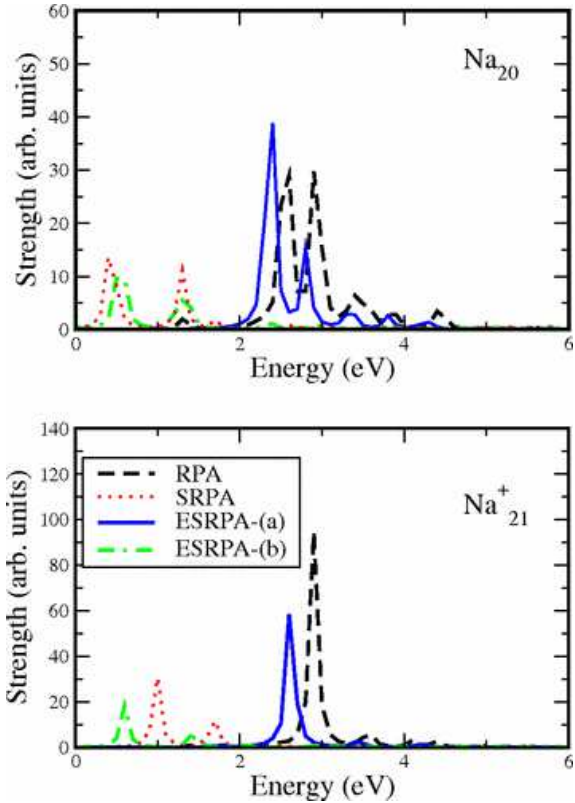,scale=0.6}
\end{minipage}
\begin{minipage}[t]{16.5 cm}
\caption{ Dipole strength distributions for the metallic clusters Na$_{20}$ (upper panel) and Na$_{21}^+$ (lower panel). RPA, SRPA, ESRPA-(a), and ESRPA-(b) results are shown. ESRPA-(a) denotes a full calculation whereas ESRPA-(b) denotes an approximation  where ground--state correlations are taken into account only in RPA submatrices.   
Taken from Ref. \cite{ga2010}. 
\label{cluster}}
\end{minipage}
\end{center}
\end{figure}

Two possible directions were indicated by Papakonstantinou to handle this drawback: the {\it{ad hoc}} subtraction procedure of Tselyaev and a full reformulation of the model based on a the use of a correlated ground state (instead of the HF state). Correlations in the SRPA ground state may indeed be very important and the violation of the Pauli principle induced by the use of the quasiboson approximation (which, in practice, amounts to employing the HF ground state instead of the correlated one) can be significant. 
However, this direction indicated in Ref. \cite{papa2014} is intended to handle only the problems related to the stability. 
If traditional functionals are used (still adjusted at the MF level), the risk of overcounted correlations is not avoided. On the other hand, the subtraction method has the advantage of handling both problems at the same time (instabilities and double counting) and seems for this reason to be more adapted for EDF--type applications.
 
The suggestion of using a correlated ground state was already applied in the past to metallic clusters. Reference \cite{ga2010} indeed illustrates an extended SRPA model applied to sodium clusters. The authors treat these clusters in the so--called jellium approximation \cite{jellium}, where a frozen ionic structure described by a positively charged and uniform background interacts with a cloud of delocalized valence electrons. The interaction between the delocalized electrons and between such electrons and the positively  charged jellium is the bare Coulomb interaction. In this case, there are no problems of overcounting of correlations.  
A correlated ground state is introduced and renormalized matrix elements appear where the renormalization factors are related to the difference of occupation numbers between hole and particle states. Owing to the use of a correlated ground state, such factors are different from (less than) 1 and produce a rescaling of the matrix elements and of the excitation spectrum. 
The dipole plasmon excitations were studied in this work for two clusters, the neutral cluster Na$_{20}$ and the positively charged cluster Na$_{21}^+$. Figure \ref{cluster} displays the dipole strength distributions for these two clusters calculated with RPA, SRPA and two extensions of SRPA (ESRPA), namely a full calculation with a correlated ground state (ESRPA-(a)) and an appoximated calculation where the ground--state correlations are partially taken into account, only in the RPA submatrices (ESRPA-(b)). One may observe that only the full scheme ESRPA-(a) is able to correct the anomalous strong shift generated by the SRPA model with respect to RPA: the corrected energy spectrum is strongly pushed upwards compared to the SRPA case.  

\begin{figure}[tb]
\begin{center}
\begin{minipage}[t]{10 cm}
\epsfig{file=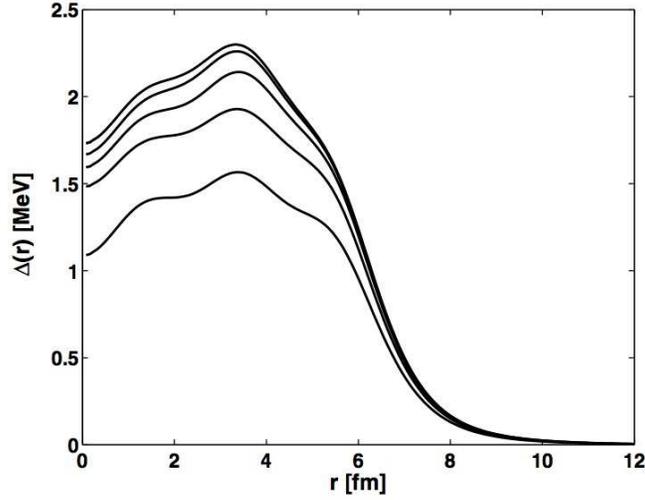,scale=0.67}
\end{minipage}
\begin{minipage}[t]{16.5 cm}
\caption{Radial profiles of the neutron pairing field for the nucleus $^{110}$Sn at different values of the energy cutoff. Starting from the lowest curve, the values of the energy cutoff are 20, 30, 35, 40, 45, and 50 MeV. The last two curves are practically superposed.  
Taken from Ref. \cite{bulgac2002}.}
\label{convebul}
\end{minipage}
\end{center}
\end{figure}

References \cite{gamba2015} and \cite{epja} show the first applications of the subtraction procedure to the SRPA model with the Skyrme and Gogny forces, respectively. 
An additional problem arises for these specific BMF applications when Skyrme and Gogny forces are used, namely a strong cutoff dependence (ultraviolet divergence) generated by the zero range of such interactions. 
It is important to stress that the problem of regularizing ultraviolet divergences related to the zero range of the used interaction does not arise only for BMF calculations. At the MF level, such an ultraviolet divergence occurs  also when superfluid systems are treated  and the HFB model is applied with a zero--range interaction in the pairing channel. One may of course handle this divergence with a cutoff regularization by choosing an energy cutoff and by adjusting  the parameters entering in the pairing interaction for that cutoff value. 

It turns out that the HFB model is also currently used for treating gases of ultracold fermionic atoms trapped by external potentials in the superfluid regime. In the context of atomic physics, the model is called Bogoliubov-de Gennes \cite{degennes}. These gases of fermions are  typically very dilute and can be described in  a satisfactory way by a contact interaction characterized by a coupling constant directly related to the scattering length. Such a scattering length may be tuned by modifying an external magnetic field around Feshbach resonances \cite{roberts} and a superfluid regime where Cooper pairs are present may be attained.  
A specific subtraction method was proposed within the Bogoliubov-de Gennes model for a gas of ultracold fermionic atoms trapped by a harmonic potential \cite{bruun}, based on the so--called pseudoptential prescription of Ref. \cite{huang}. A modified prescription was applied to the nuclear case \cite{bulgac2002} and to superfluid trapped atoms \cite{grasso2003}. Further applications to nuclei are presented in Refs. \cite{bulgac2003,bor2006}. 
Having identified the type of divergence in the pairing field ($1/r$ when $r \rightarrow 0$), a quantity is added and subtracted having exactly the same divergence and containing the Green's function related to the single--particle Hamiltonian. (It turns out that one is able to separate the regular and the diverging parts for such a Green's function). This procedure allows for a full cancellation of the diverging contribution in the pairing field and leads to cutoff--independent results. In practice, such a subtraction procedure corresponds to a renormalization of the coupling constant characterizing the used interaction. The modification introduced in Refs. \cite{bulgac2002,grasso2003} with respect to the previous work \cite{bruun} amounts to using the Thomas-Fermi approximation in the computation of the regular part of the single--particle Green's function.  This represents an important simplification of the numerical computation without altering the quality of the results. 
Figure \ref{convebul} is an example of this regularization. It shows several radial profiles of the neutron pairing field for different values of the energy cutoff. The regularized HFB calculations are performed with a Woods-Saxon potential adjusted to describe the nucleus $^{110}$Sn.  Starting from the lowest curve, the values of the energy cutoff are 20, 30, 35, 40, 45, and 50 MeV. The last two curves are practically superposed indicating that the convergence with respect to the cutoff has been attained. 

\begin{figure}[tb]
\begin{center}
\begin{minipage}[t]{8 cm}
\epsfig{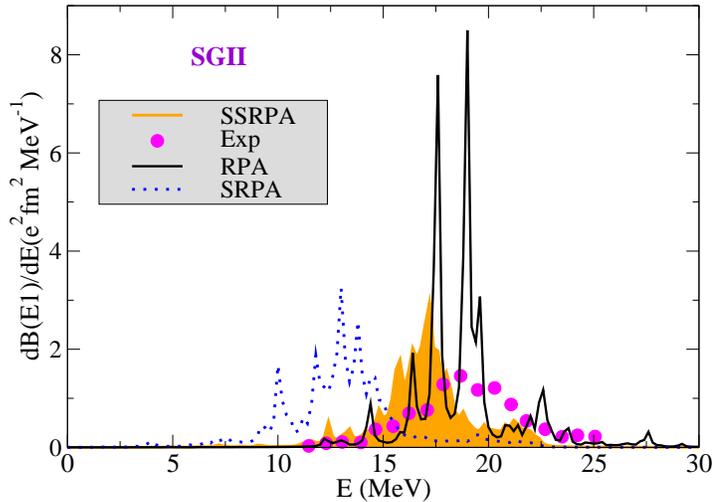}
\end{minipage}
\begin{minipage}[t]{16.5 cm}
\caption{Dipole strength distribution of $^{48}$Ca computed with RPA (solid black line), SRPA (blue dotted line), and SSRPA (orange line and area) with the Skyrme parametrization SGII. The experimental distributions (magenta circles) of Ref. \cite{osaka} are also shown.
Taken from Ref. \cite{gamba2018}. }
\label{shift}
\end{minipage}
\end{center}
\end{figure}

\begin{figure}[tb]
\begin{center}
\begin{minipage}[t]{8 cm}
\epsfig{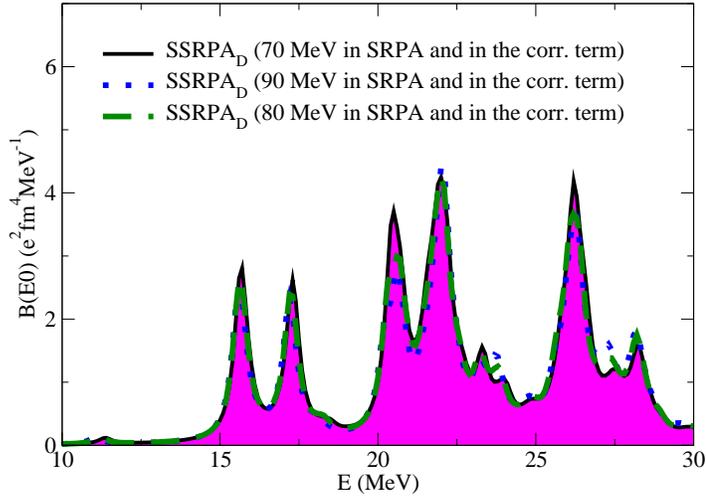}
\end{minipage}
\begin{minipage}[t]{16.5 cm}
\caption{ Isoscalar monopole response computed for $^{16}$O with cutoff values of 70 (black line and
magenta area), 80 (green dashed line), and 90 (blue dotted line) MeV (see text).
Taken from Ref. \cite{gamba2015}.
\label{convergence}}
\end{minipage}
\end{center}
\end{figure}

Coming back to the SSRPA discussion, an important result was reported in 
Ref. \cite{gamba2015} in the framework of the SSRPA model, where the authors indicated that the Tselyaev subtraction procedure  also cures the ultraviolet divergence produced by the zero--range of the employed interaction. Let us illustrate the main achievements provided by the subtraction procedure within the SSRPA model.   
Figure \ref{shift} describes the dipole strength distributions for the nucleus $^{48}$Ca computed with RPA, SRPA and SSRPA \cite{gamba2018}, compared to the experimental results \cite{osaka}. One observes the strong correction provided by SSRPA with respect to SRPA, where the spectrum was shifted downwards by several MeV compared to RPA. 
The damping properties were already well described in the SRPA model and the subtraction does not alter this specific feature. It produces only a global upwards shift of the energies.
These calculations are performed with the parametrization SGII \cite{sgii}. 
An example of the convergence of the results with respect to the energy cutoff is provided in Fig. \ref{convergence} where the isoscalar monopole response is shown for the nucleus $^{16}$O. Several SSRPA results are displayed (the subscript $D$ indicates that the subtraction term has been calculated with a diagonal approximation for the $A_{22'}$ matrix to be inverted). The same cutoff values in the 2p2h space are used for each  
SSRPA diagonalization and for the corresponding correction term, 70, 80, and 90 MeV. 
One can observe that the dependence on the cutoff has been removed by the subtraction procedure. The calculations are done with the parametrization SGII. 

\begin{figure}[tb]
\begin{center}
\begin{minipage}[t]{8 cm}
\epsfig{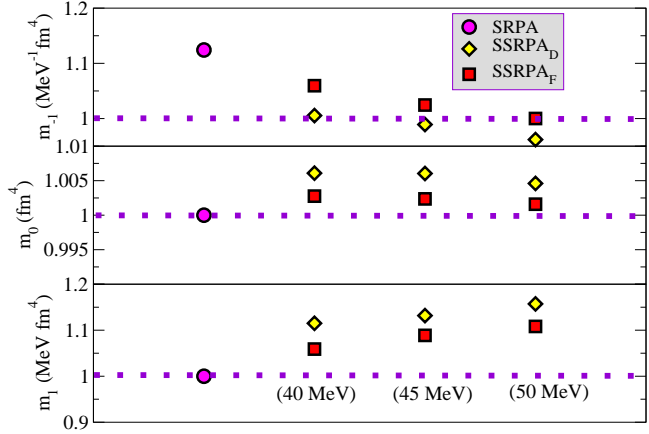}
\end{minipage}
\begin{minipage}[t]{16.5 cm}
\caption{ Ratios of the moments $m_{-1}$, $m_0$, and $m_1$ of the
quadrupole strength distribution (with respect to the RPA values) in the SRPA (magenta circles), the 
SSRPA$_F$ (red squares), and the SSRPA$_D$ (yellow diamonds) for different cutoff values 
in the correction terms, at 40, 45, and 50 MeV.
Taken from Ref. \cite{gamba2015}.
\label{moments}}
\end{minipage}
\end{center}
\end{figure} 

The first application of the same subtraction procedure to particle--vibration coupling models was illustrated in Ref. \cite{roca} for the nuclei $^{16}$O and $^{208}$Pb, with special emphasis on the analysis of the moments $m_0$, $m_1$, and $m_{-1}$ of the strength function.
Figure \ref{moments} indeed provides an overview to summarize how these moments are (or not) conserved in the cases of SRPA and SSRPA. 
The subscripts $F$ and $D$ refer to calculations where the $A_{22'}$ matrix to be inverted in the correction term is treated without approximations or in the diagonal approximation, respectively.
The example refers to the quadrupole strength distribution computed for $^{16}$O. The SRPA and SSRPA calculations are performed with a cutoff of 50 MeV for the 2p2h configurations with the parametrization SGII. 
Different cutoff values are then taken for the correction term, from 40 to 50 MeV. 
The figure shows the ratios of the moments with respect to the RPA ones. 
 One observes that the moments $m_0$ (middle ponel) and $m_1$ (lower panel) are conserved in the SRPA case \cite{ya,dro} and not conserved in the SSRPA case where they are sligthly overestimated. For the $m_1$ moment, the same situation occurs also in the particle--vibration coupling model, where this moment is conserved (slightly overestimated) without (with) the application of the subtraction procedure \cite{roca}. The $m_{-1}$ moment (upper panel), not conserved in the SRPA case, is perfectly conserved by construction when the subtraction is applied without approximations (full calculation) and the same cutoff values are taken for the 2p2h states in the matrix to diagonalize and in the correction term.     
In the case of the particle--vibration coupling, the $m_{-1}$ moment is almost conserved \cite{roca}. The fact that this moment is not exactly conserved is probably related to some approximations employed in those calculations. 

\subsubsection{\it Instabilities and pathologies related to the density dependence and to the spurious self--interaction in BMF applications \label{patho2}}
Anomalies (spurious divergences and finite steps) were identified in several BMF models based on the restoration of particle number \cite{angu2001,tajima,donau,almehed}. Such anomalies were analyzed in Ref. \cite{la} for GCM--type and configuration--mixing calculations based on projection. A correction scheme was proposed for the restricted case of a Skyrme--type interaction with an integer power of the density and applied in Ref. \cite{be} to the case of particle--number restoration, where these anomalies are more easily visible and identified. Reference \cite{du} concluded saying that this regularization procedure cannot be extended to cases where noninteger powers of the density matrices are present. Later, Ref. \cite{hupin} proposed a different regularization procedure where also cases with noninteger powers of the densities can be handled.

 It is important to stress that such anomalies are a peculiarity of EDF models and do not occur in cases where a genuine Hamiltonian is used and the ground--state energy (HF or HFB) is computed without any truncation or approximation. This means: (i) no density dependencies 	are admitted; (ii) the same interaction has to be used in the particle-hole (MF) and in the particle-particle (pairing) channels if pairing is included in the HFB model; (iii) no approximations have to be adopted for exchange terms (for instance, for the Coulomb exchange contribution). 

If one considers the simple MF--based EDF framework, the EDF functional may be written as
\be
\mathcal{E}_q \equiv \mathcal{E}_q[\rho_{qq},\kappa_{qq},\kappa_{qq}^*], 
\label{funcMF}
\ee
where $\rho$ and $\kappa$ are the normal and anomalous densities, respectively, and $q$ is a collective parameter (generator coordinate) that will be used for the restoration of broken symmetries in configuration--mixing models. In such models, the BMF state is written as the superposition  
\begin{equation}
| \Psi \rangle = \sum_q f(q) | \Phi(q) \rangle,
\label{gs}
\end{equation}   
where $\{  \Phi(q) \}$  is a set of reference states (generating functions). Now, generalizing a genuine Hamiltonian formalism, the BMF functional is written as a weighted sum (the weights are the functions $f(q)$ in Eq. (\ref{gs})) over the non diagonal kernels $\mathcal{E}_{qq'}[\rho_{qq'},\kappa_{qq'},\kappa_{qq'}^*]$ computed between all possible combinations of reference states $| \Phi(q) \rangle$. 
For calculating these kernels, the generalized Wick theorem is used \cite{balian}, the density matrices appearing in Eq. (\ref{funcMF}) are replaced by the transition density matrices, 
\begin{eqnarray}
\label{eq:transition} 
\rho_{ij}^{qq'} & \equiv & \frac{\langle \Phi(q) | a^{+}_{j} a_{i}| \Phi(q') 
\rangle}
{\langle \Phi(q)  | \Phi(q')  \rangle}, 
\nonumber \\
\kappa_{ij}^{qq'}  &\equiv&  \frac{\langle \Phi (q) | a_{j} a_{i}| \Phi (q') 
\rangle}
{\langle \Phi (q) | \Phi (q') \rangle}, \\
\kappa^{q'q \, \ast}_{ij}  &\equiv&  \frac{\langle \Phi (q) | a^{+}_{i} a^{+}_{j}| 
\Phi (q')\rangle}
{\langle \Phi(q) | \Phi (q') \rangle}, 
\nonumber
\label{dens}
\end{eqnarray} 
and the kernel  $\mathcal{E}_{qq'}$ is multiplied by $\langle \Phi (q) | \Phi (q') \rangle$. The BMF functional reads  
\begin{equation}
\mathcal{E}[\Psi] \equiv \frac{\sum_{q,q'}  f^{\ast}({q}) f(q')
        \mathcal{E}_{qq'}[\rho_{qq'},\kappa_{qq'},\kappa_{qq'}^*] \,
        \langle \Phi (q) | \Phi (q') \rangle}
       {\sum_{q,q'}  f^{\ast}({q}) f(q')  \langle \Phi (q) | \Phi (q') \rangle} 
\,  . \label{MRenergy}
\end{equation}

Let us consider the case of particle--number restoration. If $| \Phi_0 \rangle$
 is the MF state, a set of reference states is constructed by rotating in gauge space, 
\be
| \Phi (\varphi) \rangle = e^{i\varphi \hat{N}}| \Phi_0 \rangle,
\label{gauge}
\ee
where $\hat{N}$ is the particle number operator. The functional (\ref{MRenergy}) becomes in this case
\begin{equation}
\label{scalar2}
\mathcal{E}_{N}
\equiv \frac{1}{c^{2}_{N}} \int_{0}^{2\pi} \! \frac{d\varphi}{2\pi} \,
       e^{-i\varphi N} \,
       \mathcal{E}_{0\varphi}[\Phi_0 , \Phi (\varphi)] \,  
       \langle  \Phi_0 | \Phi (\varphi) \rangle
\, ,
\label{funcPNR}
\end{equation}
with 
\begin{equation}
c^2_N =  \int_{0}^{2\pi} \! \frac{d\varphi}{2\pi} \,
       e^{-i\varphi N} \,
       \langle  \Phi_0 | \Phi (\varphi) \rangle
\, ,
\label{cfactors}
\end{equation}
where $N$ is the number of particles. 

\begin{figure}[tb]
\begin{center}
\begin{minipage}[t]{8 cm}
\epsfig{file=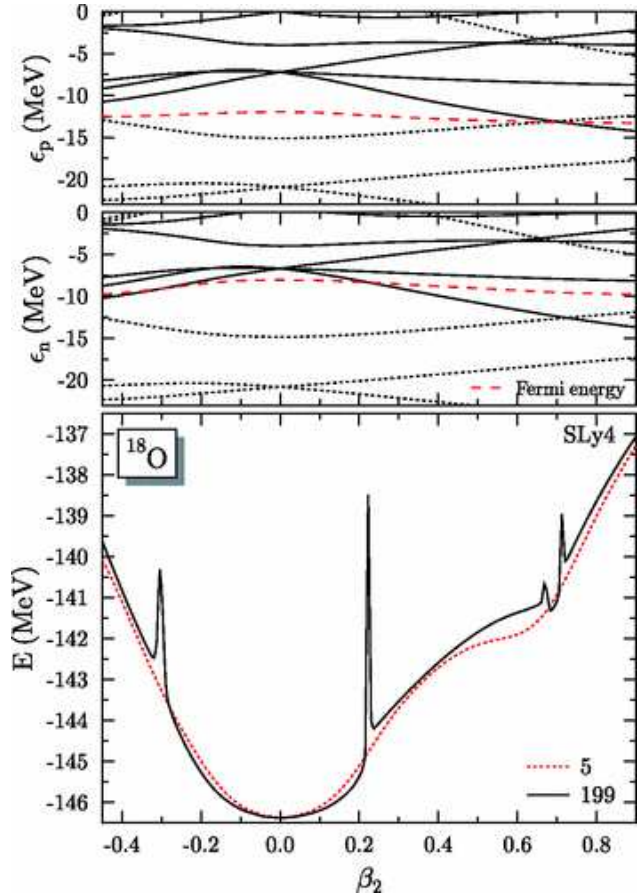,scale=0.67}
\end{minipage}
\begin{minipage}[t]{16.5 cm}
\caption{Lower panel: Particle--number restored energy surface as a function of the quadrupole deformation $\beta_2$ for two values of the discretization points, 5 and 199. 
Medium and upper panels: single--particle energies for neutron ($\epsilon_n$) and proton ($\epsilon_p$ states), respectively, as a function of the quadrupole deformation $\beta_2$.
Taken from Ref. \cite{be}.
\label{dive}}
\end{minipage}
\end{center}
\end{figure}

Figure \ref{dive} is an example of pathologies in particle--number--restored configuration--mixing calculations. The lower panel shows the particle--number restored energy surface as a function of the quadrupole deformation $\beta_2$, 
\be
\beta_2=\sqrt{\frac{5}{16\pi}} \frac{4\pi}{3R^2A} (2z^2-y^2-x^2),
\label{defo}
\ee
with $R=1.2$ $A^{1/3}$ fm. 

Two calculations are shown, corresponding to two different numbers of discretization points 
used to solve the integral (\ref{funcPNR}) over the gauge angle $\varphi$.
One observes that, by increasing the number of discretization points, the convergence is not achieved and anomalies appear corresponding to values of $\beta_2$ where a single--particle level for neutrons ($\epsilon_n$) or for protons ($\epsilon_p$) crosses the corresponding Fermi energy (medium and upper panels, respectively). The shown calculations are done for the nucleus $^{18}$O with the SLy4 Skyrme parametrization \cite{lyon1,lyon2,lyon3}. 

\begin{figure}[tb]
\begin{center}
\begin{minipage}[t]{11 cm}
\epsfig{file=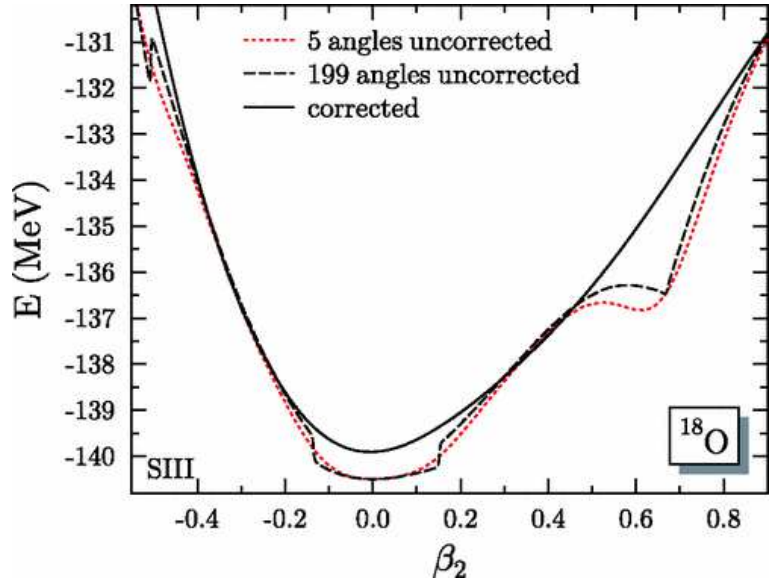,scale=0.57}
\end{minipage}
\begin{minipage}[t]{16.5 cm}
\caption{Corrected and uncorrected particle--number--restored quadrupole deformation energies as a function of $\beta_2$ for the nucleus $^{18}$O computed with the parametrization SIII.
Taken from Ref. \cite{be}. }
\label{regu}
\end{minipage}
\end{center}
\end{figure}

This kind of anomalies were found not only for particle--number--restored calculations but also in the case of angular--momentum restoration \cite{zdu}. Which are the causes? Reference \cite{la} presented a detailed analysis and identified the problematic terms in the BMF functional. Such terms are indeed related to the self--interaction problem and the associated violation of the Pauli principle that were previously mentioned in this review. In a genuine Hamiltonian formalism, such diverging terms exist but they recombine with other terms also present in the functional and the final result leads to a finite energy. In the EDF case, due to the employed questionable analogy with a true--Hamiltonian case and due to the used generalized Wick theorem, the resulting functional presents the anomaly of having only part of the diverging contributions. Recombinations are thus not possible and the pathologies become visible. The 
Bloch-Messiah-Zumino procedure \cite{bloch1,bloch2} was used in Ref. \cite{la} to identify all the diverging terms and a a simple regularizing scheme was proposed consisting in subtracting such contributions from the functional. This scheme can be applied only for integer powers of the density matrices. As an illustration of the obtained results, Fig. \ref{regu} shows, for the parametrization SIII (where the power of the density--dependent term is equal to 1), regularized particle--number--restored quadrupole deformation energies for the nucleus $^{18}$O. The two uncorrected curves obtained with a number of discretization points equal to 5 and 199 are compared to the corresponding corrected ones, which do not differ one from the other. This indicates that the regularized calculations already converged with a small number of discretization points. In addition, results are free from finite steps and divergences. 

\begin{figure}[tb]
\begin{center}
\begin{minipage}[t]{11 cm}
\epsfig{file=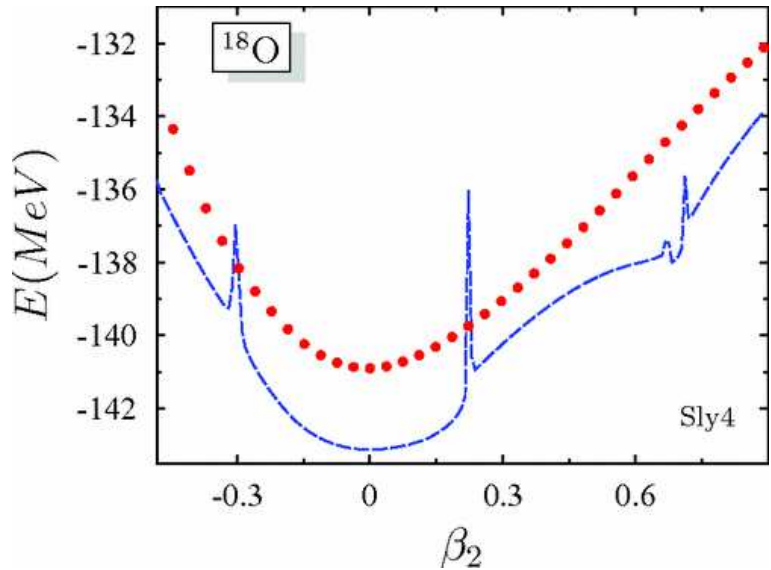,scale=0.57}
\end{minipage}
\begin{minipage}[t]{16.5 cm}
\caption{Particle--number--restored quadrupole deformation energies as a function of $\beta_2$ for the nucleus $^{18}$O computed with the parametrization SLy4. The dashed line is the result of the non regularized integral over the gauge angle using 199 points and the circles represent the corrected curve.  
Taken from Ref. \cite{hupin}.}
\label{hup}
\end{minipage}
\end{center}
\end{figure}

Another direction was explored in Ref. \cite{hupin} for the case of the particle--number restoration. A theory was proposed, called ``symmetry--conserving EDF approach" where a functional is formulated in terms of the projected one--body and two--body density matrices. In this formulation, the pathologies observed in the above--mentioned BMF models (where transition density matrices were used)  are not found. This theory can be applied also to cases where there are density--dependent terms with noninteger powers. An illustration is displayed in Fig. \ref{hup} for the nucleus $^{18}$O and the SLy4 Skyrme parametrization. 
The dashed line is the result of the non regularized integral over the gauge angle using 199 points and the circles represent the corrected curve.  

Other procedures were proposed in the literature for removing such pathologies appearing when going from a single trial state in MF--based models to several trial states in BMF models. We mention for instance a procedure where the generalized Wick theorem is not used and multiplying factors are introduced defined as a given power of the overlap between trial states \cite{satula}. 

Of course, the most general solution would be to use a genuine Hamiltonian, for example removing the two--body density--dependent part of the interaction and replacing it with a three--body interaction. This was already discussed in Subsec. \ref{3body} with two examples \cite{sadou,labe}. Another example of EDF development 
based on a finite--range three--body interaction can be found in Ref. \cite{ge10}.

\section{New generation of effective density functionals tailored for BMF models \label{newgene}}
The {\it{ad--hoc}} procedures illustrated in Sec. \ref{edfbmf} for regularizing effective density functionals, removing the divergences, and eliminating the double counting of correlations are certainly very efficient and provide clear practical ways for handling these problems in each specific case. Nevertheless, from a conceptual point of view, a more satisfactory protocol for designing functionals adapted for BMF models should be based on a generalization of the construction method of the functional. More specifically, the functional should be derived from a given interaction at the same level of approximation as the one of the employed theoretical model (for example, the same truncation in the perturbative many--body expansion). This would provide a coherent BMF EDF functional. Still remaining within an empirical procedure for constraining the functionals, the parameters entering in the BMF functional should then be adjusted performing the corresponding BMF calculations. This would eliminate by construction the risk of any double counting of correlations. Possible divergences or instabilities should then be regularized. 

We will focus in the next subsections on many--body perturbation--theory (MBPT) calculations truncated at second order for nuclear matter, which is a system that allows for an analytical derivation of the second--order EOS of infinite matter in the case for example of Skyrme--type interactions. 
The perturbative expansion provided by the Dyson equation \cite{fetter} was already analyzed long time ago. Some examples may be found for a gas composed by electrons in Refs. \cite{gellmann,sawada,nozieres,moller} and for nuclear systems with finite--range interactions in Refs. \cite{euler,huby,thouless,levinger}. 
In Ref. \cite{baranger} a procedure for computing the second--order contribution was proposed, based on an expansion of the interaction in partial waves. 

\begin{figure}[tb]
\begin{center}
\begin{minipage}[t]{20 cm}
\epsfig{file=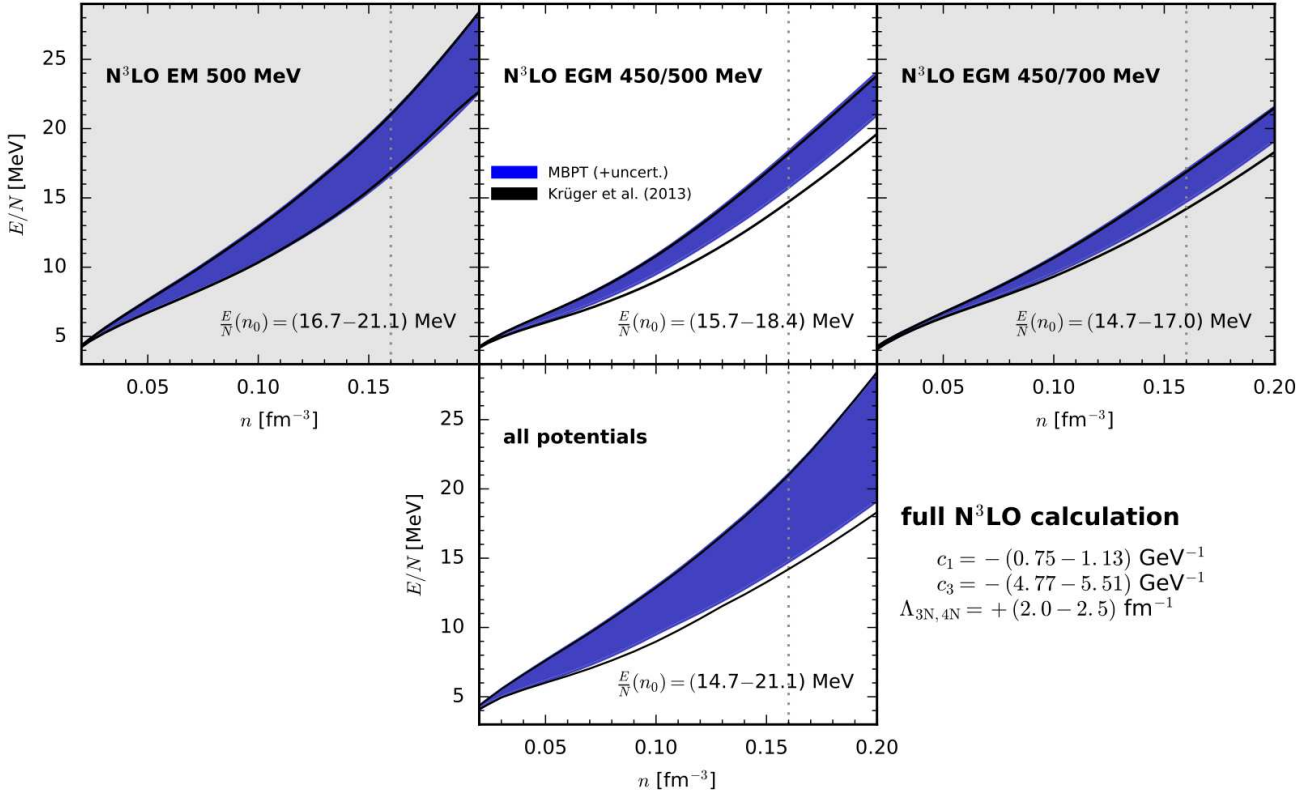,scale=0.61}
\end{minipage}
\begin{minipage}[t]{16.5 cm}
\caption{EOS of neutron matter calculated at N$^3$LO (blue bands) with three Hamiltonians, EM 500 MeV \cite{em}, EGM 450/500 MeV, and EGM 450/700 MeV \cite{egm} as a function of the density $n$ (the empirical saturation density of symmetric matter is denoted by $n_0$ and indicated by a vertical dotted line). Results from Ref. \cite{kru} (area between the black lines) are also shown. The lower row summarizes the results of the three panels of the upper row.
Taken from Ref. \cite{dri2016}. }
\label{mbpt1}
\end{minipage}
\end{center}
\end{figure}

In nuclear physics, MBPT models have been developed for treating closed--shell \cite{gun2010,roth2010,roth2006,roth2,navra2009,roth2011,tichai} and open--shell finite nuclei \cite{lang2012}.  The EOS of neutron matter computed with chiral EFT interactions and based on the MBPT is also estensively  discussed in the literature \cite{hebe2010,tews,kru,dri2016}. Also studies for asymmetric matter are available \cite{dri}. (The framework of EFTs, that is the formulation of QCD as a low--energy theory, was established in Ref. \cite{wei}. See for instance Refs. \cite{beda2002,epe2009}. Through chiral EFTs, chiral nuclear potentials are derived. See for instance Refs. \cite{epe2012,samma}).  
As an illustration, Fig. \ref{mbpt1} shows one of the MBPT results of Ref. \cite{dri2016} obtained with N$^3$LO $3N$ interactions up to third order. The EOS of neutron matter is calculated at N$^3$LO (blue bands) for the potentials EM 500 MeV \cite{em}, EGM 450/500 MeV, and EGM 450/700 MeV \cite{egm} and  is  compared to previous results published in Ref. \cite{kru} (area between the black lines). The lower row summarizes the results shown in the three panels of the upper row. The uncertainty band related to the Hamiltonian is 
generated by the cutoff $\Lambda_{3N,4N}$ (contributions at N$^3$LO from $4N$ interactions are also included in the HF approximation) and by the low--energy constants $c_1$ and $c_3$. It is observed that the uncertainty region is 
slighlty reduced compared to the previous EOSs of Ref. \cite{kru}. The empirical saturation density of symmetric matter is indicated by a vertical dotted line. 

\begin{figure}[tb]
\begin{center}
\begin{minipage}[t]{10 cm}
\epsfig{file=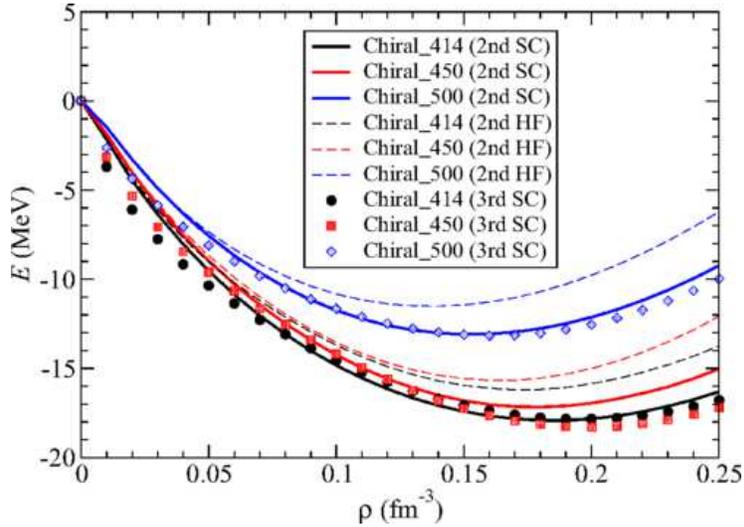,scale=0.55}
\end{minipage}
\begin{minipage}[t]{16.5 cm}
\caption{EOSs for symmetric matter calculated at second and third order in MBPT with different chiral potentials (see Ref. \cite{holt2017}). ``HF" indicates results obtained using the single--particle HF spectrum whereas ``SC" indicates results obtained with a self--consistent spectrum computed including the second--order self--energy. 
Taken from Ref. \cite{holt2017}. }
\label{hoka}
\end{minipage}
\end{center}
\end{figure}

It is important to mention that treating symmetric matter with {\it{ab--initio}} models is much more delicate. It turns out that the equilibrium point of symmetric matter is typically poorly described when chiral interactions are used. To overcome this limitation and find a way to describe in a satisfactory way the saturation point of symmetric matter, considerable effort is dedicated in the community and sophisticated calculations are performed. Reference  \cite{holt2017} represents an example of these sophisticated studies. Figure \ref{hoka}, taken from Ref. \cite{holt2017}, shows a set of EOSs computed for symmetric matter at second-- and third-- order in MBPT (see Ref. \cite{holt2017} for details on the used chiral potentials). Two types of calculations are done, with a HF spectrum and with a self--consistent (SC) spectrum obtained by including the second--order self--energy. 
One may deduce from the figure the extreme difficulty of correctly describing the saturation point of matter. Each time the energy is lowered to get closer to the empirical point, -16 MeV, the minimum of the curve is shifted to higher densities providing a saturation density which is higher than the empirical point, 0.16 fm$^{-3}$. Owing to this,  
in all cases,  either  the value of the corresponding energy,  or the value of the corresponding density, or both values are shifted with respect to the location of the empirical saturation point.      

\begin{figure}[tb]
\begin{center}
\begin{minipage}[t]{11 cm}
\epsfig{file=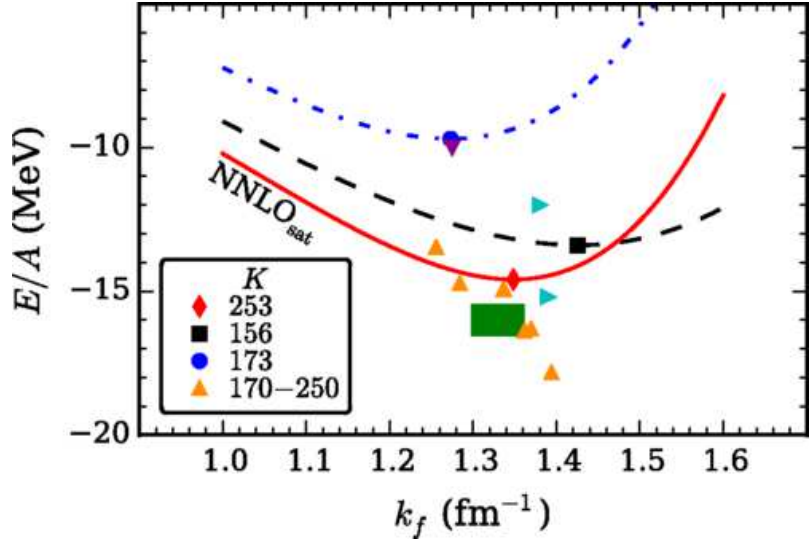,scale=0.6}
\end{minipage}
\begin{minipage}[t]{16.5 cm}
\caption{EOSs for symmetric matter obtained using different chiral potentials. The result corresponding to the N$^2$LO$_{sat}$ potential is represented by the red solid line. The blue dotted--dashed line and the black dashed line are extracted from Ref. \cite{hagen2014}. All the symbols indicate saturation points and the green area represents the empirical saturation point. Upward triangles, rightward triangles, and downward triangles are extracted from Refs. \cite{hebeler2011}, \cite{coraggio2014}, and \cite{carbone2013}, respectively. The values reported in the legend are the corresponding incompressibilities (in MeV). 
Taken from Ref. \cite{ekstrom}. }
\label{n2losat}
\end{minipage}
\end{center}
\end{figure}

A compromise consisting in adjusting the interaction on binding energies and radii of some selected C and O isotopes led in Ref. \cite{ekstrom} to an improved description of the saturation point with the so--called N$^2$LO$_{sat}$ interaction. The resulting curve is shown in Fig. \ref{n2losat} with a red solid line. The EOS is shown in the figure as a function of the Fermi momentum of matter.   
The blue dotted--dashed line and the black dashed line are extracted from Ref. \cite{hagen2014}. All the symbols indicate saturation points and the green area represents the empirical saturation point. Upward triangles, rightward triangles, and downward triangles are  results extracted  from Refs. \cite{hebeler2011}, \cite{coraggio2014}, and \cite{carbone2013}, respectively. The values reported in the legend are the corresponding incompressibilities.

\subsection{\it Nuclear matter}
The EDF framework has the advantage of properly describing the equilibrium point of matter, which is in general one of the constraints used in the adjustment protocol of the parameters. 

We report here in more detail the work developed in Refs. \cite{mo2010,yang2016} and also summarized in Ref. \cite{gra2016}, where the EOS of infinite nuclear matter is analyzed analytically up to second order with a Skyrme--type interaction and the parameters of the interaction are adjusted at second order to reproduce chosen benchmark EOSs. 
Since this work represents the heart and the starting point of several studies on EFT--inspired functionals that will be discussed later in this review, details will be provided and the main equations illustrated. 

 With second--order calculations performed with Skyrme--type interactions, two problems arise if one of the Skyrme parametrizations available on the market is employed, namely a double counting of correlations (because the parameters were adjusted performing mean--field calculations) and ultraviolet divergences (because of the zero range of the interaction). 

\begin{figure}[tb]
\begin{center}
\begin{minipage}[t]{8 cm}
\epsfig{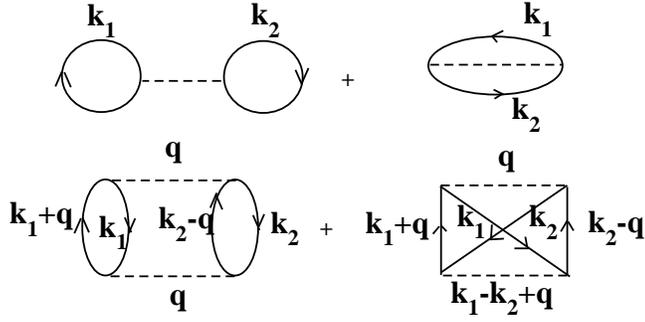}
\end{minipage}
\begin{minipage}[t]{16.5 cm}
\caption{Direct (left) and exchange (right) diagrams leading to the MF (upper row) and to the second--order (lower row) energy contributions to the EOS of nuclear matter. 
Taken from Ref. \cite{mo2010}. }
\label{dia}
\end{minipage}
\end{center}
\end{figure}

Let us consider the EOS up to second order. The standard Skyrme functional has to be extended and truncated at second order for generating this EOS. The energy contribution corresponding to the MF EOS is provided by the direct and exchange diagrams represented in the upper row of Fig. \ref{dia} (diagrams on the left and on the right, respectively). These contributions lead to the energies per particle already reported in Eqs. (\ref{eos1sm}) and (\ref{eos1nm}) for symmetric and pure neutron matter, respectively. 
For the second--order EOS one has to compute the direct and exchange terms represented by the lower row of Fig. \ref{dia} (diagrams on the left and on the right, respectively). Using a zero--range interaction, the computation of the lower row of Fig. \ref{dia} leads to an ultraviolet divergence. 

The Skyrme interaction, already shown in Eq. (1), may be written as a function of the incoming and outgoing relative momenta $\bf{k}$ and $\bf{k'}$, respectively, as 
\begin{equation}
v(\mathbf{k},\mathbf{k^{\prime
}}) =t_{0}(1+x_{0}P_{\sigma })+\frac{1}{2}t_{1}(1+x_{1}P_{\sigma })(\mathbf{k%
}^{\prime 2}+\mathbf{k}^{2}) +t_{2}(1+x_{2}P_{\sigma })\mathbf{k}^{\prime
}\cdot \mathbf{k}   
+\frac{1}{6}t_{3}(1+x_{3}P_{\sigma })\rho ^{\alpha }, 
\label{skkk}
\end{equation}
where the used convention is 
\begin{equation}
v(\mathbf{k},\mathbf{k^{\prime }})=\int \int d^{3}\mathbf{r}d^{3}\mathbf{%
r^{\prime }}e^{-i\mathbf{k}\cdot \mathbf{r}}v(\mathbf{r},\mathbf{r^{\prime }}%
)e^{i\mathbf{k^{\prime }}\cdot \mathbf{r^{\prime }}}.  \label{transfo}
\end{equation}%
The spin--orbit and tensor parts of the interaction are omitted for simplicity. They do not enter in the EOS of matter at first order, but should in principle be included when second--order calculations are performed. 

One may write the second--order contribution to the energy as 
\begin{equation}
E^{(2)} = -\frac{1}{4} \frac{\Omega^3}{(2\pi)^9} \int d^3\mathbf{k_1}
\int d^3\mathbf{k_2} \int d^3\mathbf{q} \frac{|<\mathbf{k_1} \mathbf{k_2}
|V| \mathbf{k_1^{\prime}} \mathbf{k_2^{\prime}>}|^2} {\epsilon _{1}^{\prime
}+\epsilon _{2}^{\prime }-\epsilon _{1}-\epsilon _{2}},  \label{inte1}
\end{equation}%
with $V=v/\Omega$, $\Omega$ being the volume of the box in which the normalization of the wave functions is done; $v$ is given by Eq. (\ref{skkk}), $\bf{q}$ is the transferred momentum (see Fig. \ref{dia}), and 
\begin{eqnarray}
\nn
\mathbf{k_{1}^{\prime }} =\mathbf{q}+\mathbf{k_{1}}, \\
\nn
\mathbf{k_{2}^{\prime }
}=\mathbf{k_{2}}-\mathbf{q}.
\end{eqnarray}
$\mathbf{k_{1}}$ and $\mathbf{k_{2}}$ lie inside the Fermi spheres defined by the Fermi momenta $k_{F1}$ and $k_{F2}$, respectively, 
\be
\mid \mathbf{k_{1}}\mid <k_{F1},\;\;\mid \mathbf{k_{2}}\mid <k_{F2}.  
\ee
Thus, the integrals on $\mathbf{k_{1}}$ and $\mathbf{k_{2}}$ in Eq. (\ref{inte1}) do not diverge. On the other side, one has 
\be
\mid \mathbf{q+k_{1}}\mid >k_{F1},\;\;\mid \mathbf{k_{2}-q}\mid >k_{F2},
\ee
which means that the integral on the transferred momentum $\bf{q}$ in Eq. (\ref{inte1}) diverges. 
It is useful to recall that, whereas for symmetric matter the Fermi momentum (equal for neutrons and protons) and the total density are related by $k_F=(3\pi^2\rho/2)^{1/3}$, for the case of neutron matter the relation between the neutron Fermi momentum and the density is  
$k_F=(3\pi^2\rho)^{1/3}$. For asymmetric matter the Fermi momenta for neutrons and protons are obviously not the same. 

We introduce the propagator $G$, 
\begin{eqnarray}
G &=&\frac{-1}{\epsilon _{1}^{\prime }+\epsilon _{2}^{\prime }-\epsilon
_{1}-\epsilon _{2}},\;\;\epsilon _{i}^{(^{\prime })}=\frac{\hbar
^{2}k_{i}^{(^{\prime })2}}{2m_{i}^{\ast }},  \label{prop}
\end{eqnarray}%
where $m^*$ is the effective mass of the nucleon. Equation (\ref{inte1}) reads
\begin{equation}
E^{(2)} = \frac{1}{4} \frac{\Omega}{(2\pi)^9} \int d^3\mathbf{k_1}
\int d^3\mathbf{k_2} \int d^3\mathbf{q} |<\mathbf{k_1} \mathbf{k_2}
|v| \mathbf{k_1^{\prime}} \mathbf{k_2^{\prime}}>|^2 G.  \label{integral}
\end{equation}%
One can either put a cutoff on the transferred momentum $\bf{q}$ or introduce the incoming and outgoing relative momenta (which appear in Eq. (\ref{skkk})), 
\begin{equation}
\mathbf{k}=\frac{\mathbf{k_1}-\mathbf{k_2}}{2}, \; \mathbf{k^{\prime}}=\frac{%
\mathbf{k_1^{\prime}}-\mathbf{k_2^{\prime}}}{2}= \frac{\mathbf{k_1}-\mathbf{%
k_2}}{2}+\mathbf{q}  \label{varia}
\end{equation}
and put a cutoff on the outgoing relative momentum $\bf{k'}$. The latter way of dealing with the cutoff is similar to what is done for the low--momentum interaction $V_{low-k}$ \cite{bo2003}. 

In Ref. \cite{mo2010}, the integral is performed on $\bf{k_1}$, $\bf{k_2}$, and $\bf{q}$,    the cutoff is put on the transferred momentum $\bf{q}$, and a simplified Skyrme force is used where only the $t_0$ and $t_3$ terms are taken into account ($t_0-t_3$ model). Also, the parameters $x_0$ and $x_3$ are taken  equal to zero for simplicity. The interaction in Eq. (\ref{skkk}) can thus simply be written as a density--dependent coupling constant $g$, 
\be
g=t_0+\frac{1}{6} t_3 \rho^{\alpha}.
\ee
\begin{figure}[tb]
\begin{center}
\begin{minipage}[t]{8 cm}
\epsfig{file=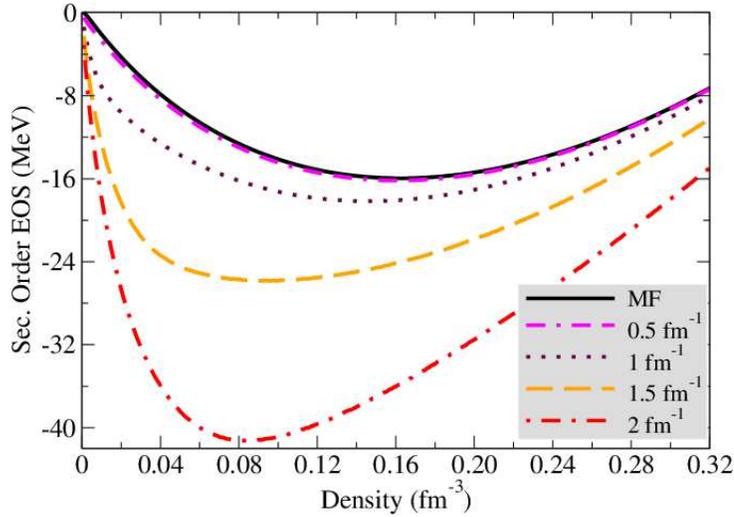,scale=0.4}
\end{minipage}
\begin{minipage}[t]{16.5 cm}
\caption{Second--order EOSs computed with the parameters of the interaction SkP for different values of the cutoff $\Lambda$ on the transferred momentum (indicated in the legend). The SkP--MF EOS is also plotted. 
Taken from Ref. \cite{gra2016}. }
\label{lindi}
\end{minipage}
\end{center}
\end{figure}
The integral leading to the second--order contribution is evaluated in Ref. \cite{mo2010} using the technique of DuBois \cite{dubois} for the region of transferred momentum where $0<q<2$.  
For such a $t_0-t_3$ model, the second--order contribution to the energy was found to have a finite part plus a cutoff--dependent contribution which is linearly diverging with respect to the cutoff. In particular, if $\Lambda$ is the cutoff put on the transferred momentum $\bf{q}$, the asymptotic behavior of the divergent part of the EOS of symmetric matter reads
\be
\frac{E_{asympt}^{(2)}}{A}=-\frac{m^* g^2 k_{F}^{4}}{8  \pi ^{4}\hbar^2}
\left[\tilde{\Lambda} -\frac{3}{35}(11-2 \ln 2)
+{\cal O}(\tilde{\Lambda}^{-1})\right],
\label{asym}
\ee
where $\tilde{\Lambda}=\Lambda/k_F$.
This divergence is shown in Fig. \ref{lindi}, where second--order EOSs for symmetric matter are shown for several values of the cutoff $\Lambda$ together with the MF EOS computed with the parametrization SkP, which represents an illustration of reasonable EOS for symmetric matter. The second--order EOSs are evaluated using the same parameters of the interaction SkP. One may observe that the resulting curves are strongly distanced from  the SkP--MF EOS  with increasing cutoff values and at all densities. The saturation density, which characterizes the equilibrium point of symmetric matter, is also strongly modified with increasing cutoff values and is shifted to lower densities. 
The SkP parametrization is chosen for producing the MF reference curve because the analysis is done here for a $t_0-t_3$ model and the SkP parametrization provides a MF EOS for symmetric matter where only the $t_0$ and $t_3$ contributions are non zero (the parameters of the velocity--dependent terms are such that these contributions are equal to zero in the SkP--MF EOS).  

Let us follow the more general derivation illustrated in Ref. \cite{yang2016}, where the full Skyrme interaction of Eq. (\ref{skkk}) is used and the analytical expressions are computed by following the procedure of Ref. \cite{baranger}. This procedure can be applied only to cases where there is a unique Fermi momentum  (either symmetric matter or pure neutron matter). The second--order EOSs for symmetric and neutron matter were deduced analytically in Ref. \cite{yang2016} whereas the EOS for asymmetric matter was computed with a numerical Monte Carlo integration \cite{mc1,mc2}. For the numerical Monte Carlo calculations, a cutoff on the transferred momentum $\bf{q}$ was used whereas the analytical derivation was done using the incoming and outgoing relative momenta and the cutoff was put in this case on the outgoing momentum. 

Let us consider as an illustration the analytical derivation for symmetric matter. By making the change of variables given by Eq. (\ref{varia}), the propagator may be written as
\begin{equation}
G=\frac{-m}{\hbar^2(k^{\prime }{}^{2}-k^{2})}.
\label{pro}
\end{equation}%
As a first step, we have taken the effective mass $m^*$ equal to the bare mass $m$.
From Eq. (\ref{integral}), by making the change of variables (\ref{varia}), writing direct and exchange terms explicitly as well as the sums over spin and isospin, and dividing by the number of particles one has, 
\bea
\nonumber
\frac{E_{sym}^{(2)}}{A} &=& -\frac{3m}{32 (2\pi)^7 \hbar^2 k_{F}^{3}}%
\sum\limits_{STM_{S}M_{S^{\prime }}}(2T+1)\int \int \int d^{3}\mathbf{K}d^{3}%
\mathbf{k}d^{3}\mathbf{k}^{\prime }  \\
&\times& \frac{|\langle X_{M_{S}}^{S}|v_{ST}(\mathbf{k},\mathbf{k}^{\prime
})-(-)^{S+T}v_{ST}(\mathbf{k},-\mathbf{k}^{\prime })|X_{M_{S^{\prime
}}}^{S}\rangle |^{2}}{(k^{\prime 2}-k^{2})},  \label{eos1}
\eea
where $S$ and $T$ denote the total spin and isospin, $X_{M_{S}}^{S}$ and $X_{M_{S^{\prime
}}}^{S}$ are two--body spin states, and 
 $M_{S}$ and $M_{S^{\prime}}$ are the projections of $S$ on the $z$ axis. One notices that $v_{ST}$ is obtained from the interaction (\ref{skkk}) after evaluation of the expectation value in the isospin state. The third variable $\mathbf{K}$ is chosen equal to $\mathbf{K}\equiv \mathbf{k}_{1}+\mathbf{k}_{2}$ as in Ref. \cite{baranger}. The integrals on $\mathbf{K}$ and $\mathbf{k}$ do not diverge whereas a diverging contribution is generated by the integral on the outgoing momentum. We use a cutoff $\tilde{\lambda}$ as a regulator and we introduce the dimensionless quantity 
$\lambda=\tilde{\lambda}/k_F$. 

We first expand the interaction in partial waves, 
\begin{equation}
v_{ST}(\mathbf{k},\mathbf{k}^{\prime })=\sum\limits_{JM_{J}ll^{\prime
}}v_{ST,ll^{\prime }}^{J}(k,k^{\prime })y_{lS}^{JM_{J}}(\widehat{\mathbf{k}}%
)[y_{l^{\prime }S}^{JM_{J}}(\widehat{\mathbf{k}}^{\prime })]^{\dag },
\label{partial}
\end{equation}
where
\begin{equation}
y_{lS}^{JM_{J}}(\widehat{\mathbf{k}})=\sum\limits_{m_{l},M_{S}}\langle
JM_{J}|lSm_{l}M_{S}\rangle Y_{lm_{l}}(\widehat{\mathbf{k}})X_{M_{S}}^{S}.
\end{equation}%
In Eq. (\ref{partial}) $l$ and $l'$ must correspond to the same parity. The antisymmetrization condition implies that 
\begin{equation}
(-1)^{l+S+T}=(-1)^{l^{\prime}+S+T}=-1,  
\end{equation}
which means that the direct and exchange terms in Eq. (\ref{eos1}) are the same. After evaluating the spin matrix elements one can obtain from Eq. (\ref{eos1})  the following 
expression:
\begin{eqnarray}
\nn
\frac{E_{sym}^{(2)}}{A} &=&-\frac{3m}{16 (2\pi)^8 \hbar^2 k_{F}^{3}}
\sum_{STJ\bar{J}LMll^{\prime}\bar{l}\bar{l^{\prime}}} \delta_{STl} \delta_{ST%
\bar{l}} (2T+1) (2J+1) (2\bar{J}+1) (2L+1)^{-1}   \\
\nn
&\times& \left[ (2l+1) (2l^{\prime}+1) (2\bar{l}+1)(2\bar{l^{\prime}}+1)%
\right]^{1/2} < L0|l\bar{l}00> < L0|l^{\prime}\bar{l^{\prime}}00> W(J\bar{J}l%
\bar{l};LS) W(J\bar{J}l^{\prime}\bar{l^{\prime}};LS)   \\
&\times& \int \int \int d^{3}\mathbf{K}d^{3}\mathbf{k}d^{3}\mathbf{k}%
^{\prime} Y_{LM}(\hat{k}) Y_{LM}^*(\hat{k^{\prime}}) v_{ST,ll^{\prime
}}^{J}(k,k^{\prime }) v_{ST,\bar{l}\bar{l^{\prime }}}^{\bar{J}}(k,k^{\prime
}) (k^{\prime 2}-k^2)^{-1},  \label{partialw}
\end{eqnarray}
where $W$ are the Racah coefficients and 
\begin{equation}
\delta_{STl}=\frac{1}{2}\left[1-(-1)^{S+T+l}\right].
\end{equation}
For our interaction, which is diagonal on $l$ and independent of $J$, one may write 
\be
v_{ST,ll^{\prime }}^{J}(k,k^{\prime })=\delta _{ll^{\prime
}}v_{S,T,l}(k,k^{\prime }).
\ee
From Eq. (\ref{partialw}), one deduces that $l$ and $\bar{l}$ must have the same parity (from the product $\delta_{STl} \delta_{ST\bar{l}}$). Second--order contributions may thus mix only odd-odd and even-even partial waves of the interaction. 
Even--wave contributions of the Skyrme force are the terms $t_0$, $t_3$, and $t_1$ ($s$--wave). There is only one odd--wave contribution that is the $p$--wave $t_2$ term. We will thus have at second order only terms proportional to $t_0^2$, $t_3^2$, $t_1^2$, $t_0 t_3$, $t_0 t_1$, 
$t_1 t_3$, and $t_2^2$. $L$ in Eq. (\ref{partialw}) may take only the values $L=0,2$ for the $p$--wave mixing and the value $L=0$ for the $s$--wave mixing.  

The squares of the interaction $v^2_{S,T,l}(k,k^{%
\prime})$ in the isovector case $T=1$ [$%
S=0$ ($P_{\sigma }=-1$) for $l=0$ and $S=1$ ($P_{\sigma }=1$) for $l=1$] read
\begin{eqnarray}
\nonumber
v_{S=0,T=1,l=0}^2 (k,k') &=&(4\pi)^2 [t_{0}^{2}(1-x_{0})^{2}+\frac{1}{4}%
t_{1}^{2}(1-x_{1})^{2}(k^{\prime 4}+2k^{\prime 2}k^{2}+k^{4})+\frac{1}{36}%
t_{3}^{2}(1-x_{3})^{2}\rho ^{2\alpha }  \\
\nonumber
&&+t_{0}(1-x_{0})t_{1}(1-x_{1})(k^{\prime 2}+k^{2})+\frac{1}{6}t_{1}
(1-x_{1})(k^{\prime
2}+k^{2})t_{3}(1-x_{3})\rho ^{\alpha }   \\
&&+\frac{1}{3}t_{0}(1-x_{0})t_{3}(1-x_{3})\rho ^{\alpha }],  \label{vsquare1}
\end{eqnarray}%
and 
\begin{equation}
v_{S=1,T=1,l=1}^{2}(k,k') =\frac{(4\pi)^2}{9}t_{2}^{2}(1+x_{2})^{2}(k^{\prime }k)^{2}.    \label{vsquare2}
\end{equation}%
For the isoscalar case, $T=0$, one has $S=1$ ($P_{\sigma
}=1$) for $l=0$ and $S=0$ ($P_{\sigma }=-1$) for $l=1$. The 
square of the interaction $v_{S=1,T=0,l=0}^{2}(k,k^{\prime })$ and $%
v_{S=0,T=0,l=1}^{2}(k,k^{\prime })$ can be deduced from Eqs. (\ref{vsquare1}%
) and (\ref{vsquare2}), respectively, with $(1+x_{i})$ instead of $%
(1-x_{i})$ in Eq. (\ref{vsquare1}) and $(1-x_{2})$ instead of $(1+x_{2})$ in Eq. (%
\ref{vsquare2}).

\begin{figure}[tb]
\begin{center}
\begin{minipage}[t]{8 cm}
\epsfig{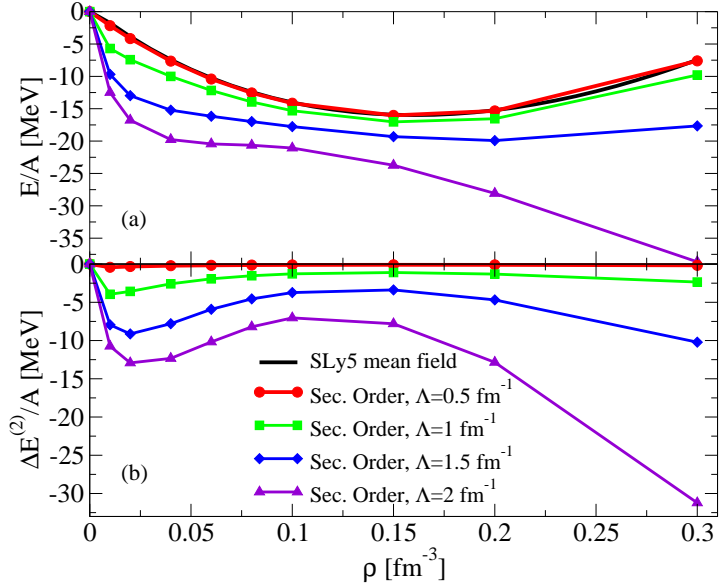}
\end{minipage}
\begin{minipage}[t]{16.5 cm}
\caption{(a) Second--order EOSs calculated for symmetric matter and at different cutoff values with the parameters of the interaction SLy5. Also the MF SLy5 EOS is plotted. (b) Second--order corrections.
Taken from Ref. \cite{yang2016}. }
\label{divefull}
\end{minipage}
\end{center}
\end{figure}

Using the change of variables 
\begin{equation}
\mathbf{y}=\frac{\mathbf{k}}{k_{F}},\;\mathbf{y^{\prime }}=\frac{\mathbf{%
k^{\prime }}}{k_{F}},\;\mathbf{x}=\frac{\mathbf{K}}{2k_{F}},
\end{equation}%
where 
\begin{equation}
|\mathbf{y}|<1,\;|\mathbf{y^{\prime }}|<\lambda ,
\end{equation}%
one can integrate over all the angles by introducing the following function: 
\begin{equation}
J_{LM}(x,y,y^{\prime })=\int \int \int d\hat{x}d\hat{y}d\hat{y^{\prime }}%
Y_{LM}(\hat{y})Y_{LM}^{\ast }(\hat{y^{\prime }}).
\end{equation}%
One can show that 
\begin{equation}
J_{LM}(x,y,y^{\prime })=16\pi ^{2}\delta _{M,0}(2L+1)A_{L}(y,x)A_{L}^{\prime
}(y^{\prime },x),
\end{equation}%
where the  functions $A_{L}^{(^{\prime })}$ are reported 
in Eqs. (3.16a)-(3.17b) of Ref. \cite{baranger}.
A new function $I$ is introduced by performing the radial integration on $x$, 
\begin{equation}
I^{(L)}(y,y^{\prime })=\int_{0}^{1}x^{2}dxA_{L}(y,x)A_{L}^{\prime
}(y^{\prime },x).
\end{equation}%
Finally, one may write two expressions for the $p$-- and $s$--wave contributions, 
\begin{eqnarray}
\nn
\frac{E_{sym (l=1)}^{(2)}}{A} &=&-\frac{18mk_{F}^{4}}{4\pi^4 \hbar^2}\sum_{ST}(2T+1)(2S+1)\delta _{ST1}\int \int
dydy^{\prime }\frac{y^{2}y^{\prime 2}}{y^{\prime 2}-y^{2}}\left[
I^{(0)}(y,y^{\prime })+2I^{(2)}(y,y^{\prime })\right] \\
\nn
&\times&
[v_{S,T,1}(k_{F}y,k_{F}y^{\prime })]^{2},   \\
\frac{E_{sym (l=0)}^{(2)}}{A} &=&-\frac{18mk_{F}^{4}}{4\pi^4 \hbar^2}\sum_{ST}\delta _{ST0}\int \int dydy^{\prime }%
\frac{y^{2}y^{\prime 2}}{y^{\prime 2}-y^{2}}I^{(0)}(y,y^{\prime
})[v_{S,T,0}(k_{F}y,k_{F}y^{\prime })]^{2}.  \label{a2}
\end{eqnarray}%
The two above expressions lead to two terms ($s$-- and $p$--wave) that have to be summed up for obtaining the EOS of symmetric matter. The second--order contribution to the EOS is thus the sum of the following two terms,

\begin{eqnarray}
\nn
 \frac{E_{sym (l=0)}^{(2)}}{A}&=&-\frac{mk_{F}^{4}}{110880\hbar^2\pi ^{4}} \Bigg\{ 
\Bigg[ 
-6534+1188 ln[2]+3564\lambda -19602\lambda ^{3}-5940\lambda ^{5}  \\ \nn
&+&(1782-20790\lambda ^{4})  ln\bigg[\frac{\lambda -1}{\lambda +1}\bigg]  
+(24948\lambda ^{5}-5940\lambda ^{7}) ln\bigg[\frac{\lambda ^{2}-1}{\lambda ^{2}}\bigg]
\Bigg] \widetilde{T}_{03}^{2}  \\ \nn
&+& \Bigg[ 
-
14696+2112 ln[2]+5280\lambda -2860\lambda ^{3} 
- 48840\lambda ^{5} - 18480\lambda ^{7} \\ \nn &+& (2640-55440\lambda ^{6}) ln\bigg[\frac{%
\lambda -1}{\lambda +1}\bigg]  
+(71280\lambda ^{7}-18480\lambda ^{9}) ln\bigg[\frac{\lambda ^{2}-1}{\lambda ^{2}}%
\bigg]%
\Bigg] k_{F}^{2}\widetilde{T}_{03}\widetilde{T}_{1}  \\ \nn
&+& \Bigg[ 
-9886+1128 ln[2]+
2520\lambda +147\lambda ^{3}-3654\lambda ^{5} 
-
35280\lambda ^{7}-15120\lambda ^{9} \\  &+& (1260-41580\lambda ^{8}) ln\bigg[\frac{%
\lambda -1}{\lambda +1}\bigg]  
+(55440\lambda ^{9}-15120\lambda ^{11}) ln[\frac{\lambda ^{2}-1}{\lambda ^{2}%
}]%
\Bigg] k_{F}^{4}\widetilde{T}_{1}^{2}%
\Bigg\}
\label{symml0}
\end{eqnarray}%
and
\begin{eqnarray}
\nn
\frac{E_{sym (l=1)}^{(2)}}{A}&=&-\frac{mk_{F}^{8}}{73920\hbar^2\pi ^{4}}\Bigg\{ \Bigg[ 
-1033+156 ln[2]+420\lambda +140\lambda ^{3}-840\lambda ^{5} 
- 5880\lambda ^{7}-2520\lambda ^{9}\\  &-& (-210+6930\lambda ^{8}) ln\bigg[\frac{\lambda
-1}{\lambda +1}\bigg] 
+(9240\lambda ^{9}-2520\lambda ^{11}) ln\bigg[\frac{\lambda ^{2}-1}{\lambda ^{2}}\bigg]%
\Bigg] \widetilde{T}_{2}^{2}\Bigg\},
\label{symml1} 
\end{eqnarray}
with, 
\begin{eqnarray}
\nonumber
\widetilde{T}_{03}^{2} &=&\left[ t_{0}(1-x_{0})+\frac{1}{6}%
t_{3}(1-x_{3})\rho ^{\alpha }\right] ^{2}+\left[ t_{0}(1+x_{0})+\frac{1}{6}%
t_{3}(1+x_{3})\rho ^{\alpha }\right] ^{2}   \\
\nonumber
\widetilde{T}_{1}^{2} &=&\frac{1}{4}t_{1}^{2}\left[
(1-x_{1})^{2}+(1+x_{1})^{2}\right] =\frac{1}{2}t_{1}^{2}(1+x_{1}^{2})
 \\
\nonumber
\widetilde{T}_{03}\widetilde{T}_{1} &=&\frac{t_{1}}{2}\left[ [t_{0}(1-x_{0})+%
\frac{1}{6}t_{3}(1-x_{3})\rho ^{\alpha }](1-x_{1})+[t_{0}(1+x_{0})+\frac{1}{6%
}t_{3}(1+x_{3})\rho ^{\alpha }](1+x_{1})\right]   \\
\nonumber
\widetilde{T}_{2}^{2} &=&[t_{2}^{2}(1-x_{2})^{2}+9t_{2}^{2}(1+x_{2})^{2}]/9 
\\
&=&\frac{2}{9}t_{2}^{2}(5+8x_{2}+5x_{2}^{2}).
\label{tnomsymm}  
\end{eqnarray}%
The asymptotic behavior of the EOS may be easily obtained. For example, for the $s$--wave part one has: 
\begin{eqnarray}
\nn
\frac{E_{sym (l=0),asympt.}^{(2)}}{A}&=&-\frac{9mk_{F}^{4}}{2\hbar^2\pi ^{4}}%
\Bigg[ 
\frac{k_{F}^{4}\widetilde{T}_{1}^{2}}{360}\lambda ^{5}+\bigg(\frac{k_{F}^{2}%
\widetilde{T}_{03}\widetilde{T}_{1}}{108}+\frac{k_{F}^{4}\widetilde{T}%
_{1}^{2}}{240}\bigg)\lambda ^{3}+\bigg(\frac{\widetilde{T}_{03}^{2}}{72}+\frac{%
k_{F}^{2}\widetilde{T}_{03}\widetilde{T}_{1}}{60}+\frac{k_{F}^{4}\widetilde{T%
}_{1}^{2}}{140}\bigg)\lambda  \\ \nn
&+& \frac{44k_{F}^{2}\widetilde{T}_{03}\widetilde{T}%
_{1}(-167+24 ln[2])+k_{F}^{4}\widetilde{T}_{1}^{2}(-4943+564 ln[2])+297%
\widetilde{T}_{03}^{2}(-11+2 ln[2])}{249480} \\ 
&-& \bigg(\frac{\widetilde{T}_{03}^{2}}{240}+\frac{k_{F}^{2}\widetilde{T}_{03}%
\widetilde{T}_{1}}{140}+\frac{k_{F}^{4}\widetilde{T}_{1}^{2}}{270}\bigg)/\lambda
+O(\lambda ^{-2})%
\Bigg]. 
\end{eqnarray}%

For neutron matter, the integrals to solve are the same with different factors. The reader may refer to Ref. \cite{yang2016} for more details and for the analytical expressions of the EOS of neutron matter. 
All the numerical calculations were performed with Monte Carlo integrations where a cutoff $\Lambda$ on the transferred momentum was used. In the numerical calculations, also asymmetric matter was treated and the case $\delta=0.5$ was chosen as an illustration, with $\delta=(\rho_n-\rho_p)/(\rho_n+\rho_p)$. 

As an example, the upper panel of Fig. \ref{divefull} shows the second--order EOS of symmetric matter computed numerically for several values of $\Lambda$ with the parameters of the Skyrme parametrization SLy5. The SLy5--MF EOS is also shown. The lower panel presents only the second--order correction. Whereas the simple $t_0-t_3$ model produces an ultraviolet divergence that depends linearly on the momentum cutoff, the inclusion of the other terms of the Skyrme interaction leads to a stronger divergent behavior.

\begin{figure}[tb]
\begin{center}
\begin{minipage}[t]{8 cm}
\epsfig{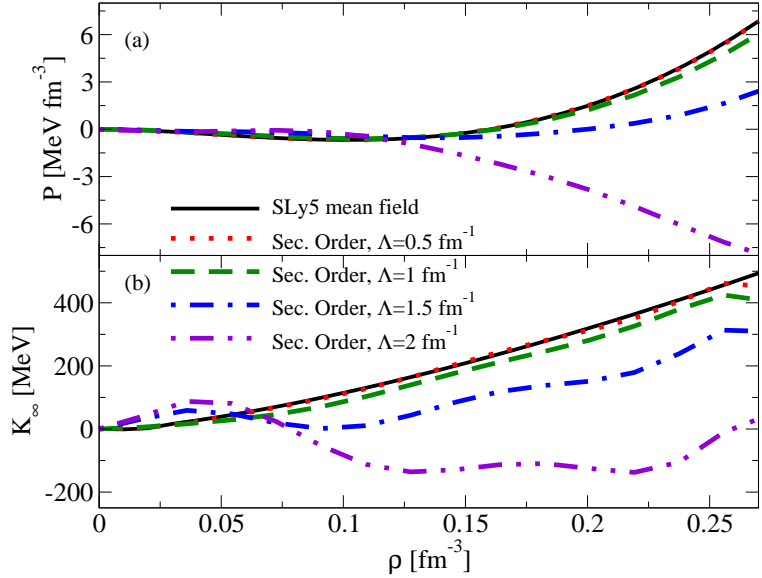}
\end{minipage}
\begin{minipage}[t]{16.5 cm}
\caption{Second--order pressure (a) and incompressibility (b) calculated with the parameters of the Skyrme interaction SLy5. 
Taken from Ref. \cite{yang2016}. }
\label{prebefore}
\end{minipage}
\end{center}
\end{figure}

\begin{figure}[tb]
\begin{center}
\begin{minipage}[t]{8 cm}
\epsfig{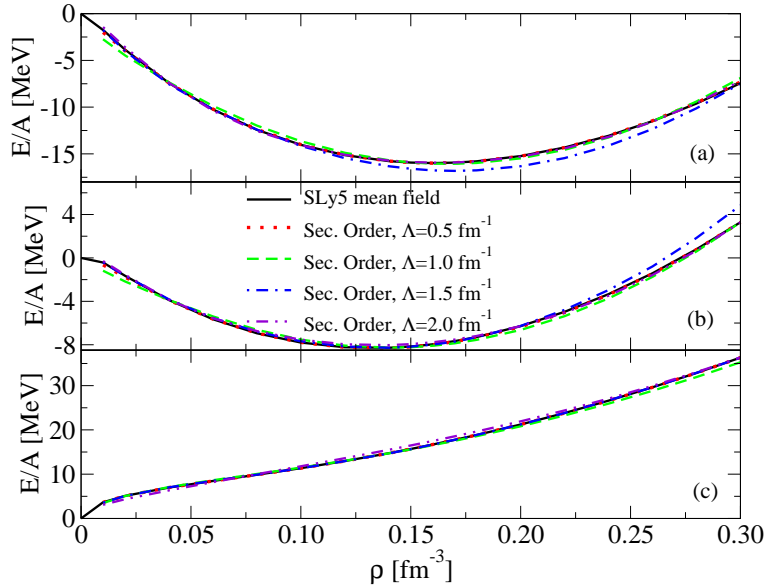}
\end{minipage}
\begin{minipage}[t]{16.5 cm}
\caption{Second--order EOS for symmetric (a), asymmetric with $\delta=$ 0.5 (b), and pure neutron (c) matter calculated with the readjusted parameters obtained with a simultaneous fit of the three EOSs (the parameters are listed in Tables \ref{simultable} and \ref{simultable1}). The benchmark data, which are the SLy5--MF corresponding EOSs are also shown in the figure. 
Taken from Ref. \cite{yang2016}. }
\label{simul}
\end{minipage}
\end{center}
\end{figure}

\begin{figure}[tb]
\begin{center}
\begin{minipage}[t]{8 cm}
\epsfig{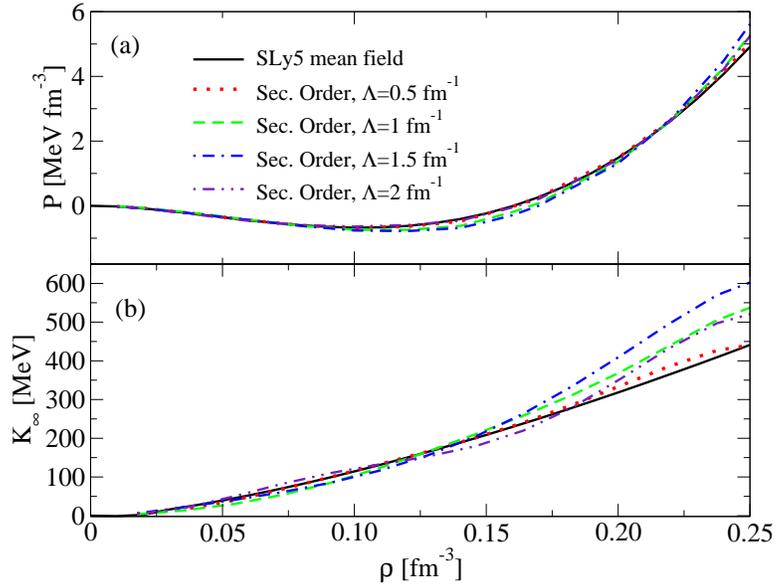}
\end{minipage}
\begin{minipage}[t]{16.5 cm}
\caption{Second--order pressure (a) and incompressibility (b) calculated with the adjusted parameters listed in Tables \ref{simultable} and \ref{simultable1} and compared with the MF SLy5 pressure and incompressibility. 
Taken from Ref. \cite{yang2016}. }
\label{preafter}
\end{minipage}
\end{center}
\end{figure}

\begin{figure}[tb]
\begin{center}
\begin{minipage}[t]{8 cm}
\epsfig{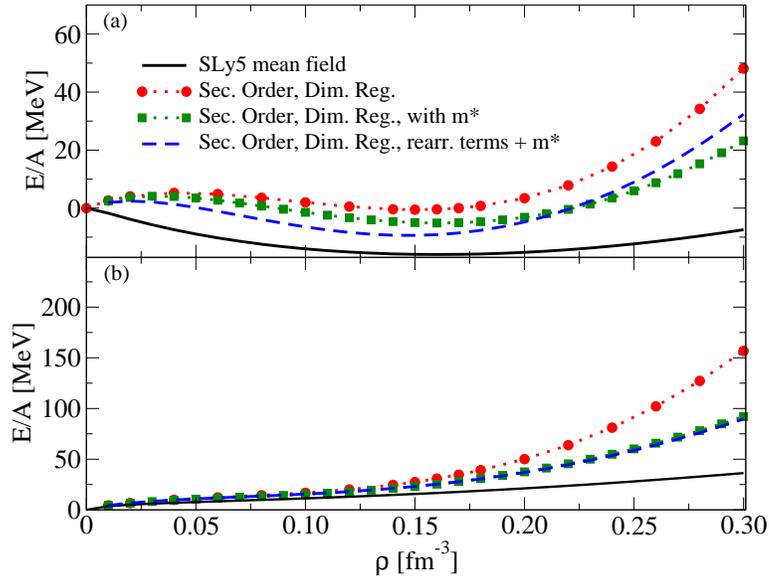}
\end{minipage}
\begin{minipage}[t]{16.5 cm}
\caption{(a) Second--order dimensional regularized EOS for symmetric matter compared with the SLy5--MF EOS and computed with the SLy5 parameters in the following cases: $m^*=m$ and no rearrangement terms (red circles and dotted line), $m^*$ equal to the MF effective mass without rearrangement terms (green squares and dotted line), and $m^*$ equal to the MF effective mass with the inclusion of rearrangement terms (blue dashed line). (b) Same as in (a) but for neutron matter.  
Taken from Ref. \cite{yang2016}. }
\label{emass}
\end{minipage}
\end{center}
\end{figure}

The corresponding second--order pressure $P(\rho,\Lambda)$ and incompressibility  
$K(\rho,\Lambda)$ calculated with the SLy5 parametrization are shown in Fig. \ref{prebefore}, where 
\begin{equation}
P(\rho,\Lambda)=\rho^2\frac{d}{d\rho}\frac{E}{A}(\rho,\Lambda)
\end{equation}
and 
\begin{equation}
K(\rho,\Lambda)=9\rho^2\frac{d^2}{d\rho^2}\frac{E}{A}(\rho,\Lambda).
\end{equation}

The parameters of the interaction may now be readjusted choosing a reasonable symmetric matter EOS 
as a benchmark curve to reproduce. This can be done also for neutron and asymmetric matter. Figure \ref{simul} illustrates the curves resulting from a simultaneous fit performed on symmetric, pure neutron, and asymmetric matter in the case $\delta=0.5$, taking the corresponding SLy5--MF EOSs as benchmark EOSs. 
One may observe the very good quality of the fit. For every value of the chosen cutoff, a BMF set of parameters may be generated so that the adjusted curve is very close to the benchmark EOS. This occurs in all cases, symmetric, neutron, and asymmetric matter. 
Tables \ref{simultable} and \ref{simultable1} show the corresponding values of the obtained parameters compared to the SLy5 parameters.  
For details on the fitting procedure and on the corresponding $\chi^2$ values the reader may refer to Ref. \cite{yang2016}.

\begin{table}
\centering
\caption{Parameters adjusted with a simultaneous fit done on the second--order  symmetric, asymmetric with $\delta=0.5$, and neutron matter EOSs for different values of the cutoff $\Lambda$ 
compared with the original set SLy5. This table shows the parameters $t_0$, $t_1$, $t_2$, and $t_3$.}\vspace{0.5cm}
\begin{tabular}{  c c c c c   }
    \hline
    \hline
   & $\quad t_0$ & $t_1$ & $t_2$ & $t_3$   \\
 & (MeV fm$^3$) & (MeV fm$^5$) & (MeV fm$^5$) & (MeV fm$^{3+3\alpha}$)  \\
SLy5 &-2484.88  &483.13     &  -549.40  &   13736.0    \\
\hline
$\Lambda$(fm$^{-1})$ & & & &   \\
0.5 &   -2245.402  &      493.322     &  -1832.783& 11961.86  \\
1.0 &  -1239.909   &    674.272  &    -387.948 &4687.107 \\
1.5 &  -803.325     &   670.917     &  -42.426   &   4854.284 \\
2.0 &-668.075   &    80.904   &   0.8980 &8779.939 \\
   \hline\hline
\label{simultable}
  \end{tabular}
  \end{table}

\begin{table}
\centering
\caption{Parameters adjusted with a simultaneous fit done on the second--order  symmetric, asymmetric with $\delta=0.5$, and neutron matter EOSs for different values of the cutoff $\Lambda$ 
compared with the original set SLy5. This table shows the parameters $x_0$, $x_1$, $x_2$, $x_3$, and $\alpha$.}\vspace{0.5cm}
\begin{tabular}{  c c c c c c  }
    \hline
    \hline
   &  $x_0$ & $x_1$ & $x_2$ & $x_3$ & $\alpha$   \\
SLy5 &  0.778        &       -0.328  &     -1.0      &   1.267&      0.16667  \\
\hline
$\Lambda$(fm$^{-1})$ & & & & &   \\
0.5 &     0.7462     & -0.3936   &  -0.9684    &    1.309   &    0.1832     \\
1.0 &  0.3649     & -0.5993  &    -1.1349    &    3.4299   &    0.5558  \\
1.5 &  0.1165     & -1.1436   &   -2.6727    &    3.4271 &1.1831   \\
2.0 & 0.1605& 0.3874     & -0.2652   &    0.0004687  &      1.4723   \\
   \hline\hline
\label{simultable1}
  \end{tabular}
  \end{table}

Figure \ref{preafter} illustrates how the pressure and the incompressibility are described with the adjusted parameters. The incompressibility has a maximum discrepancy of $\sim$ 25 MeV at the saturation density, compared to the SLy5--MF incompressibility value. These results are purely predictions since pressure and incompressibility are not used as additional constraints in the adjustment of the parameters. 

In addition to the procedure that we have described above, based on a cutoff regularization, a dimensional regularization procedure was also applied in Ref. \cite{yang2016}. The divergent part of the EOS was in this way removed and the resulting second--order EOS was readjusted on a reference curve. We recall that this regularization method was introduced in the framework of the electroweak theory \cite{dr1,dr2,dr3}: a continuous parameter $d$ is used instead of the integer dimension of the divergent integral. Once the integral is computed, one let $d$ go to the original integer value. The cutoff dependence is removed and only the finite contributions are maintained. 

Another interesting analysis discussed in Ref. \cite{yang2016} addressed the effects of the effective mass and of the rearrangement terms on the second--order EOSs. For this investigation, only symmetric and pure neutron matter were treated. For simplicity, MF effective masses were used, namely 
\begin{equation}
\frac{m^*_{S}}{m}=\left(1+\frac{m}{8\hbar^2}\rho \Theta_S \right)^{-1}
\label{msym}
\end{equation}
 for symmetric matter and 
\begin{equation}
\frac{m^*_{N}}{m}=\left(1+\frac{m}{4\hbar^2}\rho \Theta_S - \frac{m}{4\hbar^2}\rho \Theta_V \right)^{-1}
\label{mneu}
\end{equation}
for neutron matter. 

Reference \cite{car2013} indicated how to compute the rearrangement terms, by using the square of the RPA $B$ matrix. For this, the Landau parameters \cite{pa2015} were employed. As an illustration, Fig. \ref{emass} shows the effects of the effective mass and of the rearrangement terms on the dimensional--regularized EOSs of symmetric and neutron matter before the adjustment of the parameters (the SLy5 parameters are employed). The parameters of the interaction were then eventually adjusted to provide parameter sets tailored for second--order calculations. The effect of  the inclusion of a density--dependent effective mass is clearly visible. The effect of the inclusion of rearrangement terms is much weaker, especially for neutron matter. 

\subsection{\it Renormalizability and first exploratory steps towards a power counting in EDF theories}
A power--counting analysis is presently missing in the EDF framework to guide us in the construction of a given functional/interaction with an appropriate hierarchy of terms. A first step towards the establishment of a rigorous power counting should be the renormalizability analysis of a given model because a renormalizable theory has the good property of being  independent of the used regularization technique. Renormalizability was for instance required in the framework of the pionless EFT for the description of the three--body system and this led to the appearance of a three--body force at leading order \cite{bhk991,bhk992,bhk00}. 

One may require that, at each order of the MBPT, the employed theory is renormalizable. This was done in Ref. \cite{yang2017-1} where the renormalizability of the theory was analyzed in the case of a Skyrme--type interaction used to compute the EOS of symmetric matter up to second order in the MBPT. The cutoff--dependent terms appearing in the EOS were investigated in terms of their dependence on the Fermi momentum $k_F$. Those cutoff--dependent 
 second--order terms having the same $k_F$ dependence as terms generated at leading order may indeed be regrouped together and the cutoff dependence is in this way absorbed by a redefinition of the parameters to adjust. There are however in this Skyrme--based model some cutoff--dependent second--order  terms with different $k_F$ dependencies compared to MF terms. They cannot be regrouped with MF contributions and lead in general to a non renormalizable theory. However, 
by requiring that such terms are equal to zero, one may define a renormalizable theory valid under some 
specific constraints on the parameters and special density dependencies that must be imposed to cancel such contributions. This  {\it{ad hoc}}  procedure obviously ensures renormalizability  only for those restricted cases. A renormalizable theory was in this way defined in Ref. \cite{yang2017-1} for two specific density dependencies, $\alpha=$ 1/3 and $\alpha=$ -1/6, in the simple $t_0-t_3$ model. 

Nevertheless, if a more general renormalizable theory is desired, without specific constraints on the parameters, new terms should be included in the interaction. They have to be chosen so to produce at leading order contributions tailored to absorb the divergent second--order non renormalizable terms (same Fermi momentum dependence in the EOS). Of course, this certainly  renders the employed theoretical model much more complicated and, furthermore, it  introduces at second order additional divergent terms that could not be reabsorbed with the MF part of the EOS. 
Thus, the appropriate strategy should be to include such terms only at leading order whereas all the other terms are included up to second order. 

Guided by this exploratory study on the renormalizability, the authors of Ref. \cite{yang2017-2} made a first attempt to define a strategy  
for constructing a power counting in EDF theories, by the inclusion of counterterms and without restrictions on the density dependence of the interaction. Skyrme--type interactions were employed and both symmetric and pure neutron matter were treated. A density--dependent term is  necessary to ensure the correct description of the equilibrium point of symmetric matter at both first and second order with Skyrme--type interactions \cite{mo2010,yang2016,kaiser}. 

It is well known that the Skyrme MF model is able to describe in a satisfactory way several bulk properties of finite nuclei. This indicates that an EFT--like expansion is expected to exist where the MF scheme should represent the leading order. One may use renormalizability criteria as a guide to define such a leading order. 

On the other hand, it is also known that another expansion exists for very dilute Fermi systems, for example for very dilute neutron matter. It is an expansion in powers of $ak_F$ \cite{LY1,LY2,hammer,platter,fur,puglia,fur2,bishop,gali,abri} where, for neutron matter, $a$ is the $s$--wave neutron-neutron scattering length. 
Such a low--density regime is naturally properly described by EFT \cite{hammer}. 
As will be seen later in this revue, this expansion is however valid only in the very low--density regime which, owing to the very large value of the neutron-neutron scattering length ($a=$ -18.9 fm), means $\rho \le 10^{-6}$ fm$^{-3}$. 
These extremely low densities are not really relevant for describing the physics of finite nuclei, where the densities of interest are located around the saturation point of symmetric matter. 

\begin{figure}[tb]
\begin{center}
\begin{minipage}[t]{8 cm}
\epsfig{file=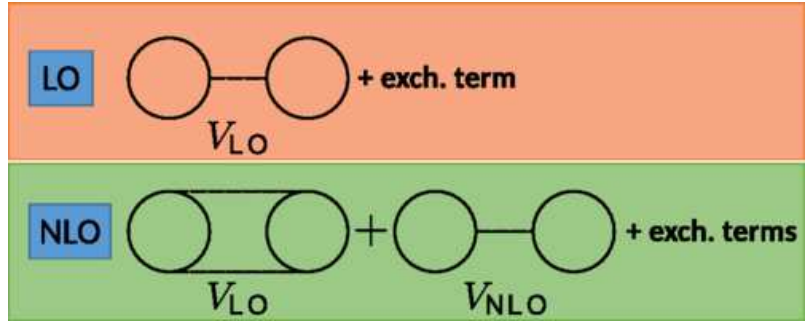,scale=0.6}
\end{minipage}
\begin{minipage}[t]{16.5 cm}
\caption{LO and NLO diagrams in the hierarchy of terms described in the text. 
Taken from Ref. \cite{yang2017-2}.}
\label{power}
\end{minipage}
\end{center}
\end{figure}

The required EFT--like expansion  should thus be different and work in the density window 0 $\sim$ 0.3 - 0.4 fm$^{-3}$. 
The idea developed in Ref. \cite{yang2017-2} is based on the following two points: (i)  renormalizability required order by order in the MBPT, and (ii) construction of a functional that is used up to next--to--leading order (NLO) and which we call NLO EDF. 
The leading order (LO) functional is chosen on the basis of the findings of Ref. \cite{yang2017-1}, namely using the fact that renormalizability leads to a $t_0-t_3$ model. The $t_0-t_3$ model thus defines what we call the LO interaction $V_{LO}$ (however, no restrictions on the density dependence are imposed, since counter terms are included in this procedure). Such an interaction used to compute the upper row of the diagrams shown in Fig. \ref{dia} provides the LO MF contribution to the EOS. It also leads to an NLO contribution when it is used to compute the diagrams shown on the second row of the figure. Let us suppose now to compute MF--type diagrams using a dressed propagator. One can imagine that this can be equivalently described by transposing this effet onto the interaction. This would define an NLO interaction that should be used to evaluate the diagram shown on the upper row of Fig. \ref{dia}. Solving the EOS up to NLO would thus correspond to evaluating the diagrams described in Fig.  
\ref{power}, which is a first very rough idea for defining a power counting with a truncation at NLO.  One could also imagine to continue with the same procedure up to the next--to--next--to--leading order (N$^2$LO). For such a calculation, one should at least evaluate the third--order contribution (in the MBPT) produced by the interaction 
$V_{LO}$ and the LO contribution (in the MBPT) generated by a given $V_{N^2LO}$ interaction introduced by 
requiring renormalizability (renormalizability must be required order by order). 

Let us briefly illustrate the procedure that is adopted in practice in Ref. \cite{yang2017-2} for addressing the computation represented by the diagrams in Fig. \ref{power}. The reader may refer to Ref. \cite{yang2017-2} for all the details. 

Guided by the renormalizability conditions investigated in Ref. \cite{yang2017-1}, the $t_0-t_3$ model is chosen to define the $V_{LO}$ interaction. The simplest form would be provided by a single coupling constant, that is the $t_0$ model. This would indeed lead to a quite reasonable EOS for neutron matter governed by a single parameter, the Bertsch parameter \cite{be2000}. Such an aspect was also discussed in Refs. \cite{la2016,la2017} where a resummed formula tailored to describe neutron matter as an expansion around the unitary limit was introduced. However, a $t_0$ model is certainly totally insufficient for correctly describing the EOS of symmetric matter with its equilibrium point. For this, a three--body part or a term that mimics a three--body contribution should be included in the interaction. An analogy can be found with the situation existing in pionless EFT where a three--body force is necessary already at LO to avoid the collapse of the triton \cite{beda1,hammerautre,beda2,gri,gri2,afnan,bar,pal}. Skyrme--type forces have for this reason a repulsive two--body density--dependent $t_3$ term that mimics a three—--body force and this term is then included in the LO interaction. 
The EOS calculated at LO contains Fermi momentum dependencies of the type $k_F^3$ and $k_F^{3+3\alpha}$. 
Going to NLO, one has to to calculate the second--order contribution coming from $V_{LO}$ and the MF contribution associated to $V_{NLO}$. Renormalizability is the criterium that is used to choose the form of the $V_{NLO}$ interaction, which must then contain all the counterterms required to cancel the second--order divergent contributions generated by the $V_{LO}$ interaction. Different scenarios may be followed. 
The finite part of the second--order contribution has Fermi momentum dependencies of the type $k_F^4$, $k_F^{4+3\alpha}$, and  $k_F^{4+6\alpha}$ in the EOS.
The divergent second--order contributions have a linear dependence on the momentum cutoff. Such terms have three types of Fermi momentum dependencies in the EOS, namely $k_F^3$, $k_F^{3+3\alpha}$, and  $k_F^{3+6\alpha}$.  
The first two types of dependencies would allow for regrouping with MF terms in the EOS. The last type of dependence does not allow however for any regrouping with MF contributions in the EOS. This indicates that, at least, a counterterm ($V_{NLO}$) expected to produce such a Fermi momentum dependence at the MF level should be included. One observes that the highest Fermi momentum dependence in the EOS is $k_F^{4+6\alpha}$. This would indicate that, for $\alpha <$ 1/6, the Skyrme velocity--dependent terms of the type $t_1$ and $t_2$ should not be included in $V_{LO}$ 
because they would contribute to the EOS with a higher power of $k_F$ (terms proportional to $k_F^5$). For $\alpha >$ 1/6, the inclusion of such terms should be rediscussed. Since $\alpha$ is a free parameter in our analysis and we do not know its value before the adjustment of the parameters, this aspect is discussed only after the fit is performed. The mass is taken equal to the bare mass. 

Three scenarios are explored in Ref. \cite{yang2017-2}. In the first scenario (called {\it{abc}}) three types of counterterms are included in $V_{NLO}$, with Fermi momentum dependencies $k_F^3$, $k_F^{3+3\alpha}$, and  $k_F^{3+6\alpha}$. In this case, there are no terms in the EOS which are regrouped with MF contributions. 
For the second scenario (called {\it{bc}}), counterterms of the type $k_F^{3+3\alpha}$, and  $k_F^{3+6\alpha}$ are included and the 
$k_F^3$ part is regrouped with MF terms. 
Finally, in the third scenario (called {\it{c}}), only a type of counterterm ($k_F^{3+6\alpha}$) is included. Both terms 
$k_F^3$ and $k_F^{3+3\alpha}$ are regrouped with MF contributions. 
The parameters were adjusted for the three scenarios (additional parameters of course appear each time a counterterm is included) and the resulting properties of matter, namely the energy and the density of the saturation point, as well as the incompressibility of symmetric matter are listed in Table \ref{matter}. For comparison, also the SLy5 MF results are shown in the table. 

\begin{table}[htbp]
\begin{center}
\begin{tabular}{c c c c c c}
\hline
& SLy5 & LO & NLO$_{abc}$& NLO$_{bc}$ & NLO$_c$ \\ \hline
$\frac{E(\rho_s)}{A}$ (MeV)& -16.18 & -16.31 & -15.93 & $-15.98\pm0.1$ & $-15.97\pm0.1$ \\ \hline
$\rho_s$ (fm$^{-3}$)& 0.162 & 0.162 & 0.16 & $0.16\pm0.003$ & $0.16\pm0.003$ \\ \hline
$K_{\infty}$ (MeV)& 232.67 & 254.64 & 236.32 & $234.3\pm3.5$ & $233.2\pm3.7$ \\ \hline
\end{tabular}%
\end{center}
\caption{Saturation density, its associated energy per particle ($\rho_s$ and $\frac{E(\rho_s)}{A}$),  and incompressibility $K_{\infty}$ produced by the SLy5 mean--field EOS, the LO model and the three different scenarios investigated at NLO. }
\label{matter}
\end{table}

\begin{figure}[tb]
\begin{center}
\begin{minipage}[t]{8 cm}
\epsfig{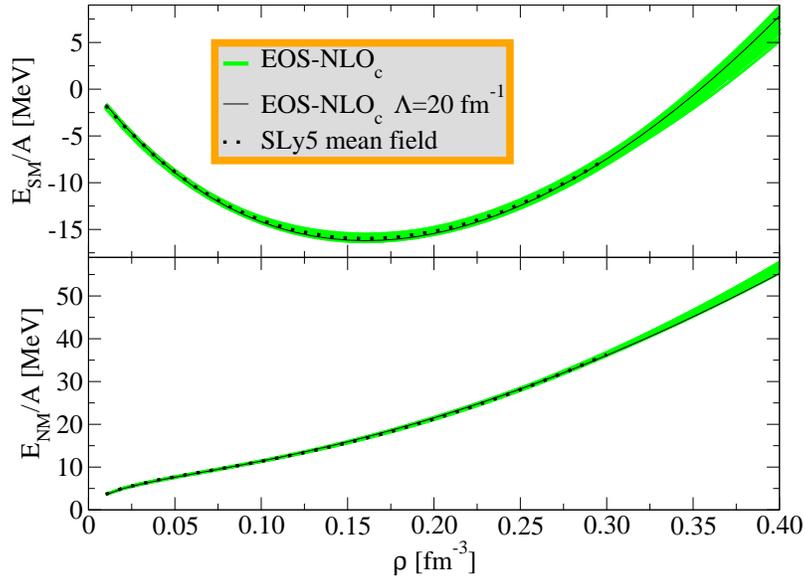}
\end{minipage}
\begin{minipage}[t]{16.5 cm}
\caption{Second--order EOS for symmetric and neutron matter after renormalization for values of the cutoff $\Lambda =1.2-20$ fm$^{-1}$ (green area) for the scenario {\it{c}}. The case of $\Lambda = 20$ fm$^{-1}$ is shown as an illustration (solid black line). The SLy5 MF (benchmark curve) EOS is also plotted (dotted black line). 
Taken from Ref. \cite{yang2017-2}. }
\label{scenario}
\end{minipage}
\end{center}
\end{figure}

Figure \ref{scenario} shows the EOSs for symmetric and neutron matter corresponding to the scenario {\it{c}}, as an illustration. The cutoff is chosen in the window $\Lambda =1.2-20$ fm$^{-1}$ and the green regions describe the results of the fits performed within this window of cutoff values. The curves corresponding to the case $\Lambda =20$ fm$^{-1}$
are also shown as an illustration (black solid line). The curves obtained with SLy5 in the MF approximation are plotted in the figure for comparison. These curves represent the benchmark curves used for the fit.  
Several points are taken in these curves for the fit in the density range from 0 to 0.3 fm$^{-3}$. 
It was found that satisfactory fits were obtained in all cases for the three chosen scenarios. One can thus conclude that NLO results are practically not affected by the regularization scheme after renormalization. This independence of the regularization is also illustrated in Table  \ref{matter}, where one can observe that the properties of matter are almost the same in all cases. For some parametrization sets, it turned out that the adjustment of the parameters provided values of $\alpha$ larger than 1/6. We included in those cases the $t_1$ and $t_2$ Skyrme terms but realized that the quality of the fit was not improved. From this, we decided to postpone the inclusion of these terms at the next order N$^2$LO. 
Work is presently in progress along this direction. 

\begin{figure}[tb]
\begin{center}
\begin{minipage}[t]{8 cm}
\epsfig{file=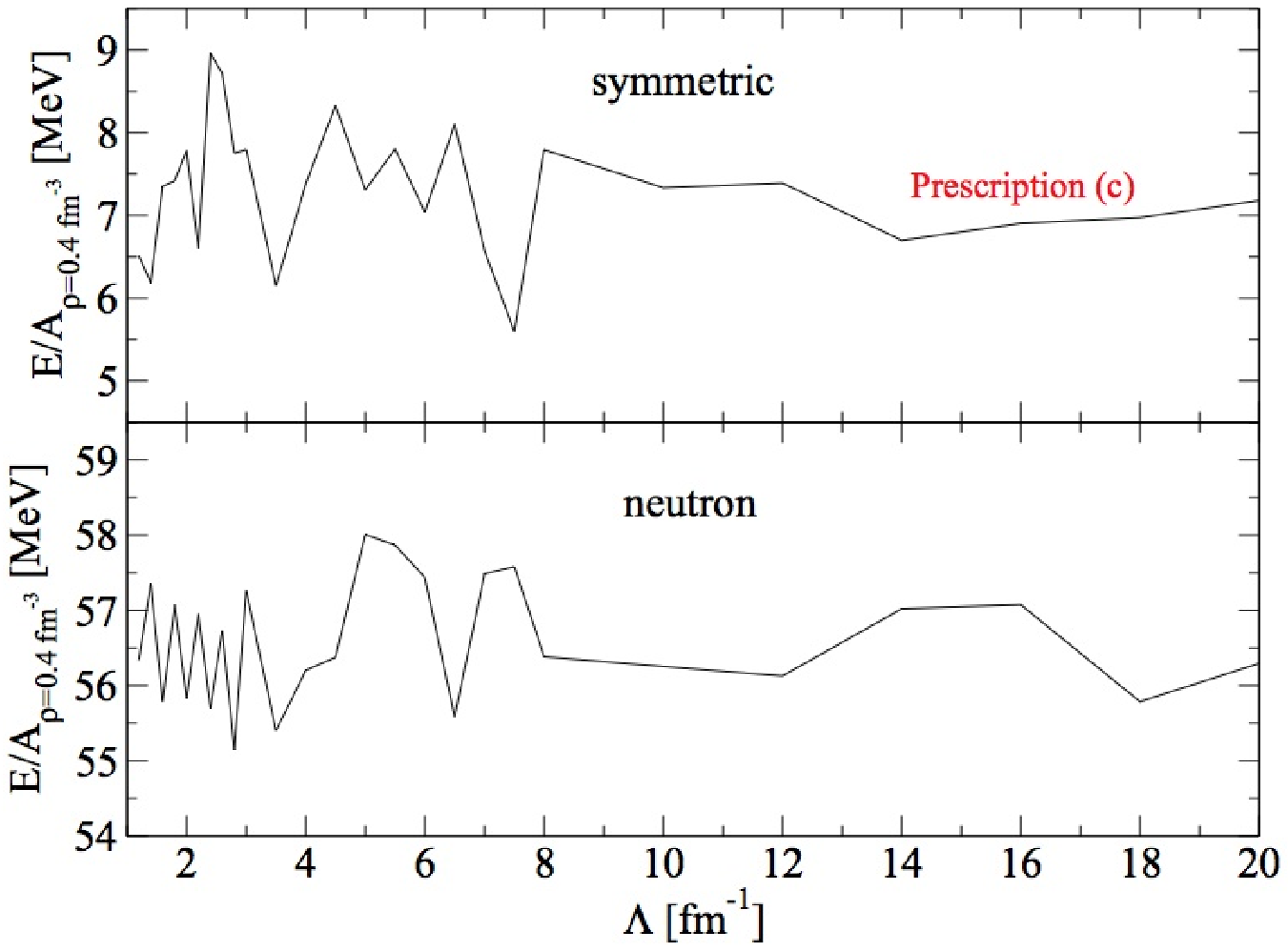,scale=0.68}
\end{minipage}
\begin{minipage}[t]{16.5 cm}
\caption{Value of the second--order EOS for symmetric and neutron matter computed at $\rho=$ 0.4 fm$^{-3}$ as a function of the momentum cutoff for the scenario {\it{c}}.
Taken from Ref. \cite{yang2017-2}. }
\label{rg}
\end{minipage}
\end{center}
\end{figure}

Finally, a renormalization group analysis was performed for the cases {\it{bc}} and {\it{c}}. The dependence on the cutoff of the value of the EOS for symmetric and neutron matter computed at the density $\rho=$ 0.4 fm$^{-3}$ 
(beyond the range of densities where the fit was performed) was analyzed. The results for the scenario {\it{c}} are  shown in Fig. \ref{rg}. One may observe that the cutoff dependence becomes smaller with increasing cutoff values. However, a more complete power counting analysis is still missing and it would be possible only going at least up to  N$^2$LO. We mention that the third--order contribution to the EOS of nuclear matter  was already computed in Ref. \cite{kaiserepja} with the Skyrme interaction based on the dimensional--regularization procedure.  

\subsection{\it First steps towards finite--size systems}
\begin{figure}[tb]
\begin{center}
\begin{minipage}[t]{10 cm}
\epsfig{file=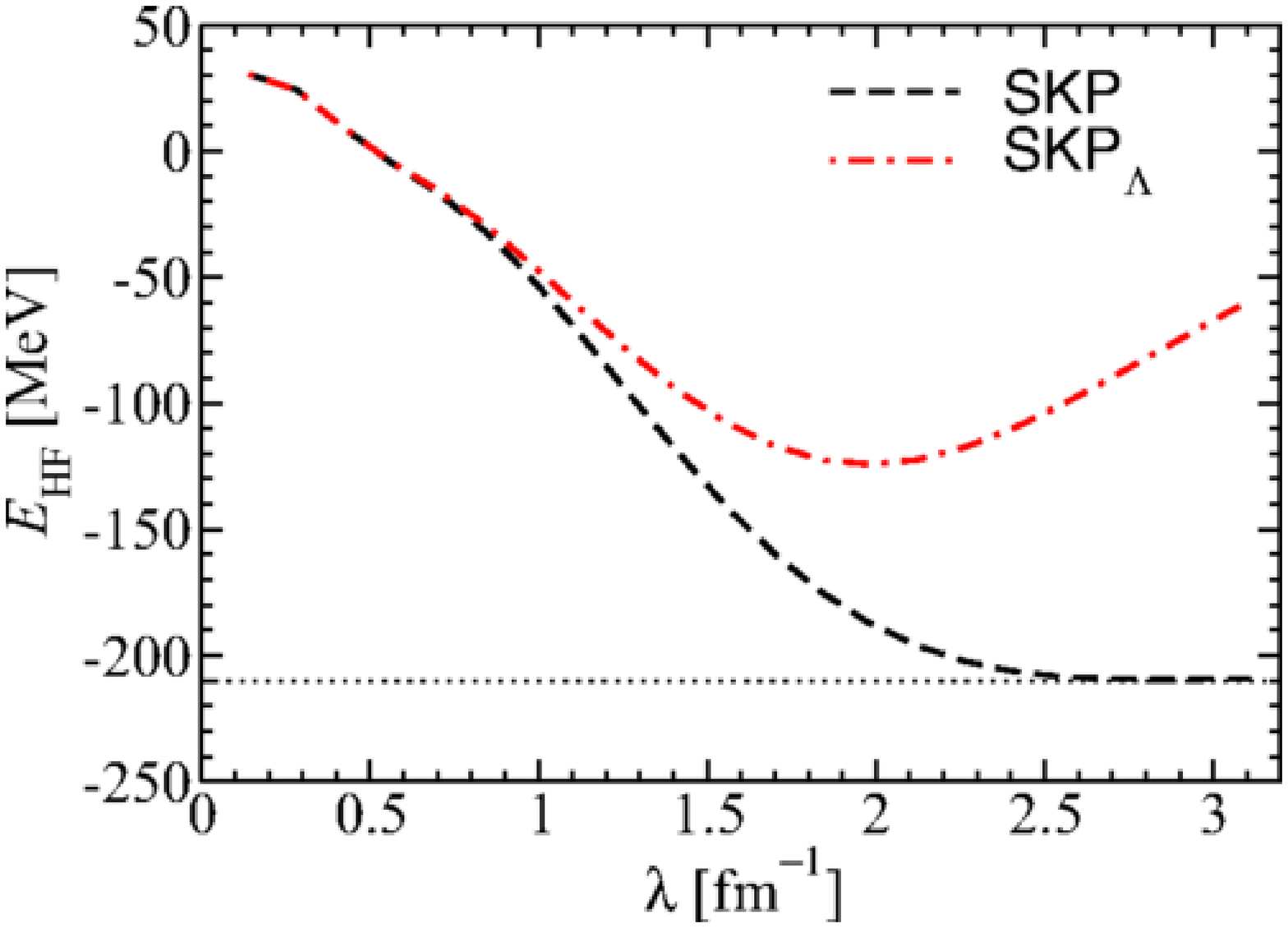,scale=0.55}
\end{minipage}
\begin{minipage}[t]{16.5 cm}
\caption{Total MF energy of $^{16}$O plotted as a function of the cutoff $\lambda$ on the relative momentum. 
The black dashed line represents the energy obtained with the original SkP interaction in a HF calculations with a cutoff $\lambda$.  The dotted horizontal line is the HF result obtained in the absence of a cutoff. The red dot--dashed curve is the HF energy computed using the interactions regularized and refitted at second order for symmetric matter \cite{mo2010}. 
Taken from Ref. \cite{brenna}.}
\label{hfenergy}
\end{minipage}
\end{center}
\end{figure}

For the simple case of the $t_0-t_3$ model a first attempt to use the parameter set introduced in Ref. \cite{mo2010} was published in Ref. \cite{brenna}. A calculation of the total energy was done for the nucleus $^{16}$O. The same Skyrme parametrization SkP as the one used in Ref. \cite{mo2010} was used as a reference. Several simplifications were adopted, 
namely the Coulomb and spin—-orbit contributions were neglected, which means that this analysis represented just a test calculation and that the resulting total binding energy of $^{16}$O is obviously very far from the correct value, already at the MF level. Instead of performing a full particle--phonon coupling calculation, the phonon was replaced by a particle--hole pair. Cutoffs on relative momenta of the initial and final states were employed, as in Ref. \cite{car2013}. The cutoff was introduced also at the MF level. The calculations were performed in an harmonic oscillator basis which allows for a separation of the center—-of—-mass and relative motion. Differently from what happens in infinite matter, in finite nuclei there is no translational invariance and this implies that two energy scales have to be considered. Such two energy scales were treated with two cutoffs, one put on the relative momentum and the second one put on the maximum particle energy. 
Figure \ref{hfenergy} displays the total MF energy of $^{16}$O plotted as a function of the cutoff $\lambda$ on the relative momentum. The black dashed line represents the energy obtained with the original SkP interaction in a HF calculations with a cutoff $\lambda$. Starting from values of $\lambda$ larger than 2.5 fm$^{-1}$, this curve converges to the full result (dotted horizontal line) that is the HF result obtained in the absence of a cutoff. One observes that, owing to the absence of the Coulomb and of the spin--orbit terms in the SkP interaction, the full HF calculation does not provide a reasonable total binding energy. We remind the reader that the experimental binding energy for $^{16}$O is -127.6 MeV \cite{audi1,audi2}.
The red dot--dashed curve is the HF energy computed using the interactions regularized and refitted at second order for symmetric matter \cite{mo2010}. For low values of the cutoff, this curve is practically superposed to the curve computed with the genuine SkP parametrization. This is reasonable because, for low values of the cutoff, the second--order correction is weak and the renormalization of the interaction is consequently negligible. 
Looking at this figure, the authors of Ref. \cite{brenna} decided to restrict their analysis to a window of cutoff values between 2 and 2.7 fm$^{-1}$, which corresponds to a range of cutoff values where the MF ground--state energy computed with the regularized and readjusted interactions is not too far from the expected full result. 

\begin{figure}[tb]
\begin{center}
\begin{minipage}[t]{14 cm}
\epsfig{file=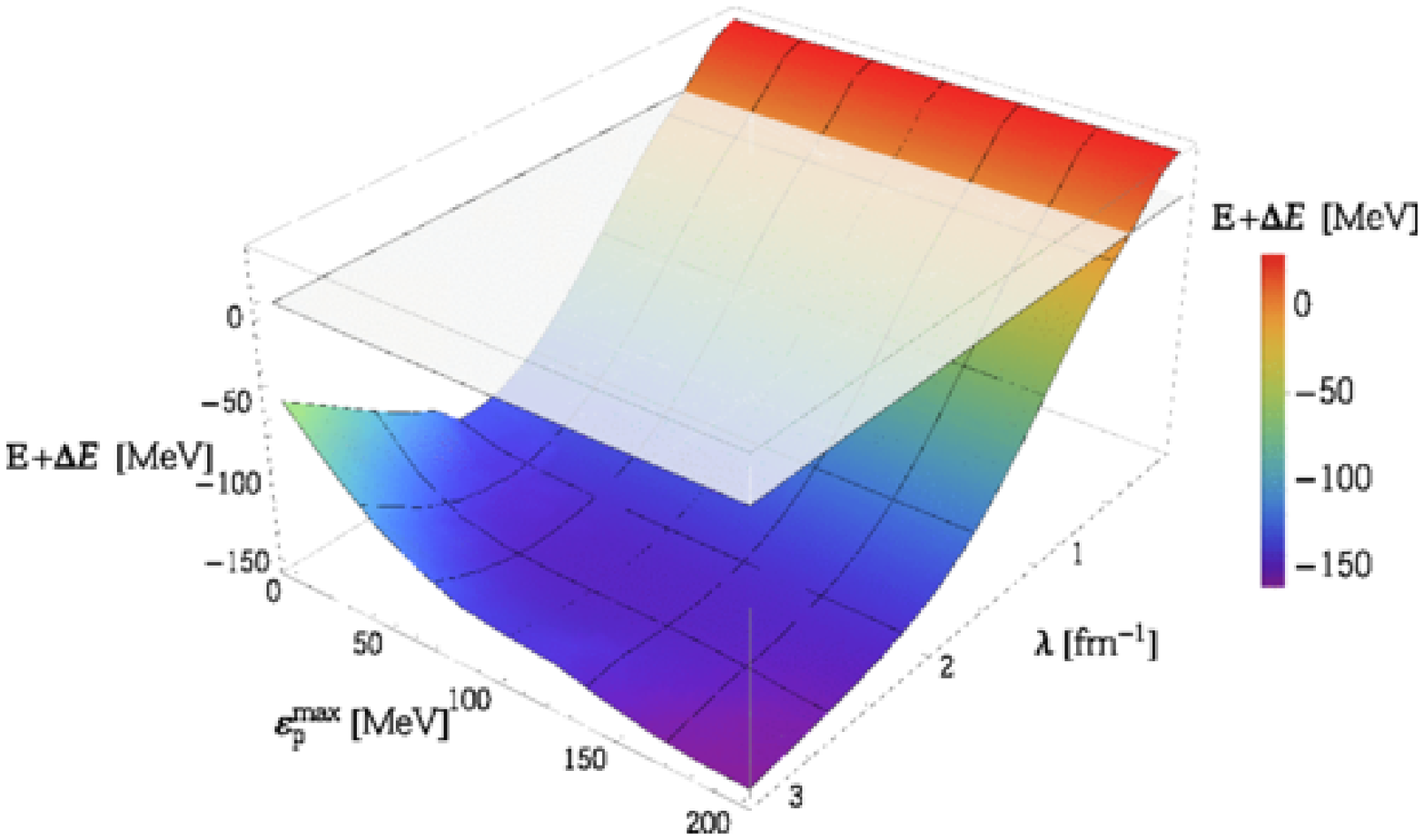,scale=0.7}
\end{minipage}
\begin{minipage}[t]{16.5 cm}
\caption{ Total second--order binding energy of $^{16}$O as a function of the cutoffs on the relative momentum $\lambda$ and on the maximum particle energy $\epsilon_p^{max}$.
Taken from Ref. \cite{brenna}. }
\label{convebrenna}
\end{minipage}
\end{center}
\end{figure}

Figure \ref{convebrenna} describes the total second--order binding energy of $^{16}$O as a function of the cutoffs on the relative momentum $\lambda$ and on the maximum particle energy $\epsilon_p^{max}$. In the selected window 
for $\lambda$, namely between 2 and 2.7 fm$^{-1}$, one observes that results are quite stable and weakly depending on the maximum particle energy. 
 This 
constituted a first encouraging result on the possibility of concretely using for finite nuclei the interactions regularized and renormalized in infinite matter. 

\section{Towards ab--initio: nuclear EDFs inspired by EFTs \label{abinitio}}

Some ideas borrowed by EFTs were used in the previous section as an illustration for constructing regularized interactions. For example, the  dimensional regularization procedure was mentioned as a way to remove ultraviolet divergences in the second--order results for the EOS of infinite matter. Also, in the same context, renormalizability issues were discussed as a preparatory study to establish a strategy for a power counting in the EDF framework. 
All these topics were discussed in connection with a BMF strategy: it is indeed desirable to formulate BMF models where the used effective interactions do not contain any double counting of correlations and are thus adjusted at a BMF level. Furthermore, these effective interactions should be regularized so to handle the cutoff dependence in those cases where ultraviolet divergences occur at the BMF level (use of zero--range interactions). This was in particular analyzed 
in the context of the MBPT with calculations truncated at second order for infinite nuclear matter and Skyrme--type interactions. 

In this section, we discuss the development of other types of bridges and links with EFT ideas and {\it{ab--initio}} benchmark data in the construction of functionals tailored this time for the leading order of the Dyson expansion  (MF--type applications) and inspired from EFT. In particular, four examples of functionals are illustrated in Subsec. \ref{matteredf}. Two of these functionals were introduced with a special focus on the behavior of the very low--density regime of neutron matter. 
One of them is the YGLO (the acronym means ``Yang-Grasso-Lacroix-Orsay") functional \cite{yglo} which is a hydrid functional containing Skyrme--like density--dependent and velocity--dependent terms, as well as a resummed formula \cite{scha,resu2,resu3,holtkaiser}. The second one is based on the tuning of a density--dependent neutron--neutron scattering length \cite{elyo}. This functional was called ELYO in Ref. \cite{bonnard}, where ELYO means ``Extended-Lee-Yang-Orsay". 
A third functional is the KIDS functional introduced in Ref.  \cite{pakids} and which is constructed as a power expansion in the Fermi momentum. KIDS is the acronym for ``Korea-IBS-Daegu-Sungkyunkwan". The three above functionals were tailored to provide not only the EOS of neutron matter but also the EOS of symmetric matter. 
Finally, with a special attention on pure neutron matter, another type of resummed expression was introduced connecting neutron matter with a Fermi gas at unitarity \cite{la2016,la2017,laboulet}.
 
The KIDS functional was introduced for nuclear matter \cite{pakids} and was recently extended to describe finite nuclei \cite{gil1,gil2,gil3}. 
The YGLO and ELYO functionals were introduced for nuclear matter \cite{yglo,elyo} and were extended to describe trapped neutron drops in Ref. \cite{bonnard}. 
The four functionals are introduced in Subsec. \ref{matteredf}, where nuclear matter applications are discussed. 
Subsection \ref{finite} illustrates the extensions of some of these functionals to finite--size systems, either nuclei or trapped  neutron drops.

Subsection \ref{dme} addresses other types of {\it{ab--initio}}--inspired functionals, some of them based on the use of the DME. In addition, an idea of constraining Skyrme functionals by using QMC results  \cite{roggero}  and an idea for deriving functionals based on a genuine DFT procedure \cite{dobaletter} are illustrated. 

\subsection{\it Functionals tailored for nuclear matter to describe the low--density regime and/or the unitary regime \label{matteredf}}

Lee and Yang introduced in the 50s an expansion tailored to describe very dilute Fermi systems \cite{fetter,LY1,LY2,bishop,gali,abri}. For very low--density systems, where the distance between particles $\sim 1/k_F$ is much larger than the range of the interaction, the details of the interparticle interaction do not have any relevant impact on the properties of the systems and their behavior may be simply described through $s$--wave scattering parameters. This is for example the case for trapped ultracold Fermi atomic gases where a contact force with a coupling constant related to the $s$--wave scattering length well describes the system \cite{bruun, giorgini,bloch,zwerger}. If the interparticle distance is larger than the scattering length $a$, an expansion in $ak_F$ can properly describe the properties of these systems in a very low--density regime. The first two terms of the Lee-Yang expansion read
\begin{equation}
\frac{E}{N}=\frac{\hbar^2k_F^2}{2m} \left[ \frac{3}{5}+ (\nu-1) \frac{2}{3\pi}(k_Fa)+ (\nu-1)
 \frac{4}{35\pi^2}(11-2ln2)(k_Fa)^2 \right],
\label{lyfirstterms}
\end{equation}
where $m$ is the mass of the particles, $N$ the number of particles, and $\nu$ the degeneracy or statistical factor, for instance equal to 2 in neutron matter and to 4 in symmetric matter. The first term in Eq. (\ref{lyfirstterms})  describes the free Fermi gas energy whereas the term linear in the scattering length corresponds to what is found with the method of pseudopotentials (generated by a contact interaction whose coupling strength is described by the scattering length)  \cite{pseudo}. 
These first terms of the expansion, up to the term quadratic in $a$, were found with different methods in Refs. \cite{LY1,LY2,gali,abri}. The following term (third term of the expansion, not reported in Eq. (\ref{lyfirstterms})) was then 
evaluated in Refs. \cite{dominicis,efimov1}. These first three terms 
are connected with two--particle low--energy scattering parameters. 
Efimov studied later higher--order terms of the expansion associated with the simultaneous interaction of at least three particles \cite{efimov2,efimov3}.

It is important to stress that, strictly speaking, this expansion is correct only for pure neutron matter at low densities. On the other side, one should be very careful in using it for symmetric matter where at very low densities clustering phenomena occur which cannot be properly described by this simple formula. A few reviews on clustering phenomena and on the associated theoretical models are for example Refs. \cite{oe2006,freer2007,hori2012,funa2015,freer2018}. 

Let us examine the case of neutron matter, with $\nu=2$. Can the associated Lee-Yang expansion be reproduced by EDF EOSs generated within the Dyson many--body expansion?  Let us start by considering the minimal Skyrme--type EOS which 
may provide a correct equilibrium point for symmetric matter at the leading order of the Dyson expansion, that is with a MF functional. This may be obtained by combining the attractive $t_0$ and the strongly  repulsive $t_3$ terms in a Skyrme--type  $t_0-t_3$ model. 
The expression for the MF Skyrme EOS for pure neutron matter was already reported in Eq. (\ref{eos1nm}) as a function of the density. We write it here as a function of the Fermi momentum $k_N$ for neutron matter in the case of the $t_0-t_3$ model. 
\begin{equation}
\frac{E_{neutr}}{N}=\frac{3}{10}\frac{\hbar ^{2}}{m} k_N^2 +\frac{1}{12 \pi^2} t_{0} (1-x_{0}) k_N^3 + 
\frac{1}{24}t_{3}(1-x_{3}) \left( \frac{1}{3\pi^2} \right)^{\alpha +1} k_N^{3\alpha+3}.
\label{eos1t0t3}
\end{equation}%
By comparing Eqs. (\ref{lyfirstterms}) and (\ref{eos1t0t3}), one observes that the term linear in the scattering length $a$ in Eq. (\ref{lyfirstterms}) may be reproduced by the $t_0$ term of the Skyrme EOS with 
\be
t_0 (1-x_0) = 4 \pi \hbar^2 a/m. 
\label{t0-a}
\ee
The following term of Eq. (\ref{lyfirstterms}) can be reproduced either by including the second  order in the Dyson expansion (second--order term proportional to $t_0^2$ generated by the $t_0$ part of the Skyrme interaction. See Sec. \ref{newgene}) or by a specific choice of the interaction that may provide such a term already at leading order.  One example is the three--body nonlocal interaction introduced in Ref. \cite{ge10}. The idea of this work was to introduce a three--body interaction of the type 
\be
H_3 = f({\bf{r}},{\bf{r\prime}}) \sum_{{\bf{\sigma}}}  \psi^{\dagger}_{\sigma_1 {\bf{r\prime}}} 
  \psi^{\dagger}_{\sigma_2 {\bf{r\prime}}} \psi^{\dagger}_{\sigma_3 {\bf{r}}}
\psi_{\sigma_4 {\bf{r\prime}}} \psi_{\sigma_5 {\bf{r}}} \psi_{\sigma_6 {\bf{r}}},   
\label{h3}
\ee
with
\be
f({\bf{r}},{\bf{r\prime}}) = \frac{\hbar^2 a^2}{m} \frac{C}{\mid {\bf{r}}-{\bf{r\prime}} \mid} .
\label{functi}
\ee
This interaction term avoids the use of a density--dependent term with a fractional density dependence providing at the same time the correct quadratic term of the Lee-Yang expansion. 

\begin{figure}[tb]
\begin{center}
\begin{minipage}[t]{8 cm}
\epsfig{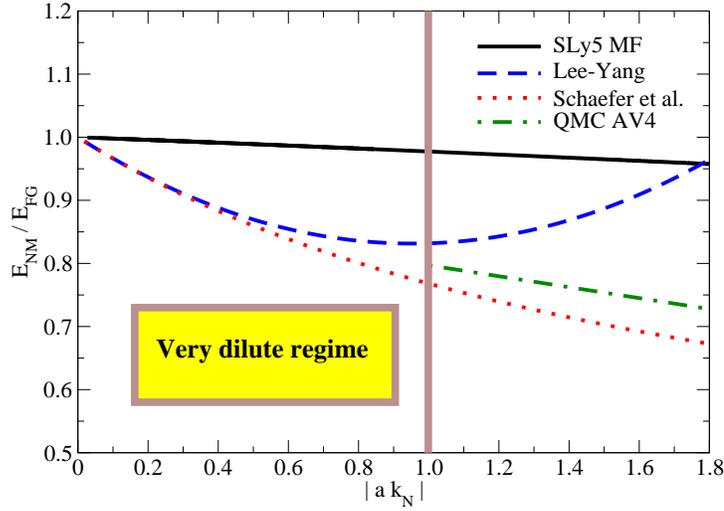}
\end{minipage}
\begin{minipage}[t]{16.5 cm}
\caption{ Energy of neutron matter divided by the energy of the free Fermi gas $E_{FG}$ as a function of $\mid a k_N \mid$. The SLy5--MF EOS (solid black line) is plotted together with the curve obtained by using the physical value of the scattering length and the first two terms of the Lee-Yang formula (blue dashed line), a resummed curve \cite{scha} (red dotted line) and the microscopic quantum Monte Carlo (QMC) results of Ref. \cite{geca} (green dot--dashed line). 
Taken from Ref. \cite{yglo}. }
\label{bassade}
\end{minipage}
\end{center}
\end{figure}

On the other side, the density--dependent $t_3$ part of the Skyrme interaction may always be chosen so to provide a $k_F^4$ dependence in the EOS at leading order, using a power for the density dependent--term equal to 1/3 \cite{bul2007,bul2012}. 
In this case, the connection between the parameters of the $t_3$ term of the Skyrme interaction and the scattering length $a$ is
\be
t_3 (1-x_3) = \frac{\hbar^2}{m}  \frac{144}{35} (3\pi^2)^{1/3} (11-2ln2)a^2, 
\label{t3-a}
\ee
with $\alpha=1/3$.
One may thus be tempted to think that a MF Skyrme--type EOS can be well adapted to describe 
low--density neutron matter, especially for those cases where Skyrme--type EOSs have been adjusted to satisfactorily describe the EOS of neutron matter at density scales around the equilibrium point of symmetric matter \cite{lyon1,lyon2,lyon3}. 
But it turns out that, even for $\alpha=1/3$, the very low--density regime of neutron matter cannot be described. 
One may indeed adjust two of the four parameters $t_0$, $t_3$, $x_0$, and $x_3$, in order to have a reasonable EOS for symmetric matter with a correct equilibrium point and constrain the remaining two parameters with Eqs. (\ref{t0-a}) and (\ref{t3-a}). However, using the physical value of the neutron--neutron scattering length, -18.9 fm, 
the resulting EOS for neutron matter is completely unrealistic outside the region $\mid a k_F \mid <1$. Owing to the very large value of $a$, this occurs starting from extremely low densities (knowing that $\mid a k_F \mid =1$ corresponds to a density of 10$^{-6}$ fm$^{-3}$). 
If then one tries to use $a$ as a free parameter and adjust it by requiring the good reproduction of the EOSs of both symmetric and neutron matter at density scales close to the saturation point, the resulting adjusted value of $a$  is very far from  -18.9 fm. A value close to -1 fm is typically found. This clearly demonstrates that the two density regimes (very low densities on one side and densities close to the saturation point on the other side) cannot be simultaneously described with a typical Skyrme EOS. 
To reconcile such two density regimes in systems having very large scattering lengths such as neutron matter, resummed expressions \cite{scha,resu2,resu3} are used in EFTs, where one requires that an infinite number of Feynman diagrams are summed at leading order. 
An illustration of this is given in Fig. \ref{bassade} where the energy of neutron matter divided by the energy of the free Fermi gas $E_{FG}$ is plotted as a function of $\mid a k_N \mid$. An example of Skyrme MF EOS is provided (SLy5) together with the curve obtained by using the physical value of the scattering length and the first two terms of the Lee-Yang expansion, a resummed curve \cite{scha} and the microscopic quantum Monte Carlo (QMC) results of Ref. \cite{geca} (based on the interaction of Ref. \cite{wiringa}) for comparison. The correct behavior in the low--density regime may be clearly identified in the figure and it is clear that the Skyrme curve does not reproduce it. 

It is also interesting to notice that, if one requires that a Skyrme--type interaction provides only terms belonging to the 
Lee-Yang expansion and not other types of terms with different $k_F$ dependencies, only density dependencies with powers equal to multiples of 1/3 may be accepted (see the discussion on the KIDS functional that will be done in the next pages). 
Namely, a power of 1/3 would provide a term of the type $k_F^4$, a power of 2/3 would generate a $k_F^5$ term (that would be provided also by velocity--dependent Skyrme terms of the types $t_1$ and $t_2$), a power of 1 would generate a term of the type $k_F^6$, and so on.  
This means also that all the Skyrme parametrizations having powers of the density lower than 1/3 lead to a term of the type $k_F^b$, where $3<b<4$, which is located between the terms of the Lee-Yang expansion  that are linear and quadratic in the scattering length $a$. This strongly alters the very first terms of the Lee-Yang formula. 

\begin{figure}[tb]
\begin{center}
\begin{minipage}[t]{11 cm}
\epsfig{file=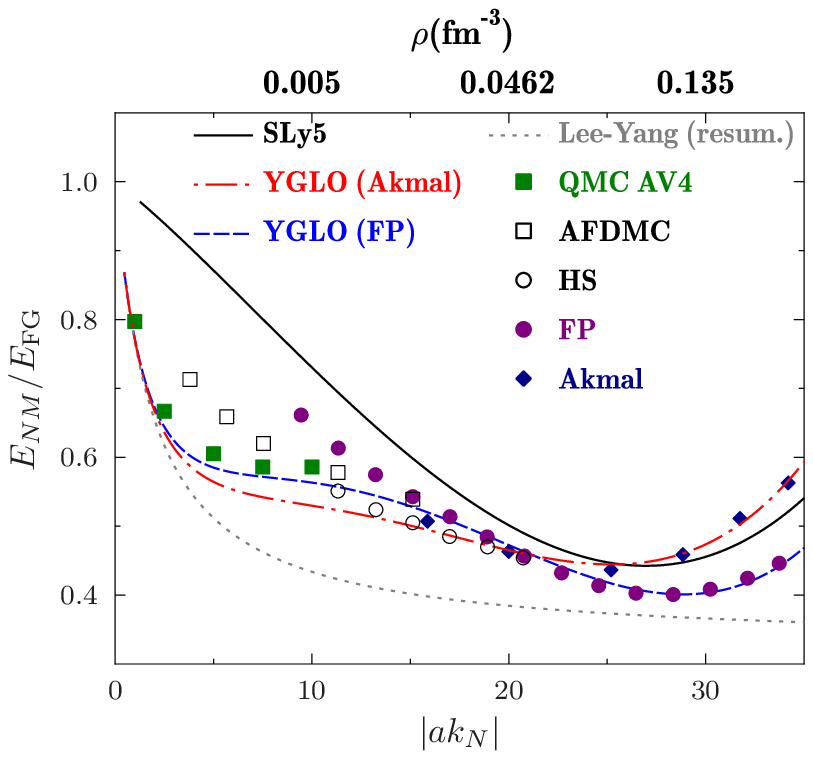,scale=1.1}
\end{minipage}
\begin{minipage}[t]{16.5 cm}
\caption{ Energy of neutron matter divided by the energy of the free Fermi gas $E_{FG}$ as a function of $\mid a k_N \mid$ (lower axis) and of the density (upper axis). The two YGLO curves, (FP) and (Akmal) are shown together with the QMC EOS of Ref. \cite{hs}, the Friedman and Pandharipande (FP) \cite{fp} and the Akmal et al. \cite{akmal} EOSs, the Skyrme SLy5 MF EOS,   the auxiliary--field results (AFDMC) of Ref. \cite{afgandolfi} and the N2LO results of Ref. \cite{hs} (HS). 
Taken from Ref. \cite{yglo}. }
\label{akfplot}
\end{minipage}
\end{center}
\end{figure}

If a MF EOS generated by a Skyrme $t_0-t_3$ model or by a full Skyrme interaction is unable to reconcile the two density scales (very low densities and densities close to the saturation point of symmetric matter) for pure neutron matter, one could 
go beyond the mean--field scheme by including the second--order diagrams of the Dyson expansion and verify whether this may cure 
such a situation. An encouraging indication is indeed what is shown in Ref. \cite{yang2016}, namely that the $t_0^2$ contribution leads exactly to the second term of the Lee-Yang expansion.
Despite this, it turns out that the found results are still analogous to what found when only the MF EOS is considered. The two density scales cannot be reconciled and an adjustment including also the scattering length as a free parameter leads also this time to values of $a$ close to -1 fm. 

Following the idea of resummation, the authors of Ref. \cite{yglo} proposed then a new hybrid functional containing a resummed formula together with two additional terms which mimic density-- and velocity--dependent Skyrme terms and which are useful for reproducing in a satisfactory way the EOSs of both pure and symmetric matter at densities around the saturation point. Such a functional provides the following EOS: 
\be
\frac{E}{A}=K_{\beta }+\frac{B_{\beta }\rho }{1-R_{\beta }\rho
^{1/3}+C_{\beta }\rho ^{2/3}}+D_{\beta }\rho ^{5/3}+F_{\beta
}\rho ^{\alpha +1},
\label{eosyglo}
\ee
where the two cases $\beta =0$ (neutron matter) and 1 (symmetric matter) are considered ($A$ becomes $N$ for neutron matter). The parameters $B_{\beta }$ and $R_{\beta }$ are not free but constrained by the first terms of the Lee-Yang formula. An analogous expansion may be used for symmetric matter \cite{fetter} despite all the already mentioned known limitations in using such types of expansions for symmetric matter at low densities. 
However, the focus of this work is not the low--density regime of symmetric matter. For symmetric matter, it is mainly required that the ordinary scales of interest for finite nuclei are well described. 
Constraining  $B_{\beta }$ and $R_{\beta }$ with the Lee-Yang formula leads to 
\be
B_{\beta }=2\pi \frac{\hbar ^{2}}{m}\frac{(\nu -1)}{\nu }a,~R_{\beta }=\frac{%
6}{35\pi }\left( \frac{6\pi ^{2}}{\nu }\right) ^{\frac{1}{3}}(11-2\ln 2)a.
\label{coparameters}
\ee
For symmetric matter, only the spin--singlet $^1S_0$ scattering length was employed and the spin--triplet $^3S_1$ neutron--proton scattering length was neglected. The adjustment of the remaining seven parameters was done as follows for neutron matter: the QMC results of Ref. \cite{geca} were used as benchmark points up to $\mid a k_N \mid=10$. For higher densities, two
curves were chosen for the adjustment, namely the Friedman {\it{et al.}} (FP) EOS of Ref. 
\cite{fp} and the Akmal {\it{et al.}} EOS of Ref. \cite{akmal}. Two parametrization sets were in this way generated called YGLO(FP) and YGLO(Akmal), respectively. All these curves are shown in Fig. \ref{akfplot}, where also the MF SLy5 EOS, the auxiliary--field results (AFDMC) of Ref. \cite{afgandolfi} and the N2LO results of Ref. \cite{hs} (HS) are reported for comparison. 
For symmetric matter, the Akmal {\it{et al.}} and the Friedman {\it{et al.}} EOSs are very similar and the EOS of Ref. \cite{fp} was employed for fitting the YGLO EOS.

Using the so--called parabolic approximation relating the EOS of symmetric matter and the EOS of asymmetric matter characterized by an asymmetry coefficient $\delta$ (previosuly defined in this revue), the symmetry energy $S(\rho)$   
may be derived simply as the difference between the EOSs of neutron and symmetric matter. Corrections beyond the parabolic approximation for the computation of the symmetry energy have been demonstrated to be negligible \cite{zuo1,zuo2}.  

\begin{figure}[tb]
\begin{center}
\begin{minipage}[t]{10 cm}
\epsfig{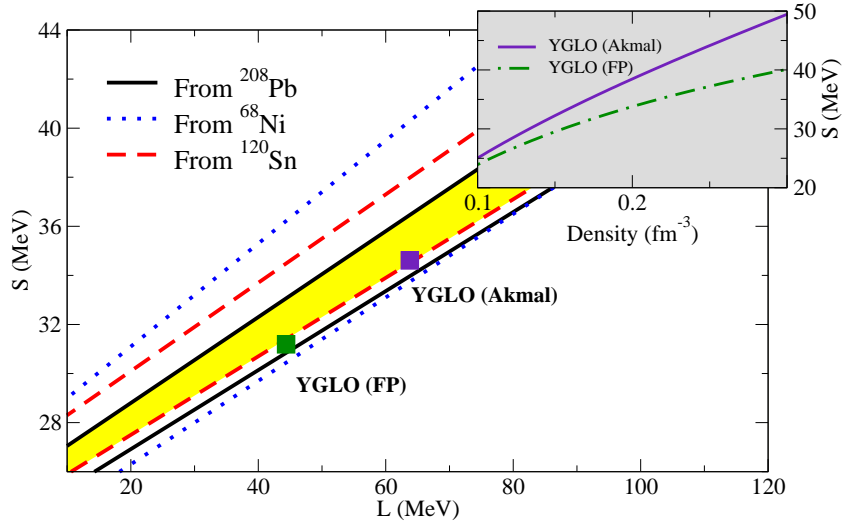}
\end{minipage}
\begin{minipage}[t]{16.5 cm}
\caption{Symmetry energy $S$ calculated at the saturation density as a function of the slope   $L$. Three empirical bands are reported, extracted from the measurements of the static dipole polarizability in the three nuclei $^{208}$Pb, $^{68}$Ni, and $^{120}$Sn (see text). The overlap region between the three areas is the yellow band. The two values corresponding to YGLO(Akmal) and YGLO(FP) are reported. The inset shows the density dependence of the symmetry energy corresponding to the two YGLO EOSs. 
Taken from Ref. \cite{yglo}. }
\label{symmetryslope}
\end{minipage}
\end{center}
\end{figure}

Many studies in the literature are focused on the analysis  of the symmetry energy and its slope  (that is, its density dependence) with a special attention devoted on the investigation of existing correlations between these quantities (defined and computed in infinite matter) and measured observables in finite nuclei, such as the neutron skin thickness in neutron--rich nuclei and the electric dipole polarizability. 
The slope $L$ of the symmetry energy is defined as
\be
L=3 \rho_0 \frac{dS(\rho)}{d\rho}\mid_{\rho=\rho_0},
\label{slope}
\ee
where $\rho_0$ is the saturation density of symmetric matter. 
 For example, a strong correlation was identified between the neutron skin thickness and the slope $L$ of the symmetry energy (see for instance Refs. \cite{centelles2009,warda2009}). 
A strong correlation was also found between the dipole polarizability times the symmetry energy and the neutron skin thickness \cite{roca2015}. Consequently, a correlation exists between the dipole polarizability times the symmetry energy and the slope of the symmetry energy \cite{roca2015}. 
From the measurements of the dipole polarizability in $^{208}$Pb \cite{pb} and $^{120}$Sn \cite{sn} at RCNP Osaka, and  
in $^{68}$Ni at GSI \cite{ni1,ni2}, three empirical regions connecting the symmetry energy calculated at saturation density and its slope could be identified \cite{roca2015}. They are displayed in Fig. \ref{symmetryslope}. The yellow band represents the overlap area between the three regions. The points obtained with the two YGLO functionals of Ref. \cite{yglo} are reported in the figure showing that they provide a very satisfactory prediction in agreement with these empirical constraints. The inset of the figure shows the density dependence of the symmetry energy obtained for the two YGLO functionals.

\begin{figure}[tb]
\begin{center}
\begin{minipage}[t]{8 cm}
\epsfig{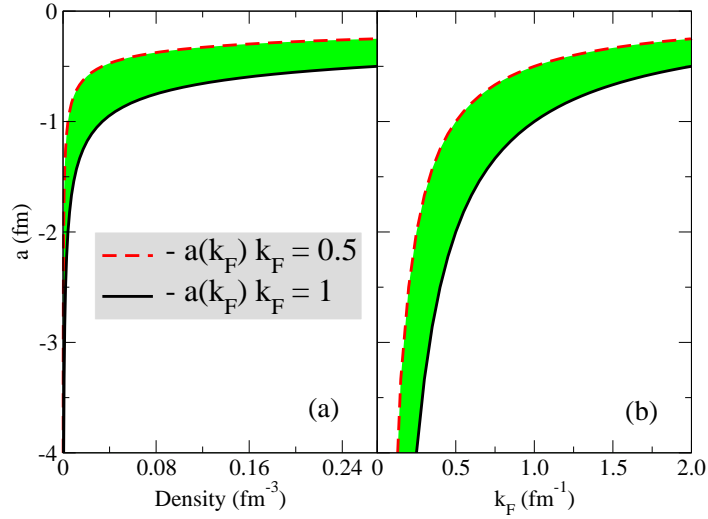}
\end{minipage}
\begin{minipage}[t]{16.5 cm}
\caption{Neutron-neutron $s$--wave scattering length versus density (a) and  Fermi momentum (b). The green region describes all the cases between 
 $|k_N \; a(k_N)|=0.5$ (red dashed line) and $|k_N \; a(k_N)|=1$ (black solid line).
Taken from Ref. \cite{elyo}. }
\label{adensi}
\end{minipage}
\end{center}
\end{figure}

Having the same objective of reconciling the two density scales in neutron matter, but without employing a resummation formula, another functional was introduced in Ref. \cite{elyo} and called later ELYO \cite{bonnard}. The good reproduction of both density scales is achieved by introducing a density--dependent neutron--neutron $s$--wave scattering length. The density dependence is automatically fixed by requiring that the low--density regime $\mid a k_N \mid <1 $ is satisfied. The condition $\mid a k_N \mid = R $, with $R \le 1$, is used to tune the scattering length for neutron matter. 
The physical value of $a$ is employed up to the value of the Fermi momentum where $\mid a k_N \mid \le R $ and the tuning is performed at higher densities using the constrain 
$\mid a k_N \mid = R $. 
This generates a profile of the scattering length as a function of the density and of the Fermi momentum as the one displayed in Fig. \ref{adensi}. 
The two cases $|k_N \; a(k_N)|=0.5$ and $|k_N \; a(k_N)|=1$ are represented by the red dashed and the black solid lines, respectively. The green area thus describes all the intermediate cases. Interestingly, this tuning of the scattering length as a function of the Fermi momentum, $1/k_N$, has strong analogies with the tuning of the scattering length as a function of the external magnetic field in ultracold trapped Fermi gases at unitarity (close to a Feshbach resonance). 

The Lee-Yang expansion, truncated for simplicity at the last term containing only 
$s$--wave parameters (the following term contains the $p$--wave scattering length) reads for neutron matter 
\begin{eqnarray}
\nonumber
\frac{E}{N}&=&\frac{\hbar^2k_N^2}{2m} \Big[ \frac{3}{5}+ \frac{2}{3\pi}(k_Na)+
 \frac{4}{35\pi^2}(11-2ln2)(k_Na)^2 \\  &+& \frac{1}{10\pi} (k_N r_s) (k_N a)^2 + 0.019  (k_N a)^3 \Big],
\label{lyelyo}
\end{eqnarray}
where $r_s$ is the effective range associated with the scattering length, whose physical value is 2.5 fm. 

\begin{figure}[tb]
\begin{center}
\begin{minipage}[t]{8 cm}
\epsfig{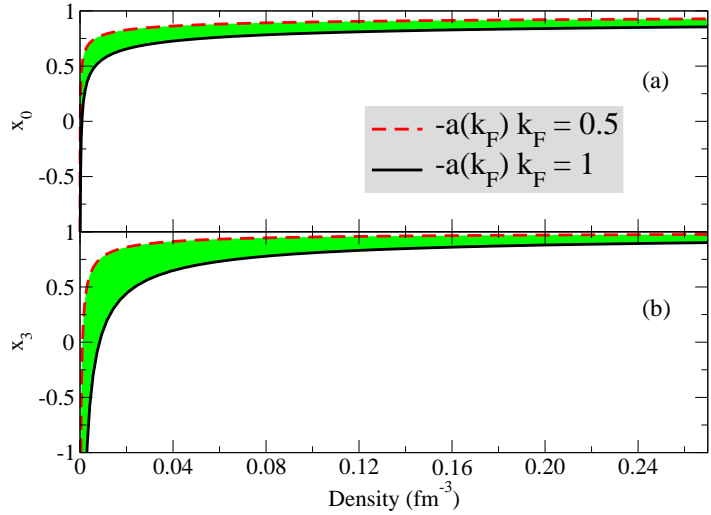}
\end{minipage}
\begin{minipage}[t]{16.5 cm}
\caption{Parameters $x_0$ and $x_3$ as a function of the density.
Taken from Ref. \cite{elyo}. }
\label{x03}
\end{minipage}
\end{center}
\end{figure}

\begin{figure}[tb]
\begin{center}
\begin{minipage}[t]{8 cm}
\epsfig{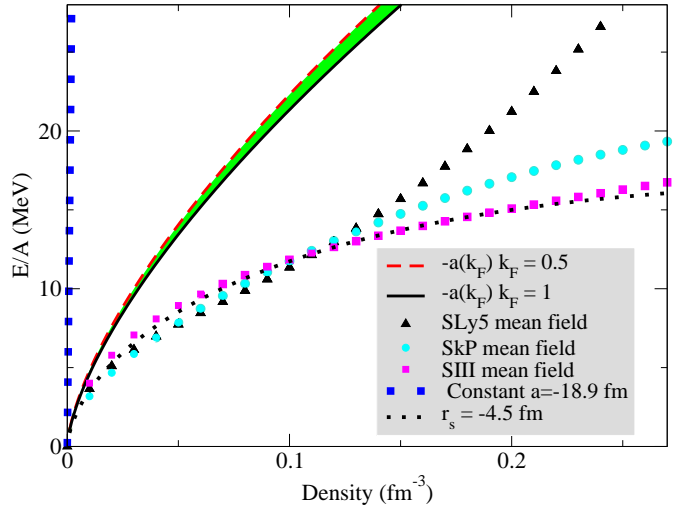}
\end{minipage}
\begin{minipage}[t]{16.5 cm}
\caption{EOSs of neutron matter. For the MF case, the SLy5 MF EOS (black triangels), the SkP MF EOS (cyan circles), and the  SIII MF EOS (magenta squares) are shown. The EOSs computed with the first two terms of the Lee-Yang expansion with $a=-18.9$ fm (blue squares) and with a density--dependent scattering length are plotted. For this latter case, the black solid line is the EOS obtained with the tuning  $|k_F \; a(k_F)|=1$ and 
the red dashed curve is the EOS obtained with the tuning $|k_F \; a(k_F)|=0.5$. The green area describes the intermediate cases. The EOS obtained with the tuning $|k_F \; a(k_F)|=1$ and by using an effective range of -4.5 fm is also shown (black dotted line).
Taken from Ref. \cite{elyo}. }
\label{eosfinal}
\end{minipage}
\end{center}
\end{figure}

It is required that the functional produces the EOS given by Eq. (\ref{lyelyo}) for neutron matter with a tuned scattering length, and provides as well a satisfactory EOS for symmetric matter close to equilibrium. A Skyrme--like functional is then introduced with only $s$--wave terms (the $t_2$ term is omitted). To reproduce Eq. (\ref{lyelyo}) for neutron matter such a functional must satisfy the following relations,  
\begin{eqnarray}
\nonumber
t_0(1-x_0)&=&\frac{4\pi\hbar^2}{m}a, \\
t_3(1-x_3)&=&\frac{144 \hbar^2}{35 m} (3\pi^2)^{1/3} (11-2ln2)a^2, \label{paratotal} \\
\nonumber
t_1 (1-x_1) &=& \frac{2\pi\hbar^2}{m}(a^2 r_s + 0.19 \pi a^3),
\end{eqnarray}
having chosen $\alpha=1/3$. One may start for example with a $t_0-t_3$ model only, adjust the parameters $t_0$ and $t_3$ to a benchmark EOS for symmetric matter, and tune the parameters $x_0$ and $x_3$ using the two upper lines of Eq. (\ref{paratotal}) and the low--density constraint for the scattering length. The resulting parameters $x_0$ and $x_3$ are plotted in Fig. \ref{x03}. 
It is interesting to note that both parameters vary very rapidly at low densities and are practically flat around the saturation density, where they have values close to typical $x_0$ and $x_3$ Skyrme values. This is indeed a clear indication that such a tuning is able to reconcile the two different density scales. The resulting EOS for neutron matter is represented by the green area in Fig. \ref{eosfinal}. 
In the figure, also the MF SLy5, SkP, and SIII EOSs are plotted. The EOS obtained by using only the first two terms of the Lee-Yang expansion and $a=-18.9$ fm is also shown. Even if the tuning of the scattering length significantly improves the EOS compared to this latter curve, one observes that neutron matter is still strongly underbound. The following terms of the Lee-Yang expansion appearing in Eq. (\ref{lyelyo}) (and, thus, the $t_1$ part of the Slyrme--like functional) are then included. The effective range is used as a free parameter to provide the missing binding and the obtained EOS is plotted in Fig. \ref{eosfinal} ($r_s=$ - 4.5 fm, for the case  $|k_F \; a(k_F)|=1$). Even if such an EOS is not the best, still it is very reasonable up to densities of $\sim$ 0.12 fm$^{-3}$ and comparable to the SIII--MF EOS at higher densities. This is indeed a very satisfactory result considering that only five free parameters have been used and, at the same time, the low--density behavior is guaranteed by construction. 

Let us now focus on another functional not motivated by the low--density regime of neutron matter. 
The authors of Ref. \cite{pakids} introduced an EFT--inspired functional formulated as a power expansion of the Fermi momentum, called KIDS. This expansion provides in general a richer and more flexible form compared to typical Skyrme--like functionals because it allows for instance for the inclusion of more than one type of density dependencies. 
The reader may refer to Refs. \cite{cochet,agrawal,xiong} to find examples of other attempts existing in the literature for including different density dependencies in Skyrme--like forces (however, using different criteria and different motivations from those associated with the KIDS functional). 
Such a choice of a power expansion in the Fermi momentum is indeed motivated by several indications. The authors were guided by: (i) the Brueckner theory for strongly interacting Fermi systems which leads to a potential energy per particle in nuclear matter written as the sum of powers of the Fermi momentum (the first term is of the type $k_F^3$ \cite{fetter}); (ii) the pionless EFT for fermionic systems at low density \cite{hammer}; (iii) the chiral perturbation theory in the three--loop approximation, where the terms $k_F^3$ and $k_F^4$ have been identified as those which provide saturation properties \cite{kaiserweise}. From all these indications, one may expect that such a low--order expansion may indeed correctly describe homogeneous matter at densities of interest for finite nuclei, around the saturation point. 
From this, the energy per particle is written in Ref. \cite{pakids} based on the following ansatz: 
\be
\varepsilon(\rho,\delta)=\frac{E(\rho,\delta)}{A}=\mathcal{T}(\rho,\delta)+\sum_{i=0}^3 c_i(\delta)
\rho^{1+i/3} + c_{ln}(\delta)\rho^2ln[\rho fm^3], 
\label{kids}
 \ee 
where the first term is the kinetic contribution, 
\begin{eqnarray}
\nonumber
\mathcal{T}&=&\mathcal{T}_p+\mathcal{T}_n, \\
\mathcal{T}_{p,n}&=&\frac{3}{5} \frac{\hbar^2}{2m_{p,n}}x^{5/3}_{p,n}(3\pi^2\rho)^{2/3},
\label{kine}
\end{eqnarray}
with $x_{p,n}=\rho_{p,n}/\rho$. In general, also higher--order powers may be included. It was found in Ref. \cite{pakids} that the logarithmic term does not have any relevant  
impact on the results. This term was then omitted. 
Equation (\ref{kids}) may be written as 
\be
\varepsilon = \mathcal{T} + \sum_{i=0}^3 \varepsilon_i + \varepsilon_{ln}. 
\label{kidsfu}
\ee
A correspondence with the terms of a Skyrme--like functional may be found (see Table \ref{parakids} where the logarithmic term is omitted). 
\begin{table}
\centering
\caption{Correspondence between Skyrme--type terms and Fermi momentum dependencies in the KIDS EOS}\vspace{0.5cm}
\begin{tabular}{  c c c   }
    \hline
 Equation (\ref{kidsfu})  &  Skyrme term & Power of Fermi momentum   \\
\hline
\hline
$\mathcal{T}$ & kinetic contribution & $k_F^2$ \\
$\varepsilon_0$ & $t_0$ & $k_F^3$ \\
$\varepsilon_1$ & $t_3$ with $\alpha=1/3$ & $k_F^4$ \\
$\varepsilon_2$ & $t_1$, $t_2$, $t^{\prime}_3$ with $\alpha^{\prime}=2/3$ & $k_F^5$ \\
$\varepsilon_3$ & $t^{\prime \prime}_3$ with $\alpha^{\prime \prime}=1$ & $k_F^6$ \\  
   \hline
\label{parakids}
  \end{tabular}
  \end{table}

\begin{figure}[tb]
\begin{center}
\begin{minipage}[t]{11 cm}
\epsfig{file=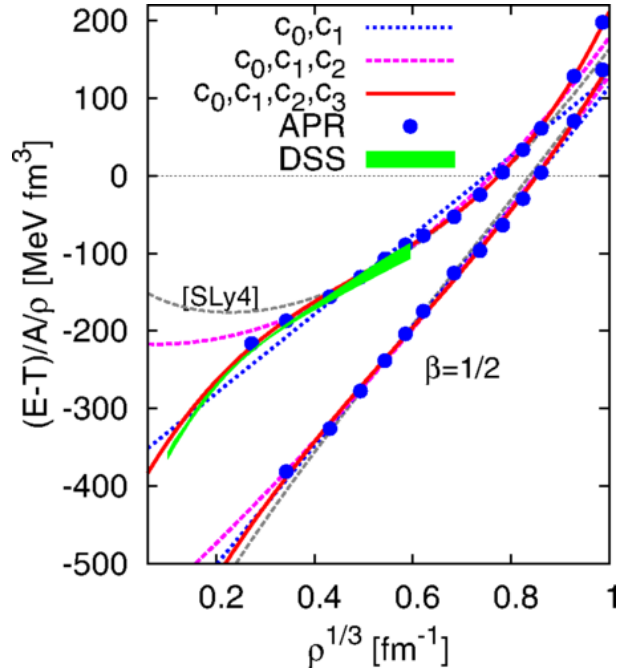,scale=0.5}
\end{minipage}
\begin{minipage}[t]{16.5 cm}
\caption{EOSs represented as the total EOS minus the kinetic contribution per particle divided by the density as a function of $\rho^{1/3}$. The case $\beta=1/2$ is considered. The EOSs for neutron (upper curves) and symmetric (lower curves) matter are plotted by increasingly including more terms in the $k_F$ expansion. Also the results from Refs. \cite{akmal} and \cite{soma} are shown. 
Taken from Ref. \cite{pakids}. }
\label{kneut}
\end{minipage}
\end{center}
\end{figure}
\begin{figure}[tb]
\begin{center}
\begin{minipage}[t]{8 cm}
\epsfig{file=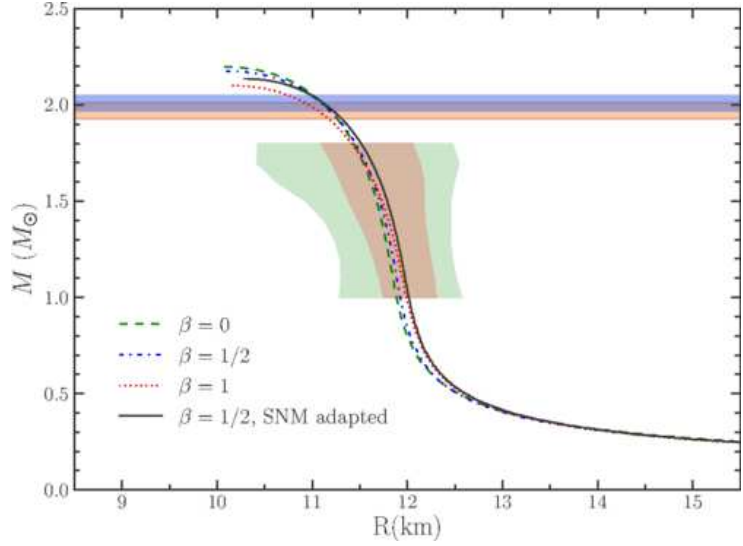,scale=0.55}
\end{minipage}
\begin{minipage}[t]{16.5 cm}
\caption{Mass/radius relation for a neutron star obtained with the parametrization sets reported in Ref. \cite{pakids}. The two horizontal bands represent the constraints provided by the observation  of two--solar--mass neutron stars \cite{solar1,solar2}. The other colored areas describe the constraints provided by x--ray burst data \cite{ray}. 
Taken from Ref. \cite{pakids}. }
\label{massradius}
\end{minipage}
\end{center}
\end{figure}
\begin{figure}[tb]
\begin{center}
\begin{minipage}[t]{10 cm}
\epsfig{file=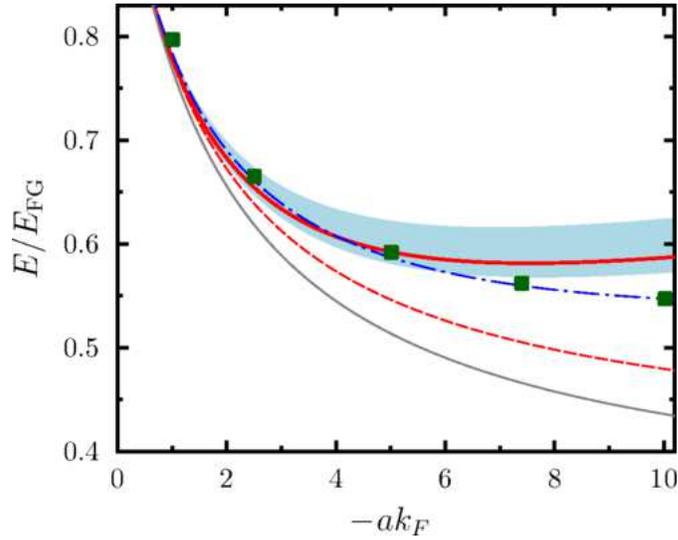,scale=0.5}
\end{minipage}
\begin{minipage}[t]{16.5 cm}
\caption{Energy of neutron matter divided by $E_{FG}$. The two red lines describe the results obtained with the new functional using the values 
$\xi_0=$ 0.3897 and $\eta_s=0.127$ taken from Refs. \cite{for1,for2} (solid line) or $\eta_s=0$ (dashed line). The colored area represents the dispersion of the results by varying $\xi_0$ from 0.37 to 0.44 ($\eta_s$ fixed). The green suares are the QMC results of Ref. \cite{geca}. 
Also the results obtained with $\xi_0=$ 0.44 and $\eta_s=0.03$ (blue dot--dashed line), and with $\xi_0=$ 0.326 and $\eta_s=0.19$ (gray line) are plotted.
Taken from Ref. \cite{la2016}. }
\label{uniene}
\end{minipage}
\end{center}
\end{figure} 
\begin{figure}[tb]
\begin{center}
\begin{minipage}[t]{10 cm}
\epsfig{file=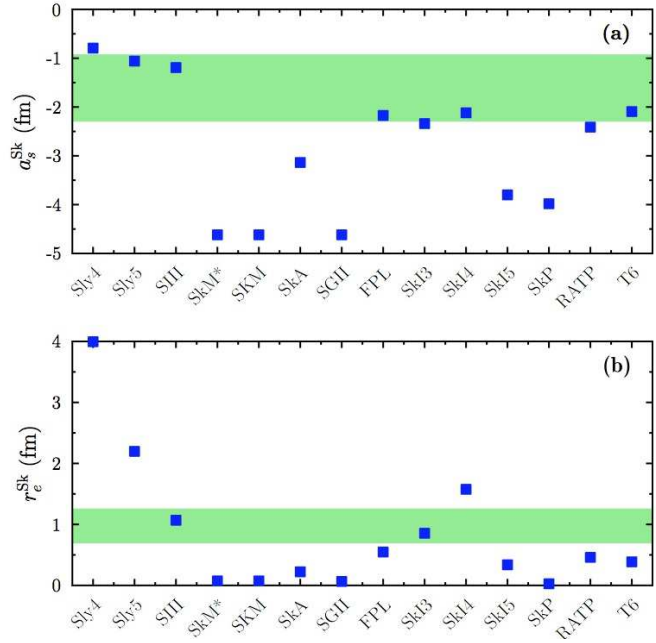,scale=0.65}
\end{minipage}
\begin{minipage}[t]{16.5 cm}
\caption{Scattering length (a) and effective range (b) dispersion obtained in the window of densities from 0.01 to 0.2 fm$^{-3}$ using the functional of Ref. \cite{la2016} (green horizontal areas). Several Skyrme values are also shown for comparison (blue squares). Adapted from Ref. \cite{la2017}}
\label{unipara}
\end{minipage}
\end{center}
\end{figure} 

\begin{figure}[tb]
\begin{center}
\begin{minipage}[t]{14 cm}
\epsfig{file=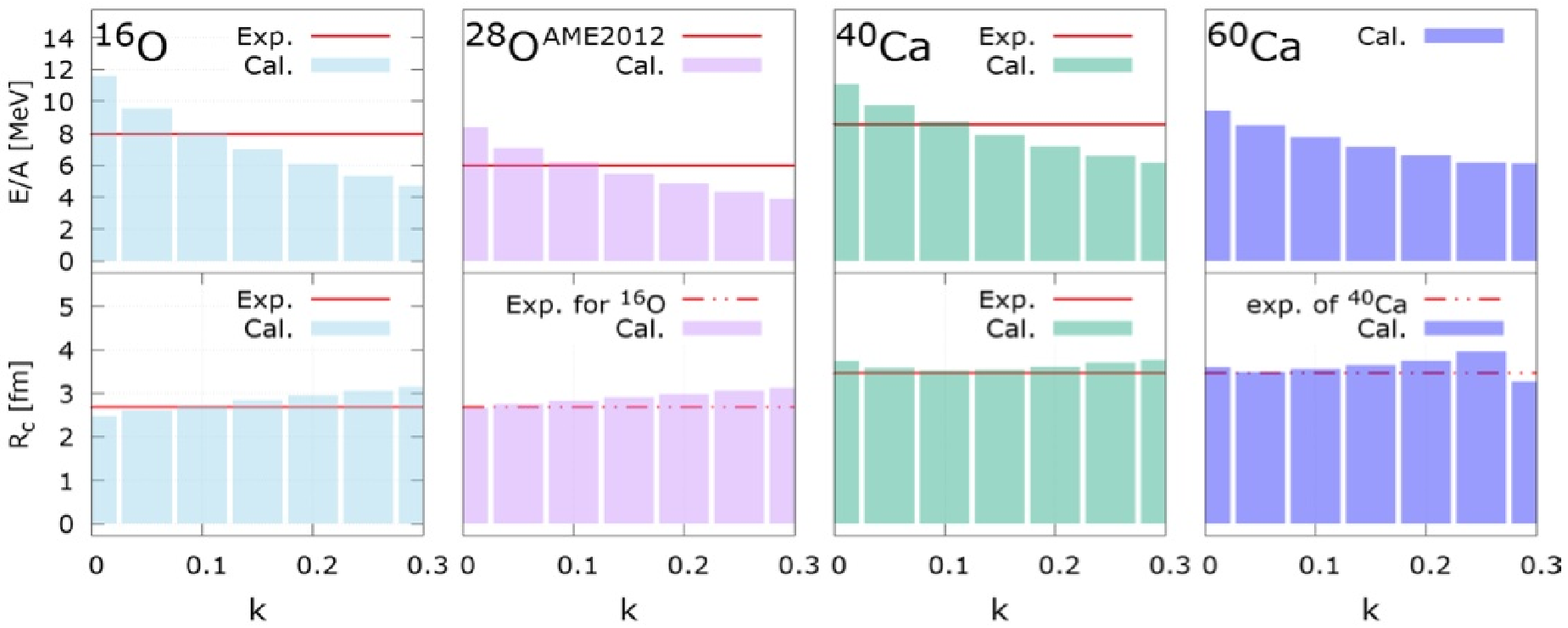,scale=0.8}
\end{minipage}
\begin{minipage}[t]{16.5 cm}
\caption{ Binding energies and charge radii for $^{16}$O, $^{28}$O, $^{40}$Ca, and $^{60}$Ca plotted as a function of the parameter $k$. Full lines describe the experimental values. The experimental binding energies are extracted from AME2012 \cite{audi1,audi2}. The dot--dashed horizontal lines represent the charge radius of $^{16}$O (in the panel corresponding to $^{28}$O) and the charge radius of $^{40}$Ca (in the panel corresponding to $^{60}$Ca)
Taken from Ref. \cite{gil1}.}
\label{masses}
\end{minipage}
\end{center}
\end{figure}

Indeed, several density dependencies may be found, including an optional density dependence associated with the term of the type $k_F^5$, which may be produced either by a velocity--dependent term or by a density--dependent term with power equal to 2/3 (or by a combination of both terms). This distinction between different contributions is not relevant for uniform matter and was analyzed only in the first applications done to study finite nuclei, where ground--state energies and radii of selected closed--shell nuclei were used for the adjustment  
\cite{gil1,gil2}. 
The adjustment of the curves was done on EOSs (pseudodata) from Ref. \cite{akmal} and from Ref. \cite{fp}, by minimizing 
\be
\chi^2(\delta)=\sum_j exp\bigg\{-\frac{\beta \rho_j}{\rho_0}\bigg\} \Bigg( \frac{\varepsilon(\rho_j)-D_j}{\mathcal{T}(\rho_j)} \Bigg)^2, \; \beta \ge 0, 
\label{chis}
\ee
with $\rho_0=$ 0.16 fm$^{-3}$ and where the available pseudodata $(\rho_j,D_j)$ were used. The parameter $\beta$ is utilized to put a weight on specific regions of the density. Figure \ref{kneut} illustrates the results obtained for the choice $\beta=1/2$ and the adjustment done on Ref. \cite{akmal} (called APR in the figure). Results for both neutron (upper curves) and symmetric (lower curves) matter are displayed by increasingly including more terms in the $k_F$ expansion. For comparison, the chiral EFT results of Ref. \cite{soma} (denoted by DSS in the figure) are also shown. 
One observes that the description of pure neutron matter requires the inclusion of terms higher than $k_F^4$ whereas, as expected, the inclusion of the first $c_0$ and $c_1$ terms is sufficient for providing satisfactory results for symmetric matter (this corresponds to a $t_0-t_3$ model). In Ref. \cite{forbes}, similar conclusions (as those for symmetric matter) were drawn for the minimal number of terms required in a functional to describe reasonably well finite nuclei.   

\begin{figure}[tb]
\begin{center}
\begin{minipage}[t]{10 cm}
\epsfig{file=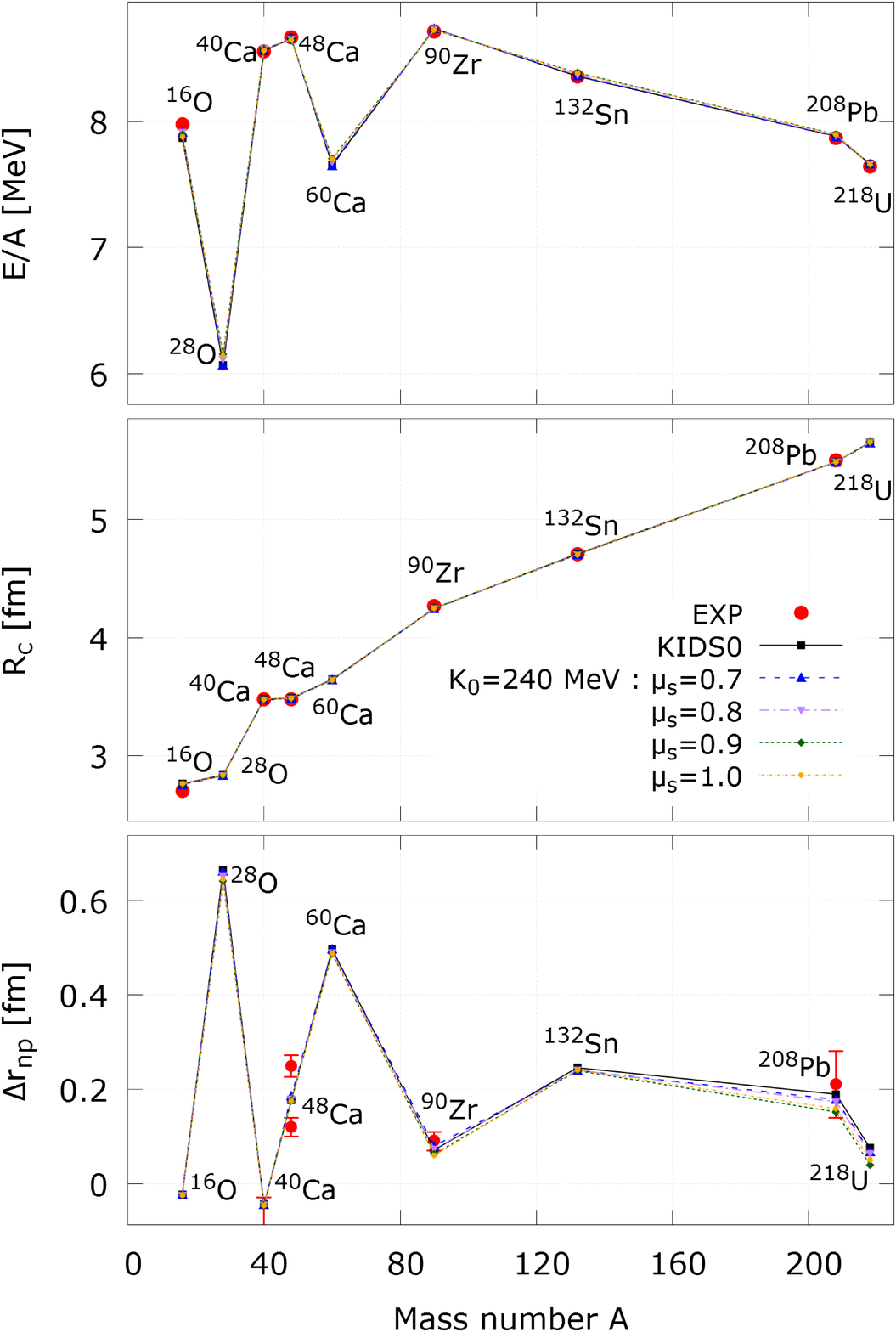,scale=0.15}
\end{minipage}
\begin{minipage}[t]{16.5 cm}
\caption{Binding energies (upper panel), charge radii (medium panel), and neutron skin thicknesses (lower panel) as a function of the number of nucleons $A$. See text for the details on the legend. Taken from Ref. \cite{gil3}. }
\label{masskids}
\end{minipage}
\end{center}
\end{figure} 
\begin{figure}[tb]
\begin{center}
\begin{minipage}[t]{10 cm}
\includegraphics[width=0.9\columnwidth]{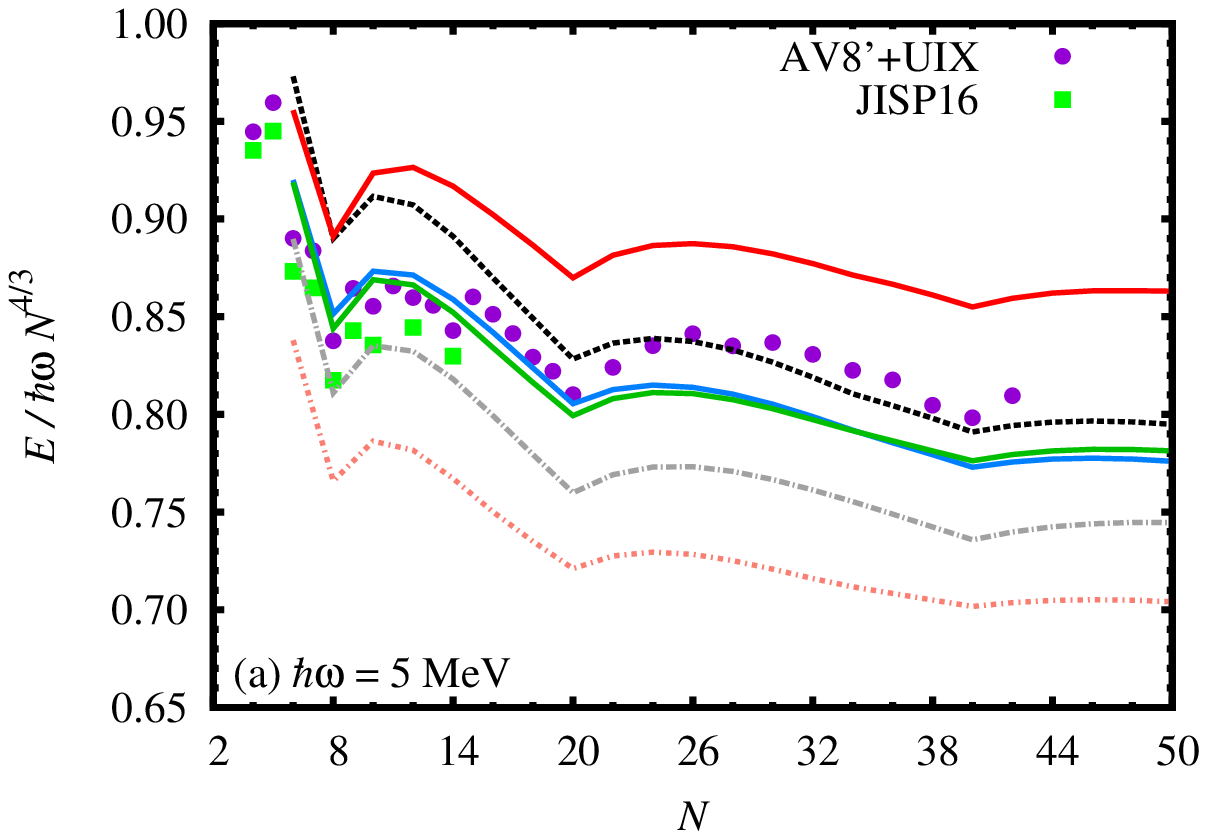}
\includegraphics[width=0.9\columnwidth]{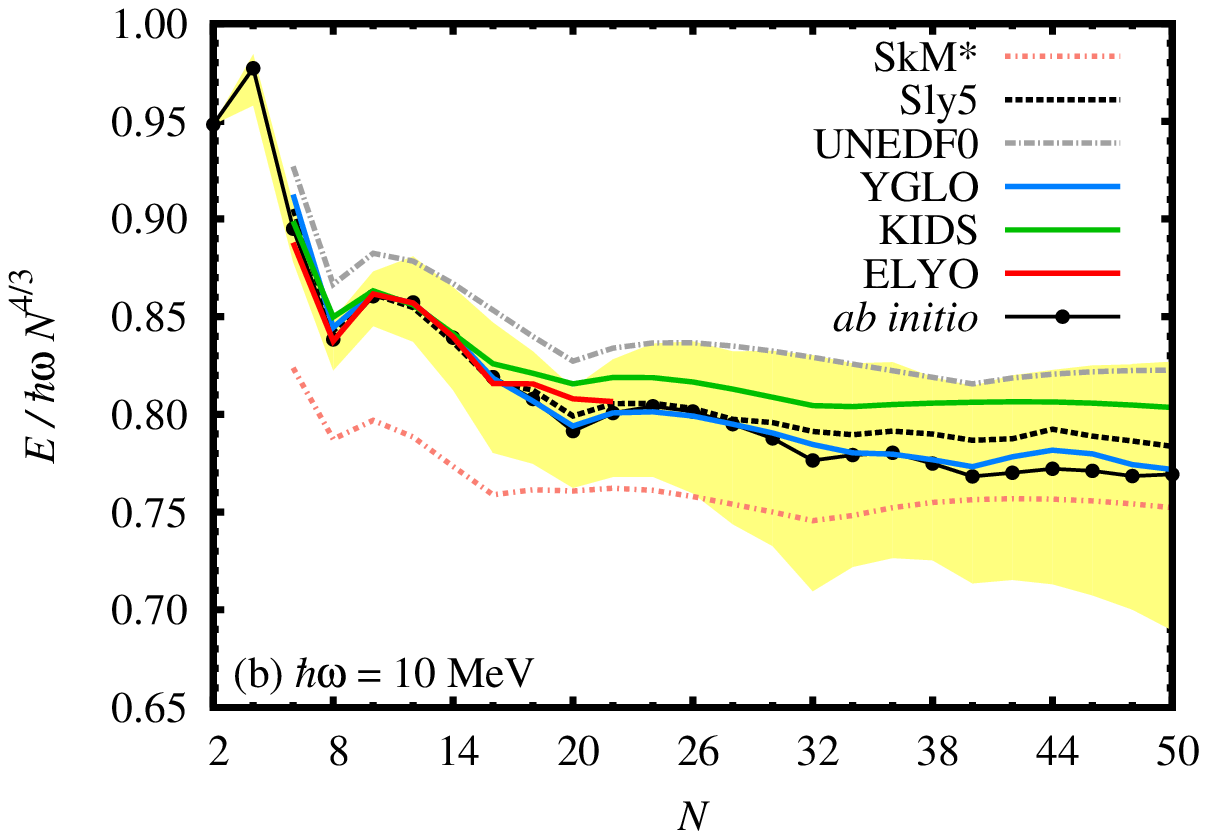}
\end{minipage}
\begin{minipage}[t]{16.5 cm}
\caption{Energies of the drops scaled by 
$\hbar \omega N^{4/3}$ as a function of $N$ for $\hbar \omega=$ 5 (a) and 10 (b) MeV. The Skyrme SLy5, SkM*, and UNEDF0 MF results are also shown. For the case $\hbar \omega=$ 5 MeV, the available AV8' + UIX and JISP16 results are plotted. The yellow area in panel (b) represents the dispersion of the {\it{ab--initio}} results that have been used to compute the average benchmark curve, denoted by {\it{ab initio}}. Taken from Ref. \cite{bonnard}  }
\label{abi}
\end{minipage}
\end{center}
\end{figure} 
\begin{figure}[tb]
\begin{center}
\begin{minipage}[t]{10 cm}
\includegraphics[width=0.9\columnwidth]{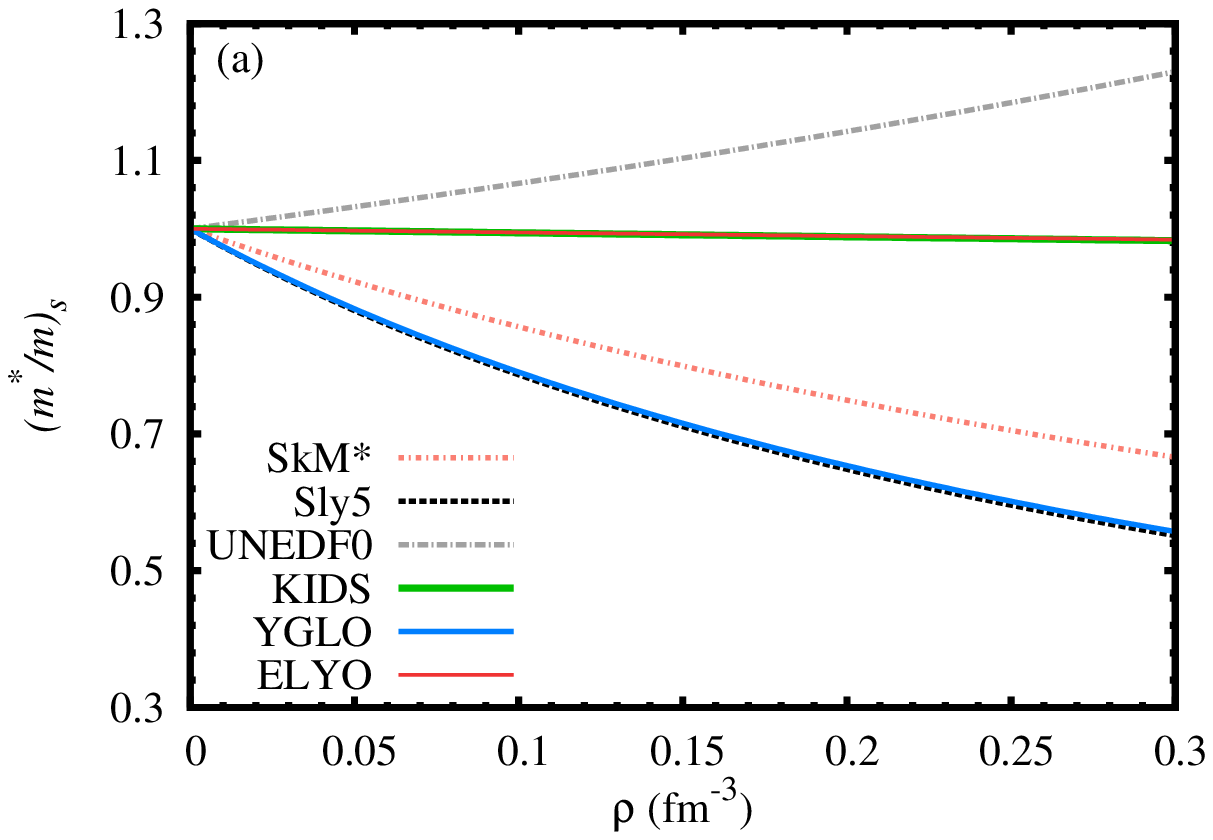}
\includegraphics[width=0.9\columnwidth]{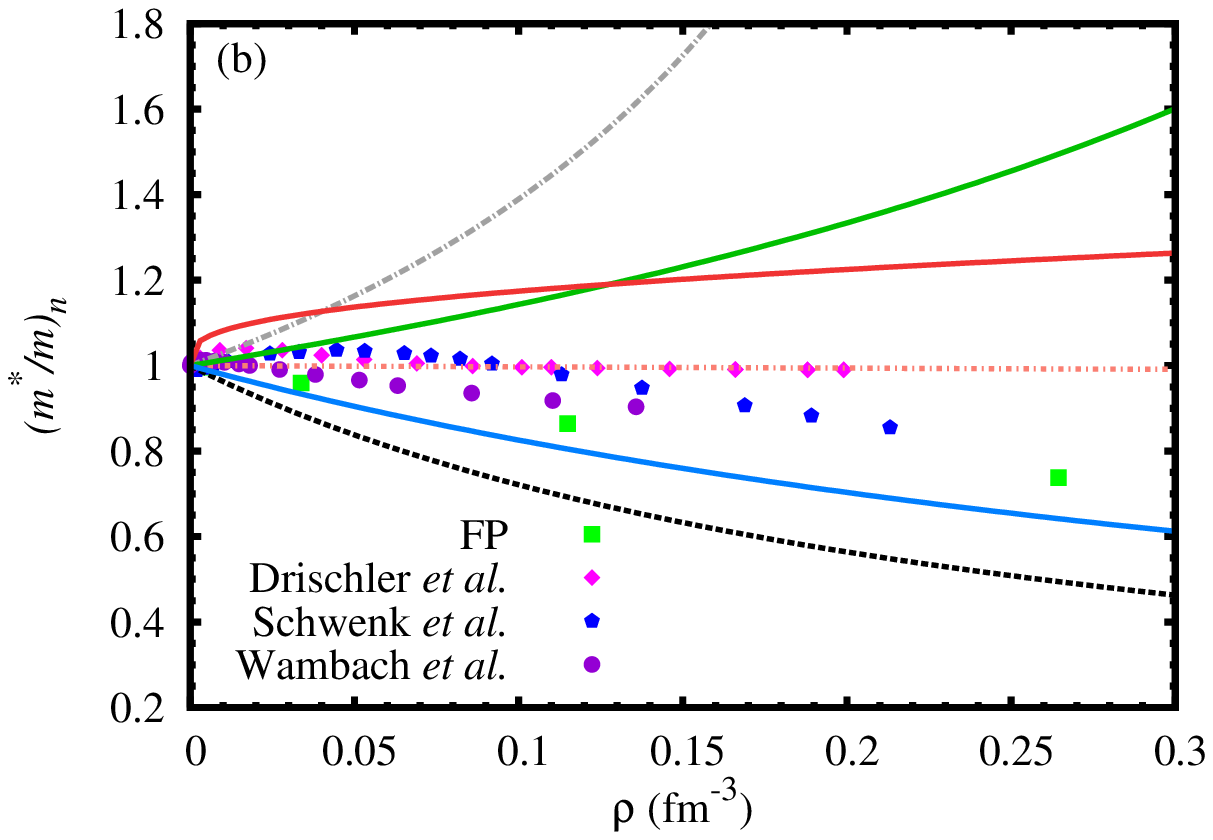}
\end{minipage}
\begin{minipage}[t]{16.5 cm}
\caption{Isoscalar (a) and neutron (b) effective masses. See text for the details on the shown curves. Taken from Ref. \cite{bonnard}. }
\label{mdrop}
\end{minipage}
\end{center}
\end{figure} 
Several sets of parametrizations were generated (see Table III of Ref. \cite{pakids}). The Tolman-Oppenheimer-Volkoff equations were solved with these parametrizations and the obtained mass/radius relation for a neutron star is shown in Fig. \ref{massradius}. 
The colored bands in the figure represent constraints provided by observation. Namely, the two horizontal bands are the constraints provided by the observation of two--solar--mass neutron stars \cite{solar1,solar2}. The other colored areas describe the constraints coming from x--ray burst data \cite{ray}. 

The three functionals described so far are constructed for describing both neutron and symmetric matter. Other criteria and motivations may be followed if one deals only with pure neutron matter.

Another functional was proposed for example in Ref. \cite{la2016} tailored for describing unitarity in Fermi gases. One may find several studies in the literature where functionals were proposed for describing the unitary regime (see, for instance, Refs. \cite{bul2007,bul2012,pape,bhatta}). Reference  \cite{la2016} proposed a different resummed formula with respect to the YGLO case, containing also the effective range $r_s$, namely 
\be
\frac{E}{N}=\frac{\hbar^2 k_F^2}{2m} \Bigg[ \frac{3}{5}+\frac{(ak_F)A_0}{1-A_0^{-1} [A_1+(r_sk_F)A_2]ak_F} \Bigg], 
\label{resuni}
\ee
for a system of $N$ fermions, where $A_0$, $A_1$, and $A_2$ are parameters. By taking the limit $a \rightarrow \infty$, one obtains the universal unitary regime where the energy of the system is proportional to the free Fermi gas energy (the factor is a universal parameter called the Berstch parameter \cite{be2000}), $E=\xi(r_sk_F)E_{FG}$, with, in the case of the functional (\ref{resuni}), 
\be
\xi (r_sk_F) = \Bigg\{ 1-\frac{5}{3}\frac{A_0^2}{A_1+(r_sk_F)A_2} \Bigg\}.
\label{bepa}
\ee
By considering a small effective range, Eq. (\ref{bepa}) may be written as the following expansion, 
\begin{eqnarray}
\nonumber
\xi (r_sk_F) &\sim& \Bigg( 1-\frac{5}{3} \frac{A_0^2}{A_1} \Bigg) + \frac{5}{3} \frac{A_0^2A_2}{A_1^2} (r_s k_F) - \frac{5}{3} \frac{A_0^2A_2^2}{A_1^3} (r_s k_F)^2 \\
&\equiv& \xi_0 +(r_s k_F) \eta_s + (r_s k_F)^2 \delta_s.
\label{expabe}
\end{eqnarray}
By constraining the values of $A_i$ from the low--density regime, one obtains $\xi_0=0.326$, $\eta_s=0.19$, and $\delta_s=-0.055$. 

Another procedure was then followed allowing for a better description of the unitary gas. It was required that the expansion of the functional (\ref{resuni}) corresponds to Eq. (\ref{expabe}) up to the linear term in $(r_sk_F)$ and that $A_0$ is equal to $2/(3\pi)$ to satisfy as well the Lee-Yang expansion up to the term linear in the scattering length. This leads to 
\be
\frac{E}{N}=\frac{\hbar^2 k_F^2}{2m} \Bigg\{  \frac{3}{5}+\frac{2}{3\pi} \frac{ak_F}{1-\frac{10}{9\pi} (ak_F)/[1-\xi(r_sk_F)]} \Bigg\}, 
\label{fununi}
\ee
where
\be
\xi(r_sk_F)=\Bigg\{ 1-\frac{(1-\xi_0)^2}{(1-\xi_0)+(r_sk_F)\eta_s} \Bigg\}.
\label{expbenew}
\ee
One may notice that a similar resummed formula was proposed in Refs. \cite{adhi1,adhi2} in the limit $r_s=0$. 
The Bertsch parameter $\xi_0$ and the parameter associated with the effective range $\eta_s$ can be taken from other studies dedicated to unitary gases. 
The advantage of the functional (\ref{fununi}) compared to previous proposed functionals 
\cite{bul2007,bul2012,pape,bhatta} is that it may describe effects related to a non zero effective range and it may also be applied to the low--density regime. 

Figure \ref{uniene} represents the energy of neutron matter divided by the free gas energy $E_{FG}$. The two red lines describe the results obtained with the new functional using the values 
$\xi_0=$ 0.3897 and $\eta_s=0.127$ taken from Refs. \cite{for1,for2} (solid line) or $\eta_s=0$ (dashed line). One may thus observe the important effect of the finite effective range. The colored area describes the dispersion of the results by varying $\xi_0$ from 0.37 to 0.44 ($\eta_s$ fixed). The comparison is done with the QMC results of Ref. \cite{geca} (green squares).
The results which perfectly reproduce the QMC trend are also shown (blue dot--dashed line). They are obtained with $\xi_0=$ 0.44 and $\eta_s=0.03$. Also the curve obtained with $\xi_0=$ 0.326 and $\eta_s=0.19$ (using the constraints from the low--density regime) is plotted (grey line). The values $a=-18.9$ fm and $r_s=2.7$ fm were used to produce these curves. 

Reference \cite{la2017} pursued the investigation of the characteristics of the functional introduced in Ref. \cite{la2016} along the objective of constructing a functional without adjustable parameters. The results of this article show that one may treat perturbatively nuclear systems as an expansion with respect to the unitary gas \cite{bira}. The limitation of the functional introduced in Refs. \cite{la2016,la2017} is that it can be utilized only for neutron matter and provides satisfactory results only up to densities $\rho \sim$ 0.01 fm$^{-3}$ (extending by the way considerably the domain of validity of the Lee-Yang expansion). If one requires that also densities larger than 0.01 fm$^{-3}$ are satisfactorily well described, then adjustable parameters are necessary. 
It is possible also for this functional to extract a tuning of the scattering length (as done for ELYO by using the low--density regime) and of the effective range by using this time criteria from unitarity. It was found in Ref. \cite{la2017} that, in the window of densities $0.01<\rho<0.2$ fm$^{-3}$, the scattering length and the associated effective range do not vary a lot. Their values are represented by the green areas in Fig. \ref{unipara}. In the same figure, values provided by several Skyrme parametrizations are also shown indicating that the obtained tuned values indeed are very close to typical Skyrme values where the parametres are empirically adjusted on properties of finite nuclei and matter close to the equilibrium density.

\subsection{\it Finite--size effects. Extensions of YGLO, ELYO, and KIDS to nuclei and neutron drops \label{finite}}

The KIDS functional was already extended for applications to finite nuclei in Refs. \cite{gil1,gil2}. A parameter $k$ was introduced to put a weight between possible velocity--dependent and density--dependent terms for the $\varepsilon_2$ term of the functional in Table \ref{parakids}. An adjustment was done on the binding energies and the charge radii of some selected closed--shell nuclei. Figure \ref{masses} is extracted for example from Ref. \cite{gil1}. It represents the binding energies and the charge radii of $^{16}$O, $^{28}$O, $^{40}$Ca, and $^{60}$Ca, obtained by varying the parameter $k$. 
The full horizontal lines describe the experimental values. Experimental binding energies are extracted from AME2012 \cite{audi1,audi2}. The dot--dashed horizontal lines represent the charge radius of $^{16}$O (in the panel corresponding to $^{28}$O) and the charge radius of $^{40}$Ca (in the panel corresponding to $^{60}$Ca). The calculations were done for the KIDS-ad2 parametrization of Ref. \cite{pakids}. For simplicity, the parameters $x_1$ and $x_2$ of the corresponding Skyrme functional were set equal to zero. 
From this figure, a value of $k \sim 0.1$ was extracted, corresponding to an effective mass almost equal to the bare mass. Also the spin--orbit coupling strength was adjusted. 

More recently, the effective mass was used as a criterium to adjust the KIDS functional for applications to finite nuclei \cite{gil3} without altering the original  EOS for matter. 
The EOS provided by the parametrization KIDS-ad2 of Ref. \cite{pakids} was used as a starting point. Differently from what done in Refs. \cite{gil1,gil2}, the parameters $x_1$ and $x_2$ were not set equal to zero. The adjustment of the parameters was done on the binding energies and the charge radii of the nuclei $^{40}$Ca, $^{48}$Ca, and $^{208}$Pb (the latter two nuclei were used for fitting in particular the spin--orbit coupling strength). Several parametrizations were  generated, each of them corresponding to some specific chosen values for the isoscalar and isovector effective masses and predictions were done for other nuclei different from those entering in the fitting procedure. It was found in Ref. \cite{gil3} that static and bulk properties of nuclei such as masses, charge radii, and neutron skin thicknesses are not affected by the different considered values of the effective mass.
For example, Fig. \ref{masskids} shows binding energies, charge radii, and neutron skin thicknesses as a function of the number of nucleons $A$ for several nuclei. Esperimental binding energies and charge radii are taken from Refs. \cite{angeli}, whereas the experimental values related to neutron skins are taken from Refs. \cite{jas,mah,pre}. KIDS0 in the legend of the figure denotes results which are obtained with the parametrization of Refs. \cite{gil1,gil2}. The other results were calculated in Ref. \cite{gil3} with parametrizations associated with an incompressibility of 240 MeV and different values of the isoscalar effective mass, denoted by $\mu_s$ in the figure. Only the neutron skin thickness of the heaviest nuclei is slightly impacted by the different values of the isoscalar effective mass. 

As a perspective, the authors of Ref. \cite{gil3} mentioned the possibility of using in future this effective--mass--based tuning of the parameters for performing an adjustment on dynamical properties, such as collective excitations \cite{blaizot,bohigas,grasso2018,bao}.  

The YGLO and ELYO functionals were recently extended to the treatment of neutron drops \cite{bonnard}. These systems are not self--bound. To localize them, one needs to apply an external trap, for instance an harmonic oscillator potential. Neutron drops represent ideal systems which may be very useful for analyzing finite--size effects and properties related to neutron--rich systems. For instance, the authors of Ref. \cite{zhao} identified a linear correlation  existing between neutron drop radii and the neutron skin thickness of $^{48}$Ca and $^{208}$Pb. Several studies are devoted to neutron drops in the literature (see, for instance, Refs. \cite{zhao,pud,maris,gando,potter}). 

Reference \cite{bonnard} presents the application to neutron drops of the functional YGLO, ELYO, and KIDS. For the latter, the same parametrization published in Refs. \cite{gil1,gil2} for finite nuclei was used. On top of this, pairing correlations were included and a pairing strength was adjusted. For the cases YGLO and ELYO, new parameters were taken into account (in addition to those already set for infinite matter), namely a parameter to disentangle between velocity-- and density--dependent (with $\alpha=2/3$) terms (as in the version of KIDS published in Refs. \cite{gil1,gil2}), a spin--orbit strength, and a pairing strength. A mixed surface/volume contact pairing interaction was adopted and HFB calculations were performed by applying an external harmonic potential characterized by a frequency $\hbar \omega$. 
A benchmark curve was adopted for the adjustment of the parameters constructed, for the case $\hbar \omega=10$ MeV, as the average of: three QMC calculations \cite{maris,gando} obtained using the AV8' two--body force \cite{pud2} and the AV8' plus three--body UIX \cite{pud3} and IL7 \cite{pieper} forces; results obtained with configuration--interaction calculations \cite{maris} with the JISP16 force \cite{shi}; no--core shell--model and coupled--cluster results with a chiral EFT interaction \cite{potter}. This average is called {\it{ab initio}} in panel (b) of Fig. \ref{abi}. The fit is done for $\hbar \omega=$ 10 MeV and a number of neutrons $N$ equal to 8, 12, 14, 16, and 20. For $N>20$ and for other oscillator frequencies, results may be thus regarded as genuine predictions. The two panels of Fig. \ref{abi} show the energies of the drops scaled by 
$\hbar \omega N^{4/3}$ as a function of $N$ for $\hbar \omega=$ 5 (a) and 10 (b) MeV. For comparison, also the Skyrme SLy5, SkM*, and UNEDF0 \cite{une} MF results are shown. For the case $\hbar \omega=$ 5 MeV, the available AV8' + UIX and JISP16 results are plotted. The yellow area in panel (b) represents the dispersion of the {\it{ab--initio}} results that have been used to compute the average benchmark curve. 

For the case $\hbar \omega=$ 10 MeV, one observes that the YGLO results are those which better follow the {\it{ab--initio}}  curve. The ELYO results could be obtained numerically in this case only up to $N=22$. It turns out that $N=22$ corresponds (for this frequency) to $\rho \sim$ 0.2 fm$^{-3}$, where the ELYO EOS of neutron matter starts to be significantly different from the others (neutron matter is quite strongly overbound). Also, one notices that the KIDS functional slightly underbinds the drops for $N \ge$ 15. One observes that, whereas two Skyrme parametrizations, SkM* and UNEDF0, are both quite far from the {\it{ab--initio}} curve, the SLy5 MF results are indeed very satisfactory. 
Panel (a) shows that the YGLO and KIDS results are close to one another, whereas for this case the ELYO functional systematically underbinds the drops. This is due to the slight underbinding of neutron matter that the ELYO EOS produces at densities lower than 0.1 fm$^{-3}$, which are the densities at play in this case. 

All these results  indicate that the ELYO EOS for neutron matter at ordinary densities (outside the low--density regime) should be indeed improved. Work is in progress along this direction with the inclusion of the $p$--wave contribution in the Lee-Yang expansion \cite{bonnard2}. 

The distinction between density-- (with $\alpha=2/3$) and velocity--dependent terms does not make any difference in the EOS of matter where a $k_F^5$ term describes their global effect. Such a distinction has on the other side a strong impact on the effective mass. 
Figure \ref{mdrop} shows the isoscalar 
\be
\Bigg( \frac{m^*}{m} \Bigg)_S^{-1}=1+\frac{m}{8\hbar^2} \Theta_S \rho
\label{isom}
\ee
and the neutron 
\be
\Bigg( \frac{m^*}{m} \Bigg)_n^{-1}=1+\frac{m}{4\hbar^2} (\Theta_S - \Theta_V) \rho
\label{neum}
\ee
effective masses as a function of the density. For the neutron effective mass, also the FP \cite{fp}, Drischler et al. \cite{dri52}, Schwenk et al. \cite{sch66}, and Wambach et al. \cite{wam67} results are shown. Without velocity--dependent terms the three EFT--inspired functionals YGLO, ELYO, and KIDS would have effective masses equal to 1 because the calculations which are done with these functionals are truncated at the leading order. By introducing a splitting parameter that distinguishes between density-- and velocity--dependent contributions and by adjusting it, the results of Fig. \ref{mdrop} are obtained. One observes that the isoscalar effective mass is almost independent of the splitting parameter for the ELYO and KIDS cases, whereas it is affected by the splitting in the case of YGLO (effective mass $\sim$ 0.7 around the saturation density). For the neutron effective mass, the impact of the splitting parameter is quite relevant in all cases. 

\subsection{\it Other {\it{ab--initio}}--inspired EDFs
  \label{dme}}

\begin{figure}[tb]
\begin{center}
\begin{minipage}[t]{10 cm}
\epsfig{file=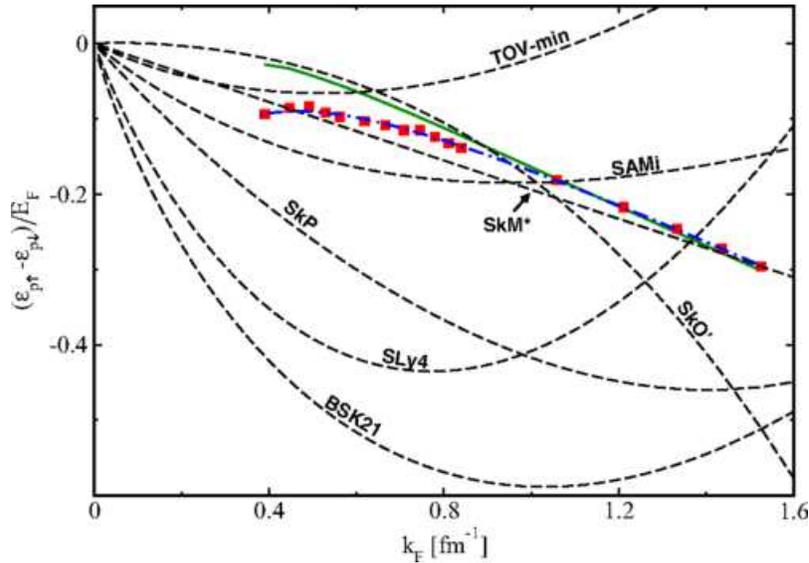,scale=0.6}
\end{minipage}
\begin{minipage}[t]{16.5 cm}
\caption{Difference between the energies of the two proton impurities (spin--up and spin--down) scaled by the Fermi energy of the uniform system composed of spin--up neutrons. Several curves obtained with Skyrme functionals are compared with configuration--interaction QMC results (red squares). The green line represents second--order perturbation--theory results and the blue dot--dashed line is obtained with the adjusted curve, Eq. (\ref{fitformula}). Taken from Ref. \cite{roggero}. }
\label{impurity}
\end{minipage}
\end{center}
\end{figure}

A way for constructing less phenomenological functionals with a strong {\it{ab--initio}} connection may be based on benchmarking their coupling constants on pseusodata obtained with microscopic 
{\it{ab--initio}} calculations. This strategy was discussed for example in Ref. \cite{roggero} where, in particular, a procedure was suggested for contraining the time--odd part of a functional, that is the part of the functional depending on the time--odd densities. 
The time--odd parts of available  functionals are in general not well constrained. Constraining them correctly may have an important impact in the description of  several properties of finite nuclei, such as the binding energies of odd nuclei \cite{satula1} and the properties of deformed rotating nuclei \cite{dobadudek,post}. 
This strategy was applied to a system composed of polarized neutron matter (with spin--up) containing one impurity that can be constituted by a spin--down neutron, a spin--up proton, or a spin--down proton. This system is strongly analogous to the polaron case in atomic gases or bulk solid--state systems \cite{chevy,massignan}. Another work analyzing polarized matter with an impurity in the nuclear case is for example Ref. \cite{forbespol}

A Skyrme--type EDF may be written in the following way for matter, 
\be
\varepsilon=\varepsilon_{kin} + \sum_{t=0,1} (C_t^{\rho} \rho_t^2 + C_t^{\tau} \rho_t \tau_t +
C_t^s s_t^2 + C_t^T s_tT_t), 
\label{funcpolaron}
\ee
where $\rho_0$, $s_0$, $\tau_0$, and $T_0$ are the isoscalar density, spin density, kinetic density, and spin kinetic density, respectively. The corresponding quantities with $t=1$ represent isovector quantities. The last two terms in Eq. (\ref{funcpolaron}) constitute the time--odd part. Starting from the EDF given by Eq. (\ref{funcpolaron}), the energy of the impurity is calculated as 
\be
\epsilon_{\tau\sigma}= \frac{\partial \varepsilon}{\partial \rho_{\tau \sigma}}|_{\rho_{\tau \sigma \rightarrow 0}}, 
\label{eneimpu}
\ee
where $\tau$ indicates neutrons $n$ or protons $p$ and $\sigma$ describe the up and down spin states $\uparrow$ and $\downarrow$.
One may thus derive the difference between the energies of the spin--up and of the spin--down proton impurities, scaled by the Fermi energy $E_F$ of the spin--up neutron matter, 
\be
\frac{\epsilon_{p \uparrow}-\epsilon_{p \downarrow}}{E_F}= \frac{4m(C_0^s-C_1^s)}{3\pi^2\hbar^2}k_F -  \frac{2m(C_0^T-C_1^T)}{5\pi^2\hbar^2}k_F^3.
\label{diffeimpu}
\ee
This difference depends only on time--odd quantities. 
This indeed defines the benchmarking strategy. One may for example constrain these quantities provided by a Skyrme--type functional by adjusting them on QMC results. In Ref. \cite{roggero}, these quantities calculated with several Skyrme functionals, BSK21 \cite{goriely}, SLy4, SkP, SkO’ \cite{sko}, SkM*, SAMi \cite{rocamaza2012}, and TOV-min \cite{erler}, were compared with results obtained with configuration--interaction QMC calculations \cite{muk,rog1,rog2}. The energy of the impurity in QMC calculations was extracted by computing the difference between the energy of the homogeneous system and the energy of the system containing the impurity. 
The QMC calculations were performed with a chiral EFT Hamiltonian \cite{ekstrom2}. 
Figure \ref{impurity} shows that the employed Skyrme functionals do not reproduce at all the microscopic results. It was shown in Ref. \cite{roggero} that the QMC results are well reproduced by an adjusted curve of the form
\be
\frac{\epsilon_{p \uparrow}-\epsilon_{p \downarrow}}{E_F}=A-\frac{B}{k_F|a_s|}-Ck_F r_s,
\label{fitformula}
\ee
with the numerical values $A=0.17 \pm 0.01$, $B=1.4 \pm 0.1$, and $C=0.101 \pm 0.001$, having used for the scattering length $a_s$ and the effective range $r_s$ the values -23.75 and 2.75 fm, respectively. This curve is shown in Fig. \ref{impurity} as a blue dot--dashed line.  

Another idea was proposed in Ref. \cite{dobaletter} based on the definition of a functional where a 
pure DFT procedure is followed. Since the Hohenberg-Kohn theorem is demostrated for a system placed in an external potential (which is necessary for binding systems which are not self--bound), the idea is to replace this external potential by the terms of the interaction that one wants to introduce to produce a given functional. Being the functional the heart of this procedure (and not the interaction terms), Dobaczewski calls these interaction terms generators of the functional.  The functional which one wishes to construct is not the exact functional, but a reasonable and well--defined functional which must be well justified in the range of energies and densities where it has to be applied. The idea is then to constrain this functional with 
{\it{ab initio}} benchmark data. Being this functional chosen of a given form and type, the variational problem may be written as a constrained problem with a given number $m$ of Lagrange multipliers $\lambda^i$, 
\be
\delta E= \delta \langle \Psi | H - \sum_{j=1}^m \lambda^j V_j | \Psi \rangle=0,
\label{minilagrange}
\ee
where $V_j$ are the generators of the functional. The solution of the minimization problem provides the energy of the system. It may then be required that this energy reproduces exact benchmark energies $E^{exact}(\lambda^j)$, 
\be
E^{exact}(\lambda^j)=T ( \rho^{exact}_{\lambda^j})+ \sum_{i=1}^m C^i V_i (\rho^{exact}_{\lambda^j}),
\label{exactenergy}
\ee
where the coupling constants $C^j$ of the functional  
are adjusted on the left--hand--side of Eq. (\ref{exactenergy}), which represents the exact result. 
The article \cite{dobaletter} contains a proof of principle of this procedure. The generators are chosen as the eight terms corresponding to the central and the tensor parts of the Skyrme interaction. To produce the exact energy, the Gogny D1S interaction is used. (By the way, the links between the functionals produced by the two interactions had been already investigated by using the DME in Refs. \cite{car2010,dob2012}). Both Skyrme and Gogny functionals contain the same types of spin--orbit and density--dependent terms. These contributions are left  unmodified.  Thus, the average value of the Gogny interaction (providing the exact energy) is computed only using the finite--range part of this interaction. Eight doubly magic nuclei are chosen and the eight Lagrange multipliers corresponding to the eight  Skyrme--type generators are varied in chosen ranges of values. The adjusted functional is called S1Sd. It was found in Ref. \cite{dobaletter} that the adjusted coupling constants reproduce in a very satisfactory way the exact energies calculated with the Gogny functional, which constitutes indeed a proof that such a procedure works well and may be applied in future using microscopic {\it{ab--initio}} data available for finite nuclei. 

Finally, the idea of constructing an {\it{ab--initio}} functional directly derived from a microscopic Hamiltonian is extensively discussed in the literature and is based on the use of the DME \cite{car2008,car2010,car2010-2,doba2010,gebre2010,gebre2011,stoi2010,erler2010,rai2011}. The introduction of this procedure was motivated in the work of Negele and Vautherin \cite{negele72,negele75} by the will of establishing a connection between the Skyrme functional that had been recently proposed at that time by Vautherin and Brink and the microscopic nucleon--nucleon force. The density is expanded around the local density so that the expectation value of a given potential is a functional of the local density plus gradient terms.  
A generalization of this is illustrated in Ref. \cite{drut} where the formulation is developed using orbital--based functionals. 

The DME was for example used in Refs. \cite{holtkaiser,kaiser2011,kaiser2012} to derive a nuclear EDF from chiral two-- and three--nucleon interactions. 

The recent work published in Ref. \cite{navarro} reports on the construction of a DME--obtained functional using local chiral potentials and on its calibration on several properties of nuclei. The authors of Ref. \cite{stoi2010} had indeed presented several numerical drawbacks and instabilities in the practictal use of a DME--generated  functional that was obtained from a (non regulated) chiral potential. The recent study of Ref. \cite{navarro} handles these problems and overcome them. Several microscopic functionals are obtained using local chiral potentials from LO up to N2LO with (and without) the inclusion of three--body forces and with (and without) the inclusion of the $\Delta$ excitations \cite{piarulli1,piarulli2}. The DME of local chiral potentials of Ref. \cite{dyh} is employed. The same protocol as for UNEDF2 is used for the optimization. Numerous applications are discussed in the article 
 to check the quality of the functional for the study of 
matter properties, masses, shell--structure, and deformations. 

\begin{figure}[tb]
\begin{center}
\begin{minipage}[t]{12 cm}
\epsfig{file=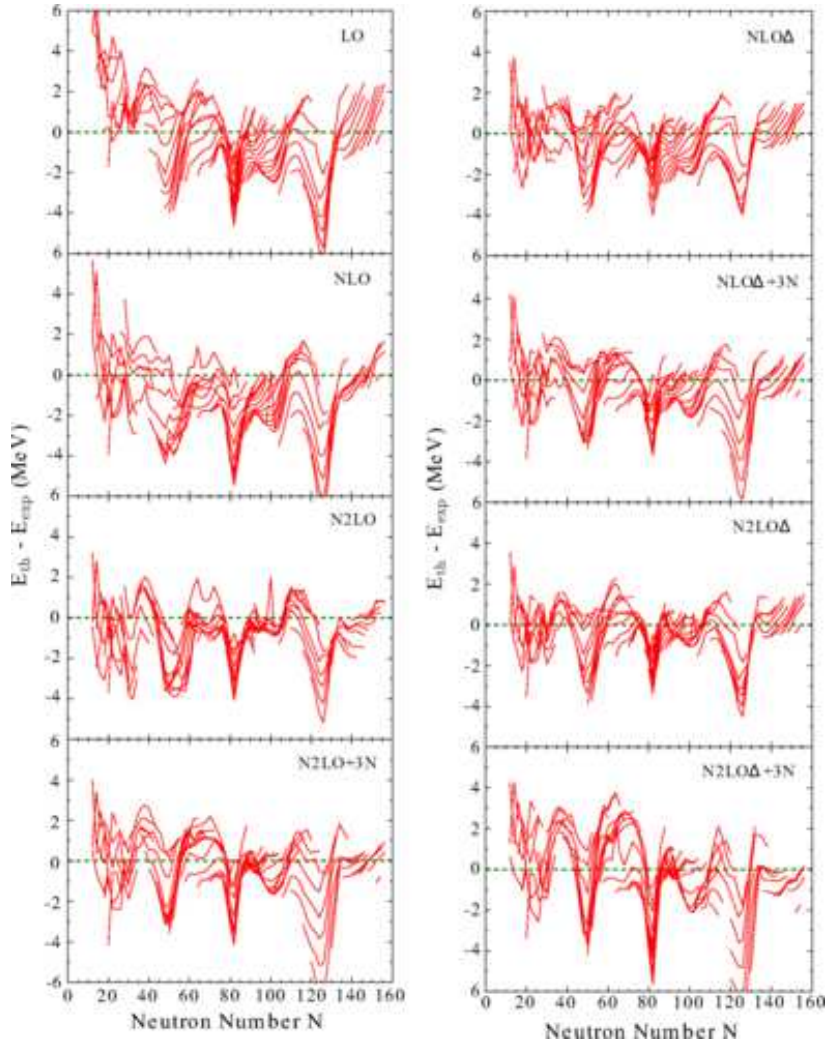,scale=0.78}
\end{minipage}
\begin{minipage}[t]{16.5 cm}
\caption{Difference between theoretical and experimental binding energies. Experimental masses are extracted from \cite{mass1,mass2}. Panels on the left (right) refer to the results obtained without (with) the inclusion of $\Delta$ excitations. Taken from Ref. \cite{navarro}. }
\label{massabinitio}
\end{minipage}
\end{center}
\end{figure} 

Figure \ref{massabinitio} shows for example the difference between the theoretical and the experimental binding energies obtained with the different obtained potentials. The LO case is actually not very different from UNEDF2. Indeed, one observes at LO and NLO that the differences between theoretical and experimental masses are larger for lighter nuclei. This is due to the fact that, as in UNEDF2, the center--of--mass correction is not included. Going to higher orders, this effect seems to be corrected but the reason for this attenuation remains unclear since the center--of--mass correction is omitted in all cases. 
Interestingly, at least when only two--body forces are included, one may observe that the quality of the results is improved going to higher orders.

\section{Concluding remarks \label{conclu}}
This review focused on some illustrations of density functionals tailored for BMF models. This discussion represented the central part 
of the manuscript where examples of BMF functionals were provided, designed for infinite nuclear matter. 
Before this, an introductory part was dedicated to traditional phenomenological functionals whose coupling constants are adjusted with MF calculations on properties of finite nuclei, such as masses and radii, and on properties of infinite matter, such as the equilibrium point of symmetric matter. Several directions have been suggested over the years to extend and generalize these functionals following different criteria. Some of such directions were presented here, in particular ideas for dealing with density--dependent terms, for introducing higher--order derivatives of the density, or for introducing functionals in the genuine spirit of the DFT. 
After having underlined the importance of going beyond traditional functionals and of constructing functionals tailored for BMF calculations (where overcounting of correlations, instabilities, and ultraviolet divergences are overcome) some examples of these BMF--designed functionals were discussed. In particular, a work was presented where second--order (within the MBPT) EOSs of infinite matter were computed with Skyrme--type interactions. Regularization and renormalization, as well as a first step towards the construction of a power counting in EDF theories were illustrated. First simplified applications to finite nuclei were also presented. 
This led to the last part of the manuscript dedicated to EDFs inspired by EFTs. In particular, different possible directions were explored. Constructing functionals with less free empirical parameters and with constraints coming from the low--density regime and/or the unitary regime, constructing functionals whose form in powers of the Fermi momentum is inspired by EFTs, and using pseudodata produced by  {\it{ab--initio}} models for adjusting the parameters are the main directions which were discussed here for rendering EDFs less empirical. The DME as a way to build functionals starting from a microscopic potential was also discussed. 

Of course, this is not (and cannot be) an exhaustive review. Other studies have been carried out in the community, which are not mentioned here. The illustrations and studies that have been presented deal with some specific selected aspects on which the accent was put. 

Constructing functionals for BMF models and constructing bridges with {\it{ab--initio}} models and EFTs are challenging objectives which are extremely important to be attained in the present context of nuclear theory. It turns out that sophisticated BMF models are available and represent powerful formal and numerical tools for addressing complex correlations and complicated phenomena in structure and reaction theories. On the other side, EDFs constitute nowadays a unique and the most versatile way for describing nuclei in both structure and reaction studies, practically covering the whole nuclear chart. However, current EDFs can be coherently utilized only for MF applications, 
owing to the procedures utilized to build them. The adjustment of the parameters carried out at the MF level and the occurrence of instabilities and divergences make 
questionnable their use in BMF models.
Generalizations of them or {\it{ad--hoc}} procedures (such as subtractions) are necessary for a proper use in a BMF framework. Also, current EDFs are essentially phenomenological and this may have a strong impact for example on their predictive power in the regions of the nuclear chart which are still far from observation. 
At the same time, the last years have seen a tremendous progress and remarkable achievements in the development and in the applications of {\it{ab--initio}} models to nuclei, not only the lightest ones and not only closed--shell nuclei \cite{hagenco,hergert,carlsonrev,barrett,lapoux,gebreru,gebreru1,forssen,genna,quaglio,capel,fukui}. 
Merging competences and expertise coming from both EDF and {\it{ab--initio}} practitioners may definitely bring in a close future to important advances in the construction of a coherent, versatile, and fully microscopic description of nuclei for structure and reaction studies, as well as, in nuclear astrophysics, for the description of nuclear systems located in the crust of neutron stars.

\section{Aknowledgements \label{akno}}
I would like to acknowledge the work (mentioned or quoted in this review) that was carried out with past and present collaborators, Marta Anguiano, J\'er\'emy Bonnard, Antoine Boulet, Francesco Catara, Gianluca Col\`o, Giampaolo Co', Viviana De Donno, Jonathan Engel, Danilo Gambacurta, Bira van Kolck, Denis Lacroix, Antonio Lallena, Haozhao Liang, Miguel Moreno-Torres, Xavier Roca-Maza, Michelangelo Sambataro, Michael Urban, Nguyen Van Giai, Olivier Vasseur, Jerry Yang. Some of the directions and ideas explored together have been illustrated in this review. 

I thank Denis Lacroix for a careful reading of this manuscript. 

Part of the work presented here has received funding from the European Union Horizon 2020 research and innovation program under Grant No. 654002.

\end{document}